\documentclass[traditabstract]{aa}  
\usepackage{txfonts}
\usepackage{epstopdf}
\usepackage[switch]{lineno}
\usepackage{graphicx,longtable,lscape,natbib,amssymb,amssymb,amsmath}
\bibpunct{(}{)}{;}{a}{}{,} 
\newcommand{\ltsima} {$\; \buildrel < \over \sim \;$}  
\newcommand{\gtsima} {$\; \buildrel > \over \sim \;$}  
\newcommand{\lta} {\lower.5ex\hbox{\ltsima}}  
\newcommand{\gta} {\lower.5ex\hbox{\gtsima}}  
\newcommand{\ha} {H$\alpha$}  
\newcommand{\hb} {H$\beta$}  

\newcommand{\ergs}{\>{\rm erg}\,{\rm s}^{-1}}

\newcommand{\kms}{$\rm{\,km \,s}^{-1}$}

\newcommand{\forb}[2]{\mbox{$[{\rm #1\, #2}]$}}
\newcommand{\oiii}{\forb{O}{III}}
\newcommand{\oi}{\forb{O}{I}\,}
\newcommand{\sii}{\forb{S}{II}\,}
\newcommand{\nii}{\forb{N}{II}\,}

\usepackage{color}

\begin{document}

\title{The MURALES survey. II.} \subtitle{Presentation of  MUSE  observations  of 20  3C low-z  radio
galaxies and first results.}

\author{Barbara Balmaverde\inst{1} 
		\and Alessandro Capetti\inst{1}
		\and Alessandro	Marconi\inst{2,3}
		\and Giacomo Venturi\inst{3,15}
		\and M. Chiaberge\inst{4,5}
		\and R.D. Baldi\inst{6} 
		\and S. Baum\inst{8,13}
		\and R. Gilli\inst{7}
		\and P. Grandi\inst{7}
		\and E. Meyer\inst{11}
		\and G. Miley\inst{9}
		\and C. O$'$Dea\inst{8,12}
		\and W. Sparks\inst{14}
		\and E. Torresi\inst{7} 
		\and G. Tremblay\inst{10}}
\institute {INAF - Osservatorio Astrofisico di Torino, Via Osservatorio 20, I-10025 Pino Torinese, Italy
\and Dipartimento di Fisica e Astronomia, Universit\`a di Firenze, via G. Sansone 1, 50019 Sesto Fiorentino (Firenze), Italy
 \and INAF - Osservatorio Astrofisico di Arcetri, Largo Enrico Fermi 5, I-50125 Firenze,Italy
 \and Space Telescope Science Institute, 3700 San Martin Dr., Baltimore, MD 21210, USA
\and Johns Hopkins University, 3400 N. Charles Street, Baltimore, MD 21218, USA
 \and Department of Physics and Astronomy, University of Southampton, Highfield, SO17 1BJ, UK
\and INAF - Osservatorio di Astrofisica e Scienza dello Spazio di Bologna, via Gobetti 93/3, 40129 Bologna, Italy
\and Department of Physics and Astronomy, University of Manitoba, Winnipeg, MB R3T 2N2, Canada
\and Leiden Observatory, Leiden University, PO Box 9513, NL-2300 RA, Leiden, the Netherlands
\and Harvard-Smithsonian Center for Astrophysics, 60 Garden St., Cambridge, MA 02138, USA
\and University of Maryland Baltimore County, 1000 Hilltop Circle, Baltimore, MD 21250, USA
\and School of Physics \& Astronomy, Rochester Institute of Technology, Rochester, NY 14623
\and Carlson Center for Imaging Science, Rochester Institute of Technology, Rochester, NY 14623
\and SETI Institute, 189 N. Bernado Ave Mountain View,CA 94043
\and Instituto de Astrofisica, Facultad de Fisica, Pontificia Universidad Catolica de Chile, Casilla 306, Santiago 22, Chile
}
\offprints{balmaverde@oato.inaf.it} 

\date{} 

\abstract{We present observations of a complete sub-sample of 20 radio
  galaxies from the Third Cambridge Catalog (3C) with redshift $<$0.3
  obtained from VLT/MUSE optical integral field spectrograph. These
  data have been obtained as part of the survey MURALES  (a MUse RAdio
  Loud Emission line Snapshot survey) with the main goal of exploring
  the Active Galactic Nuclei (AGN) feedback process in a sizeable sample of the most powerful
  radio sources at low redshift. We present the data analysis and, for
  each source, the resulting emission line images and the 2D gas
  velocity field. Thanks to their unprecedented depth (the median
    3$\sigma$ surface brightness limit in the emission line maps is
    6$\times$10$^{-18}$ erg s$^{-1}$ cm$^{-2}$ arcsec$^{-2}$), these
  observations reveal emission line structures extending to several
  tens of kiloparsec in most objects. In nine sources the gas velocity
   shows ordered rotation, but in the other cases it is
  highly complex. 3C sources show a connection between radio
  morphology and emission line properties. Whereas, in three of the four
  Fanaroff and Riley Class I radio galaxies (FR~Is), the line
  emission regions are compact, $\sim$ 1 kpc in size; in all but one
  of the Class II radiogalaxies FR~IIs, we detected large scale
  structures of ionized gas with a median extent of 17 kpc. Among the
  FR~IIs, those of high and low excitation show extended gas
  structures with similar morphological properties, suggesting that
  they both inhabit regions characterized by a rich gaseous
  environment on kpc scale.}


\titlerunning{The MURALES survey} 
\authorrunning{B. Balmaverde et al.}
 \maketitle

\section{Introduction}
\label{intro}

Radio galaxies are among the most energetic manifestations of active
galactic nuclei and harbor the most massive black holes (SMBHs) in the
Universe, typically hosted in the brightest galaxies at center of
clusters or groups. They are therefore extraordinarily relevant to
address important unknowns related to the interaction between SMBHs
and their environment \citep{gitti12}. Significant progress in understanding the
fueling and evolution of the activity of radio loud AGNs, the
triggering process of the radio emission, and its impact on the
environment, has been made by studying the third Cambridge catalog
of radio galaxies (3C, \citealt{spinrad85}). The 3C  is the
premiere statistically complete sample of powerful radio galaxies; it
includes  variety of extended radio morphologies, optical
classes, and environmental properties.

The study of radio-loud AGN has become particularly important due to
their role in the so-called feedback process, that is the exchange of
matter and energy between AGN, their host galaxies,
and clusters of galaxies. The evidence of kinetic AGN feedback mode is
often witnessed in local radio-galaxies, showing 
the presence of cavities inflated by the radio emitting gas in the X-ray images.  However,
little is known about the coupling between radio-jets and ionized gas,
whether the jets are able to accelerate the gas above the host escape
velocity \citep{mcnamara07}, and we also lack clear observational
evidence on whether jets enhance or quench star formation (positive or
negative feedback, e.g., \citealt{fabian12}). Furthermore, the mechanical
luminosity released by the AGN is not well-constrained because the
cavity expansion speed is estimated using indirect and model-dependent
approaches. Finally, the radiative output from the AGN can produce
fast outflows of ionized gas, which also affect the properties of the
ambient medium (e.g., \citealt{wylezalek16,carniani16,cresci18}).

A study of the optical-line-emitting gas properties in low redshift
radio galaxies (mostly from the 3C) has been performed by
\citet{baum88}, \citet{baum89}, and extended to higher redshift by
\citet{mccarthy95}.  The analysis of narrow band images centered on
the \ha\ +[N~II] or \oiii\ emission lines, reveals that extended
optical-line-emitting gas is common in powerful radio galaxies. The
ionized gas is often distributed along filamentary structures on
scales of 40-100 kpc, which eventually connect the host radio galaxy to
possible companions. \citet{baum88} noted that the extended
emission-line gas is preferentially observed along the radio source
axis: this suggests that the distribution and/or the ionization of this
gas is influenced by the radio source. It has also been  suggested
that tidal interactions and mergers could be related to the formation
of a radio jet, since many radio galaxies show morphological and/or
kinematic evidences for a recent encounter. However, no firm
conclusion can be drawn about the origin of the extended optical
 line emitting gas: it can not be excluded that the emission line gas
has cooled out from the hot intergalactic medium or that it has an
internal origin (i.e., a merger has stirred up and redistributed the
gas that was already present in the host galaxy before the
interaction).

To investigate the mechanism of ionization in the extended emitting
gas regions, \citet{baum90,baum92} used diagnostic diagrams (BPT
diagrams, \citealt{baldwin81,veilleux87}) based on emission line
ratios.  They found line-ratio changes within individual sources along
the elongated structures, but much higher variations from source to
source, suggesting different ionization mechanisms.  Separating the
objects according to their kinematical properties in rotators, calm
non rotators and violent rotators, they found that in rotators (mostly
FR~II, \citealt{fanaroff74}) the forbidden lines (as \nii, \sii, \oi) appear to be weak
compared to \ha. This indicates that photoionization from the nuclear
continuum is the dominant ionization mechanism. Instead, in calm non
rotators (mostly FR~I) they usually observed high [N~II] to \ha\,
ratios, which they interpreted to be produced by heating from
cosmic rays. Other models can not be ruled out, such as clouds that
condense out of the hot (10$^7$ K) gas and are ionized by soft X-ray
photons, or gas with super-solar abundances, photoionized by the AGN.

The emission line regions in radio galaxies have also been extensively
studied with the Hubble Space Telescope (HST). The Wide Field
Planetary Camera-2 images of 80 3CR radio sources up to z=1.4
\citep{privon08} show that the radio and optical emission-line
structures present a weak alignment at low redshift (z $<$ 0.6), which
becomes stronger at higher redshift. They found a trend for the
emission-line nebulae to be larger and more luminous with increasing
redshift and/or radio power. \citet{baldi19} present Advanced Camera
for Surveys emission line images of 19 low z 3C radio galaxies. They
generally show extended [O~III] emission, a large [O~III]/\ha\ scatter
across the galaxies, and a radio-line alignment effect. The line
morphologies of high and low excitation galaxies (HEG and LEG,
  respectively, \citealt{hine79}) are different; the former is
brighter and more extended.

Integral field spectroscopic data provide us with a unique
  opportunity to study of the impact of AGN feedback in galaxies,
  mapping the distribution and the kinematics of the ionized gas with
  respect to the radio jet.  \citet{couto17} observed the radio galaxy
  3C~033 with GEMINI-GMOS/IFU (over a field of view of $4\times6$
  kpc$^2$) revealing complex motions, with signatures of inflow and
  outflows.  In NGC~3393, a nearby Seyfert 2 galaxy with nuclear
  radio jets and a nuclear bar, \citet{finlez18} with
  the same spectrograph found motions of rotation and outflows, traced
  by narrow emission line and broad component along the radio lobes
  and perpendicularly to them. With respect to these studies, the
  wider MUSE field of view (1$^\prime\times1^\prime$) allows us to
  trace the distribution of the ionized emitting gas at distances of
  $\sim$110-270 kpc (at z=0.1 and 0.3, respectively), exploring the
  effect of AGN feedback at different spatial scales.
 
 The physical processes that distribute the feedback energy of the AGN
 shaping the ICM is a subject of intense theoretical efforts.  Up to
 date cosmological hydrodynamic simulation of a galaxy cluster succeed
 in matching observations \citep{tremmel19}. However, the
 micro-physical processes that distribute the feedback energy remain
 uncertain and a subject of intense theoretical efforts.  Multi
 wavelength observations have started to unveil the complex
 multi-phase structure of early-type elliptical galaxies (ETGs) at the
 center of groups and clusters. In optical, extended \ha\ and
 \nii\ filaments up to 10 kpc from the center, are tightly correlated
 with soft X-ray and cold molecular gas \citep{mcdonald10,russsel19}.
 Theoretical models predict that, at least in cool core clusters, cold
 clouds can condense out of a hot, turbulent and rarefied atmosphere
 and rain down to the black hole \citep{gaspari17,voit15}.

We have started MURALES (MUse RAdio Loud Emission lines Snapshot)
project, a program aimed at observing the 3C radio sources with the
integral field spectrograph MUSE at the VLT \citep{bacon10}. Our main
goals are to study the feedback process in a sample of the most
powerful radio sources at low redshift, to constrain the coupling
between the radio source and the warm gas, to probe the fueling
process, and to estimate the net effect of the feedback on star
formation.

The MUSE data will enable us to 1) obtain deep line emission images
and to compare them with the X-ray structures, exploring the spatial
link between the hot and warm ionized ISM phases, and with the radio
outflows, 2) derive spatially resolved emission lines ratios maps and
explore the gas physical conditions, 3) map and characterize the full
2D ionized gas velocity field, 4) obtain the 2D stellar velocity field
that will be compared with that observed in the gaseous component, 5)
detect star forming regions, in search of positive feedback with young
stars form along the jets path and/or around the radio lobes.

An example of the capabilities of MUSE comes from the observations of
3C~317, a radio-galaxy located at the center of the Abell cluster
A2052 \citep{balmaverde18a}. A complex network of emission lines
filaments enshrouds the whole northern cavity. The ionized gas
kinematics show the hallmarks of a shell expansion, with both blue-
and red-shifted regions, with a velocity of $\sim$ 250 km/s (a factor
of $\sim 2$ lower than previous indirect estimates based on X-ray
data) leading to an estimate of the cavity age of 1.1$\times 10^7$
years. We did not detect any star-forming regions from the emission
line ratios.

With the MUSE observations we also found a dual AGN associated to
3C~459 \citep{balmaverde18} whose host shows the signatures of a
recent merger, i.e., disturbed morphology and a young stellar
population. The line emission images show two peaks separated by
$\sim$ 4 kpc with radically different line profiles and ratios, and a
velocity offset of $\sim$ 300 km/s. The secondary AGN has properties
typical of a highly obscured QSO, heavily buried at the center of the
merging galaxies, producing a high ionization bicone extending more
than 70 kpc.

Here we present the results of the observations of the first 20 3C
sources obtained in Period 99. The paper is organized as follows: in
Sect. 2 we present the sample observed, provide an observation log,
and describe the data reduction. In Sect. 3 we present the resulting
emission line images, line ratio maps, and 2D velocity fields. We also
provide a description of the individual sources. In Sect 4 we study
the spectra of the extended emission line regions. The results are
discussed in Sect. 5 and summarized in Sect. 6.

We adopt the following set of cosmological parameters: $H_{\rm o}=69.7$
\kms\ Mpc$^{-1}$ and $\Omega_m$=0.286 \citep{bennett14}.

\begin{table*}
\caption{Main properties of 3C sub sample observed with MUSE and observations log}
\begin{tabular}{l l c l c c c l c l r}
\hline
Name      &  z     & FR  & Class & L$_{178}$             & L.A.S radio & r       & Obs. date & Seeing   &  Weather & Depth\\
          &        &     &       &[erg s$^{-1}$ Hz$^{-1}$]&  [kpc] &  [kpc] &            & [$\arcsec$]   &  &  \\
\hline
3C~015    & 0.073 & I    & LEG & 33.30 &   70 &    1.5  &    Jun 30 2017   &  0.65   &  TN           & 5.6 \\ 
3C~017    & 0.220 & II   & BLO & 34.44 &   54 &   14.9  &   Jul~ 20 2017  &  0.49   &   TN/CL     & 5.9 \\  
3C~018    & 0.188 & II   & BLO & 34.27 &  178 &   24.8  &    Jun 30 2017  &  0.53   &  TN          & 5.9 \\ 
3C~029    & 0.045 & I    & LEG & 32.84 &  125 &    0.9  &   Jul~ 20 2017   &  0.51   &   TN/CL    &  3.2 \\ 
3C~033    & 0.060 & II   & HEG & 33.65 &  288 &   11.2  &    Jun 30 2017  &  0.63   &   TN         & 7.4 \\ 
3C~040    & 0.018 & I    & LEG & 32.29 &  440 &    1.2  &   Jul~ 22 2017   &  0.40   &    TN/CL.  & 7.5\\ 
3C~063    & 0.175 & II   & HEG & 34.21 &   56 &   38.0  &   Jul~ 21 2017   &  0.49   &    TN        & 7.5 \\ 
3C~318.1  & 0.045 & --   &  -- & 32.72 &      &         &     Jun 22 2017          &  1.38   &  CL          & 6.3 \\ 
3C~327    & 0.105 & II   & HEG & 33.98 &  469 &   19.5  &    Jun 30 2017  &  0.70   &   TN         & 6.3 \\  
3C~348    & 0.155 & I    & ELEG& 35.35 &  153 &   34.0  &   Jul~ 20 2017 &  1.76   &   CL         & 3.5 \\
3C~353    & 0.030 & II   & LEG & 33.69 &  111 &   17.2  &    Jun 29 2017   &  1.30   &   CL         & 4.1 \\ 
3C~386    & 0.017 & II   &  -- & 32.18 &   77 &   11.1  &    Jun 03 2017        &  0.61   &  TN         & 7.3 \\
3C~403    & 0.059 & II   & HEG & 33.16 &  101 &    8.3  &    Jun 30 2017   &  0.54   &   TN         & 5.5  \\   
3C~403.1  & 0.055 & II   & LEG & 32.98 &  236 &    5.6  &    Jun 30 2017   &  0.80   &    TN        & 4.5 \\  
3C~424    & 0.127 & II   & LEG & 33.78 &   28 &   32.9  &   Jul~ 01 2017   &  0.98   &    TN         & 1.2 \\  
3C~442    & 0.026 & II   & LEG & 32.39 &  286 &    3.8  &    Jun 30 2017   &  0.61   &    TN         & 11.3\\  
3C~445    & 0.056 & II   & BLO & 33.26 &  483 &   18.5  &   Jul~ 01 2017  &  1.48   &    TN/CL    & 5.2\\  
3C~456    & 0.233 & II   & HEG & 34.23 &   24 &         &    Jun 30 2017     &  1.27   &    TN          & 8.8 \\   
3C~458    & 0.289 & II   & HEG & 34.58 &  943 &  111.3  &   Jul~ 22 2017 &  0.50   &    TN/CL     & 5.6\\   
3C~459    & 0.220 & II   & BLO & 34.55 &   31 &   76.0  &   Jul~ 22 2017   &  0.43   &    TN/CL     & 19.3\\   
\hline
\end{tabular}
\\ Column description: (1) source name; (2) redshift; (3 and 4) FR and
excitation class; (5) radio luminosity at 178 MHz from
\citet{spinrad85}; (6) largest angular size of the radio source; (7)
largest distance of emission line detection in kpc units; (8) date of
the observation; (9) mean seeing in the V band at zenith during the
observation; (10) sky conditions during the observations: (TN)
  thin cirrus clouds (CL) clear night; (11) surface brightness limit
  of emission lines at 3$\sigma$ in units of 10$^{-18}$ erg s$^{-1}$
  cm$^{-2}$ arcsec$^{-2}$.
\label{tab1} 
\end{table*}

\begin{figure*}  
\centering{ 
\includegraphics[width=2.\columnwidth]{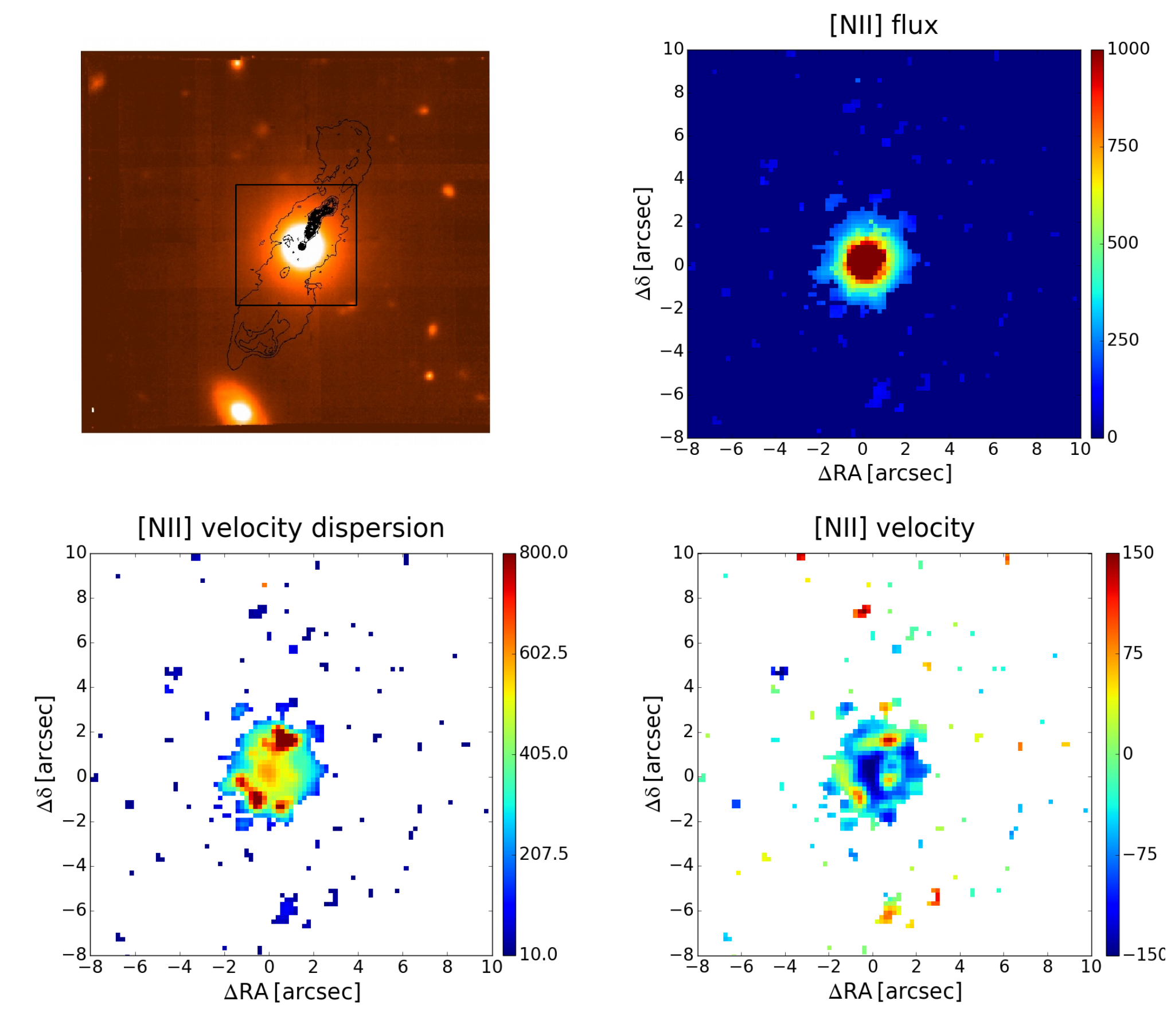}
\caption{3C~015,  FR~I/LEG, 1$\arcsec$ = 1.40 kpc. Top left: radio contours
  (black) overlaid onto the Muse  optical continuum image  in the 5800 and
  6250 \AA\ rest frame range. The size of the image is the whole MUSE
  field of view, 1$^\prime \times 1^\prime$. Top right: [N~II]
  emission line image extracted from the black square marked in the
  left panel. 
Surface brightness is in
  $10^{-18} {\rm erg}\,{\rm s}^{-1}\,{\rm  cm}^{-2} {\rm arcsec}^{-2}$. Bottom: velocity field and velocity dispersion
  for the  \nii\ line . Velocities are in \kms\ units.}}
\label{3C015}
\end{figure*}  

\begin{figure*}  
\centering{ 
\includegraphics[width=2.\columnwidth]{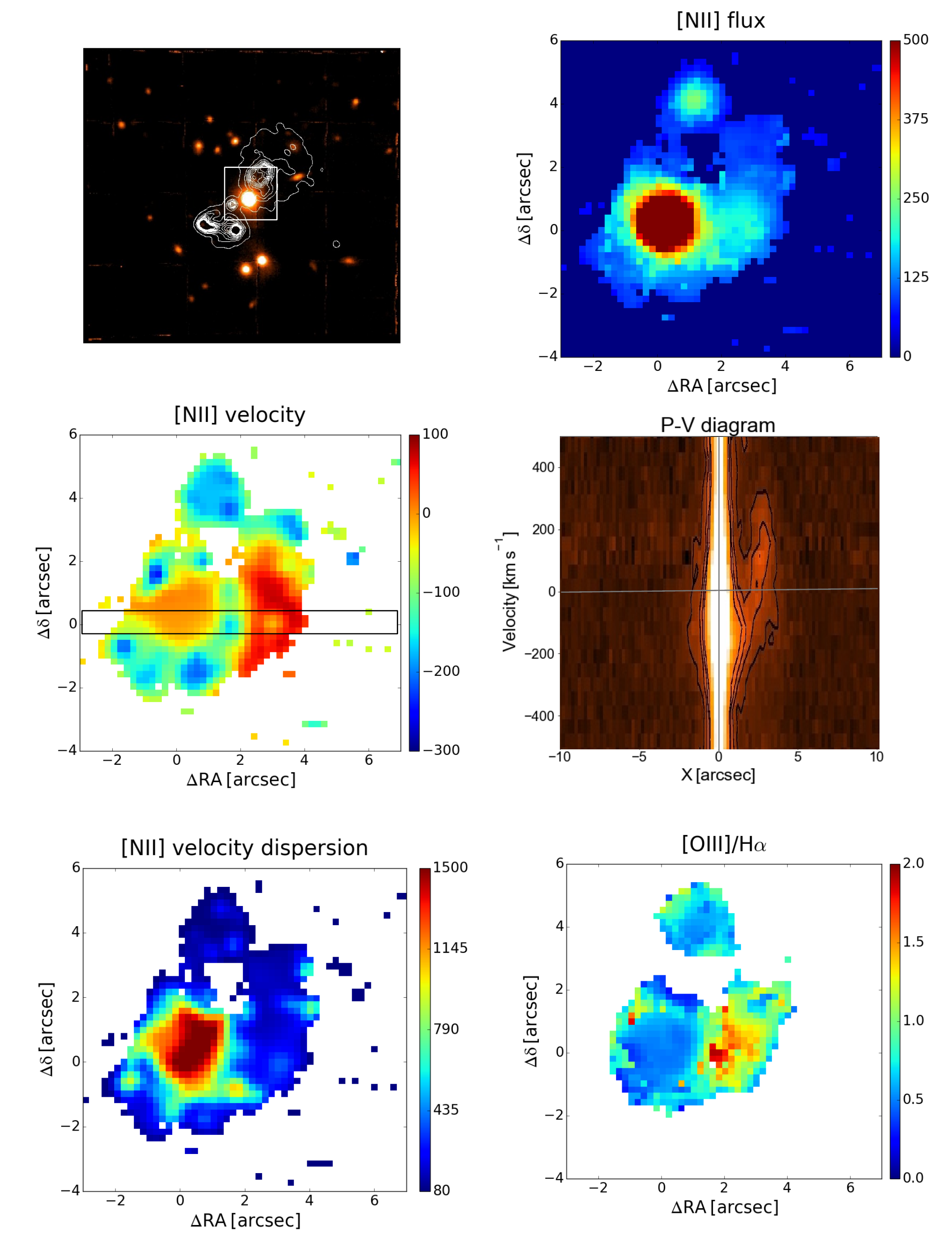}
\caption{3C~017, FR~II/BLO, 1$\arcsec$ = 3.58 kpc.  Top left: radio
  contours (in white) overlaid onto the Muse  optical continuum image  over
  the MUSE field of view. Top right: [N~II] emission line image
  extracted from the white square marked in the left panel. Middle
  left: gas velocity obtained from the [N~II] line. Middle right:
  position-velocity diagram extracted from the synthetic slit shown
  overlaid onto the velocity field. Bottom: velocity dispersion and
  \nii\ map of the \oiii/\ha\ line ratio.  Surface brightness is in
  $10^{-18} {\rm erg}\,{\rm s}^{-1}\,{\rm cm}^{-2} {\rm arcsec}^{-2}$,
  velocities are in \kms\ units.}}
\label{3C017}
\end{figure*}  

\begin{figure*}  
\centering{ 
\includegraphics[width=2.\columnwidth]{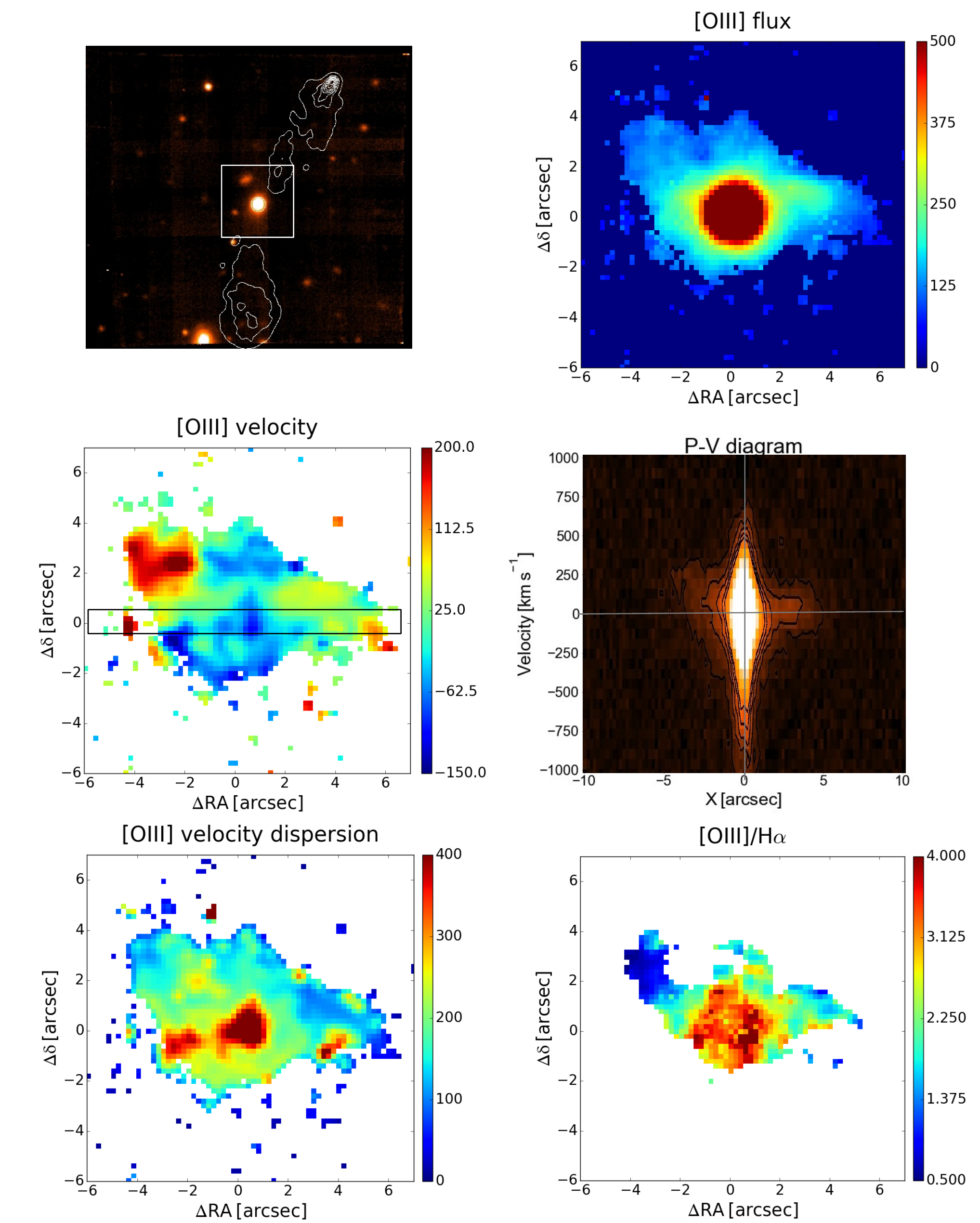}
\caption{3C~018, FR~II/BLO, 1$\arcsec$ = 3.17 kpc. Top: radio contours
  overlaid onto the Muse  optical continuum image  and [O~III] emission line
  image extracted from the white square in the top left panel.  Middle: velocity field from the [O III] line and
  position-velocity diagram extracted from the synthetic slit shown
  overlaid onto the velocity field (represented by the region centered
  on the nucleus, has a width of 5 pixels and oriented at an angle of
  0$^\circ$ measured from the X axis).  Bottom: velocity dispersion
  and \oiii/\ha\ ratio.}  }
\label{3C018}
\end{figure*}  

\begin{figure*}  
\centering{ 
\includegraphics[width=2.\columnwidth]{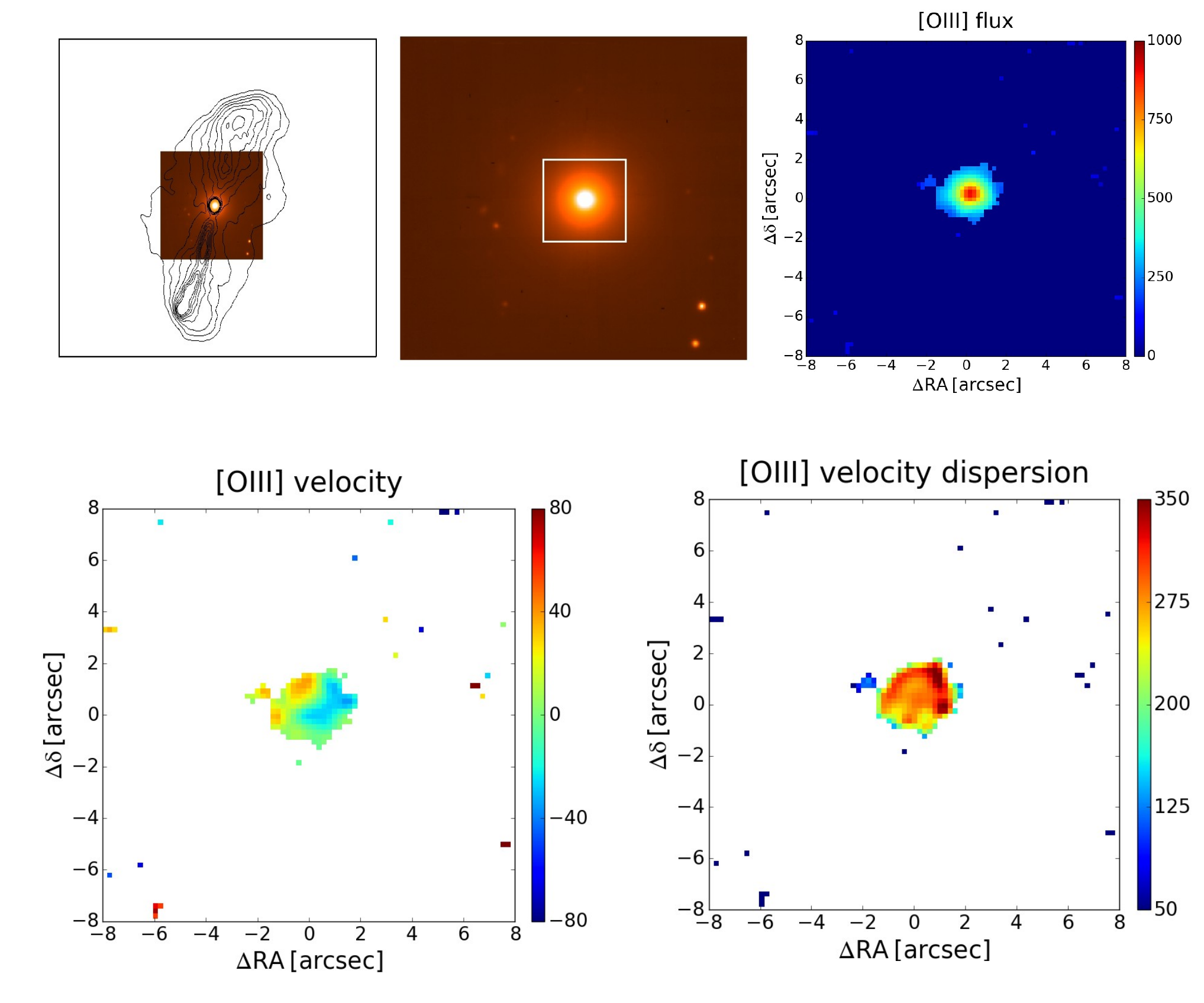}
\caption{3C~029, FR~I/LEG, 1$\arcsec$ = 0.89 kpc. Top left: radio contours
  overlaid onto the Muse optical continuum image . Top center: Muse optical continuum image . Top right: [O~III]
  emission line image extracted from the white square in the top center panel. Bottom: velocity field from the \oiii\ line and
  velocity dispersion. 
Surface brightness is in $10^{-18} {\rm erg}\,{\rm
    s}^{-1}\,{\rm cm}^{-2} {\rm arcsec}^{-2}$, velocities are in
  \kms\ units.}}
\label{3C029}
\end{figure*}  

\begin{figure*}  
\centering{ 
\includegraphics[width=2.\columnwidth]{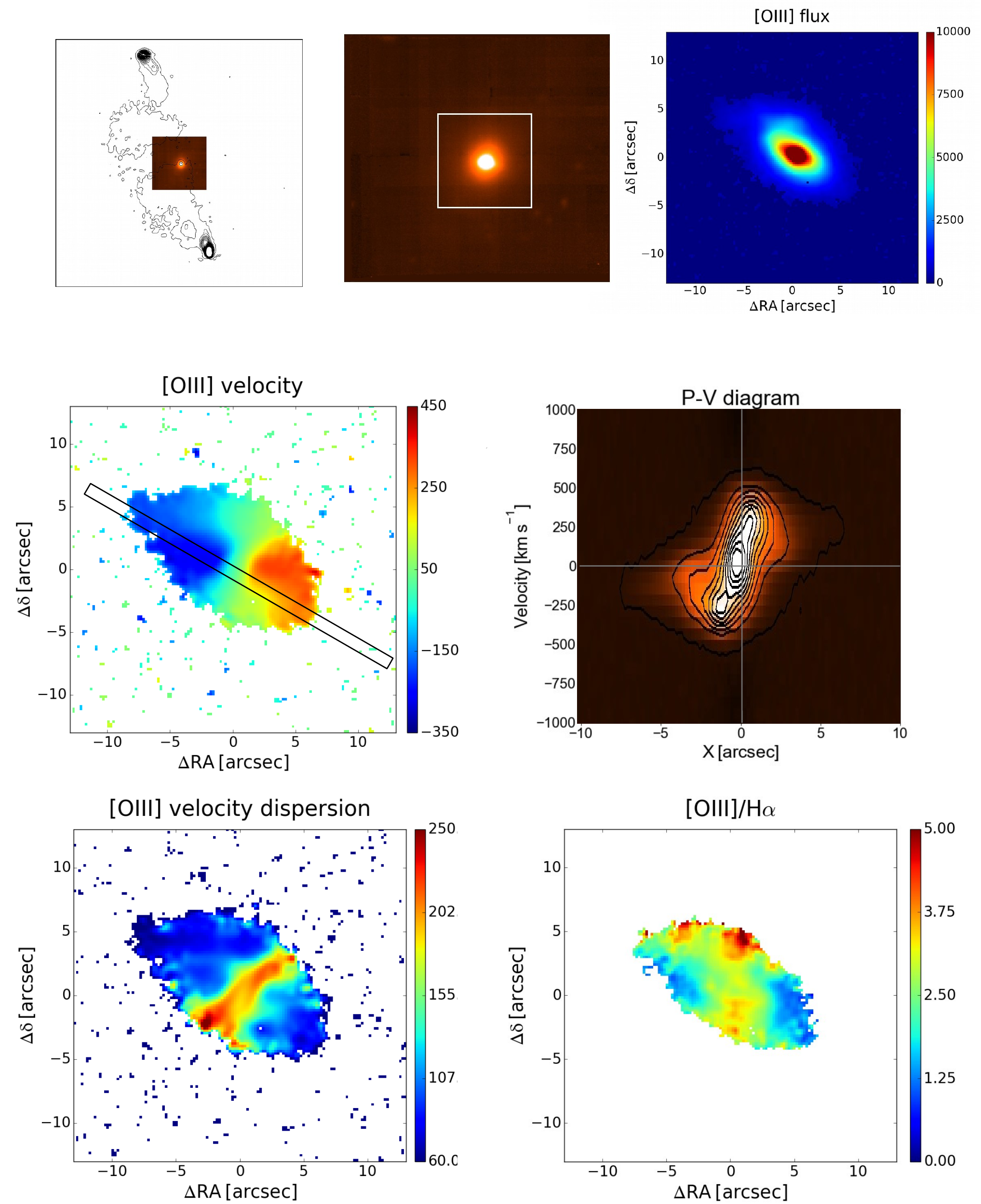}
\caption{3C~033, FR~II/HEG, 1$\arcsec$ = 1.17 kpc. Top left: radio contours
  overlaid onto the Muse optical continuum image . Top center: Muse  optical continuum image  .  Top right: [O~III]
  emission line image extracted from the white square in the top center panel.  Middle: velocity field from the [O III] line and
  position-velocity diagram extracted from the synthetic slit shown
  overlaid onto the velocity field (the slit is centered on the nucleus, has a width of 5 pixels and it is oriented at an angle of -60$^\circ$ from the X axis.).  Bottom: velocity dispersion
  and \oiii/\ha\ ratio. }}
\label{3C033}
\end{figure*}  

\begin{figure*}  
\centering{ 
\includegraphics[width=2.\columnwidth]{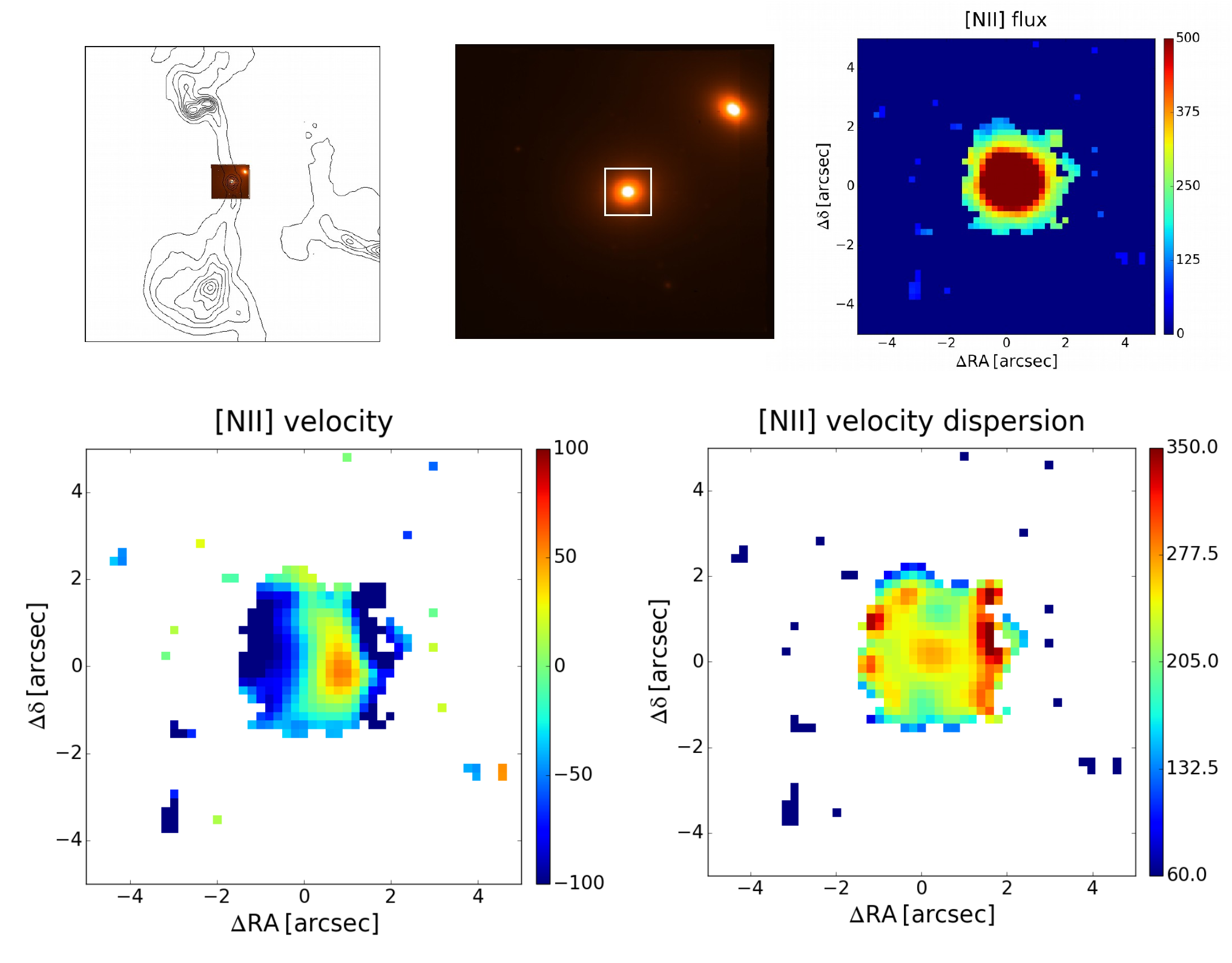}
\caption{3C~040, FR~I/LEG, 1$\arcsec$ = 0.37 kpc.  Top left: radio contours
  overlaid onto the Muse  optical continuum image  .Top center: Muse optical continuum image  . Top right: [N~II]
  emission line image extracted from the white square in the top center panel.  Bottom: velocity field from the \nii\ line and
 \nii\ velocity dispersion.}}
\label{3C040}
\end{figure*}  

\begin{figure*}  
\centering{ 
\includegraphics[width=2.\columnwidth]{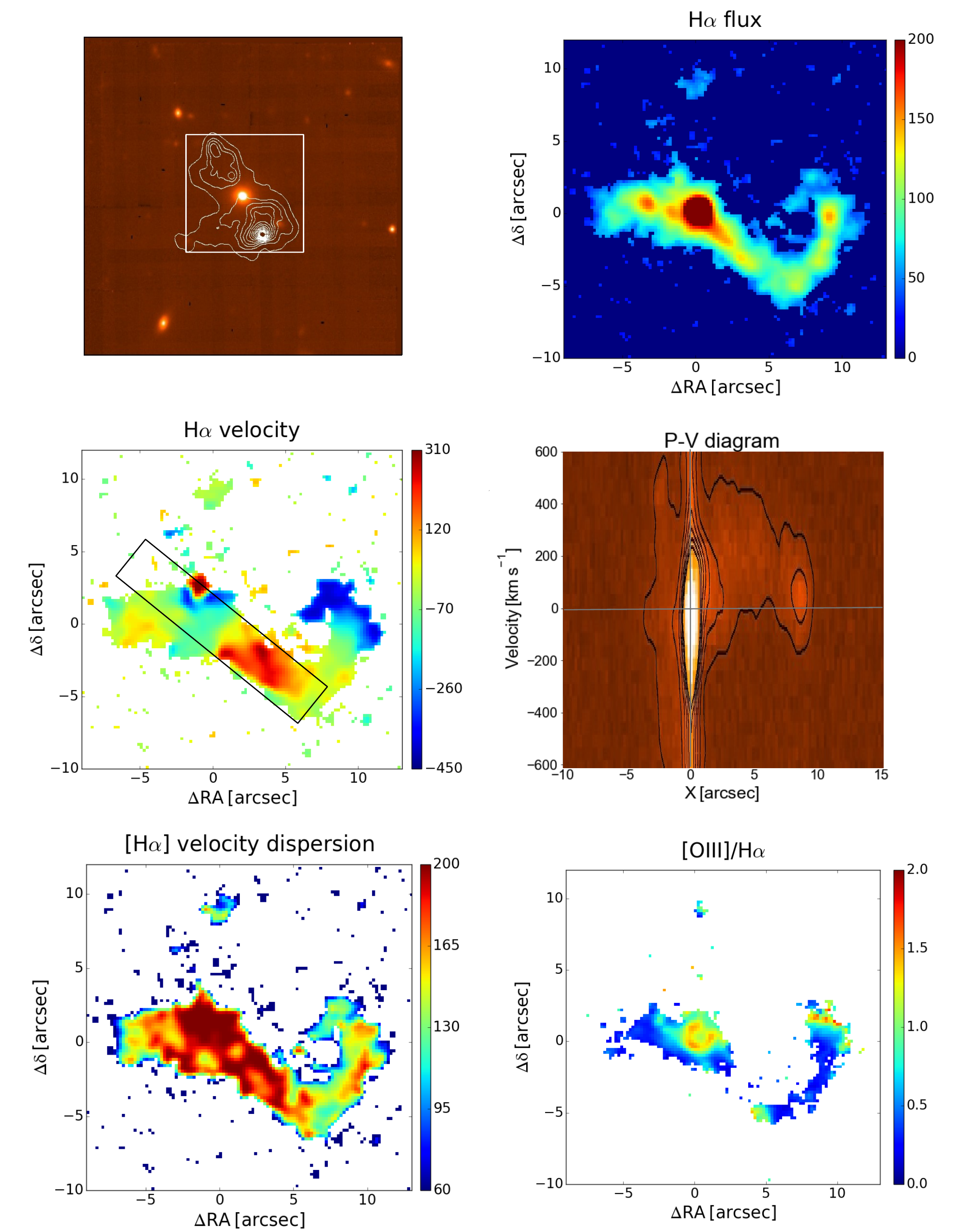}
\caption{3C~063, FR~II/HEG, 1$\arcsec$ =   2.99 kpc. Top left: radio contours overlaid onto the Muse optical continuum image . 
Top right: \ha\ emission line image extracted from the white square in the top left panel. .  Middle: velocity field from the \ha\ line and PV diagram (the synthetic slit  is centered on the nucleus, has a width of 5 pixels and it is oriented at an angle of -60$^\circ$ from the X axis). Bottom: \ha\ velocity dispersion and \oiii/\ha\ ratio.}}
\label{3C063}
\end{figure*}  

\begin{figure*}  
\centering{ 
\includegraphics[width=2.\columnwidth]{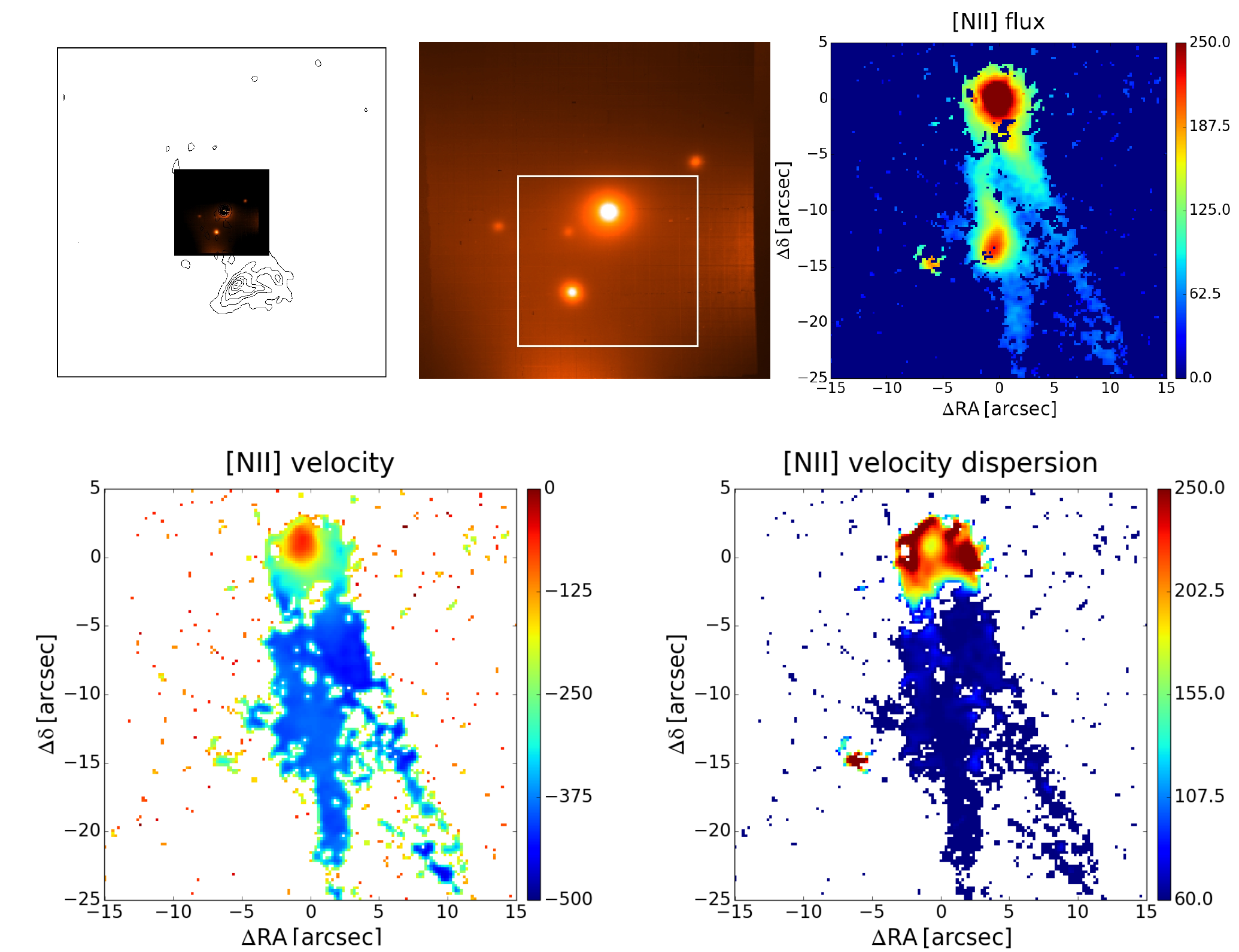}
\caption{3C~318.1, complex radio morphology, no optical spectroscopic
  classification, FR~II/HEG, 1$\arcsec$ = 0.90 kpc. Top left: radio
  contours overlaid onto the Muse optical continuum image .  Top center: Muse optical continuum image . Top right: [N~II]
  emission line image extracted from the white square in the top center panel.   Middle: velocity field and velocity
  dispersion from the\nii\ line.}}
\label{3C318}
\end{figure*}  

\begin{figure*}  
\centering{ 
\includegraphics[width=2.\columnwidth]{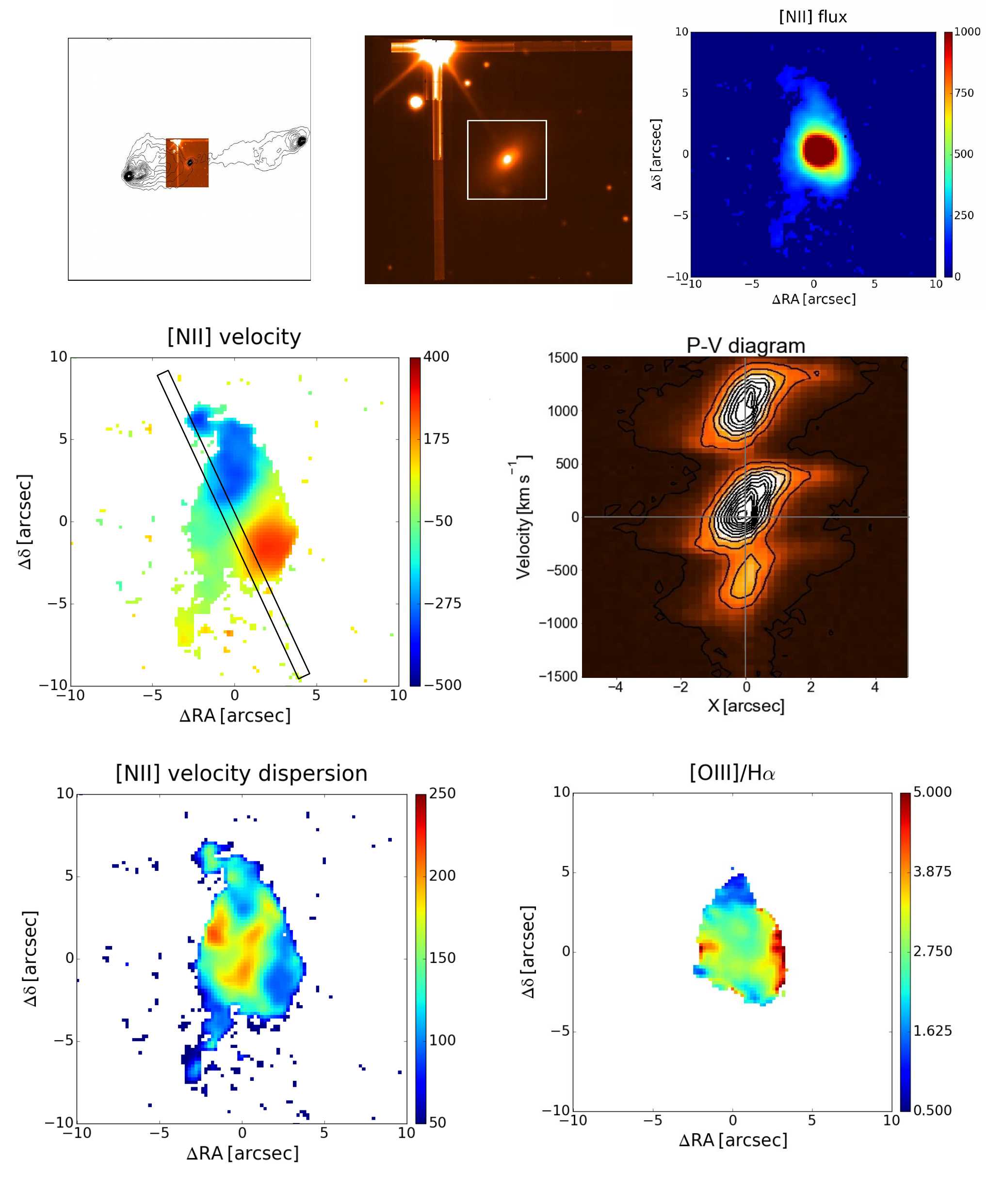}
\caption{3C~327, FR~II/HEG, 1$\arcsec$ = 1.94 kpc. Top left: radio
  contours overlaid onto the Muse optical continuum image .  Top center: Muse optical continuum image . Top right: [N~II]
  emission line image extracted from the white square in the top center panel.  Middle: velocity field from
  the\nii\ line and PV diagram (the synthetic slit is centered on
    the nucleus, has a width of 10 pixels and it is oriented at an
    angle of -52$^\circ$ from the X axis). Bottom: \nii\ velocity
  dispersion and \oiii/\ha\ ratio. }}
\label{3C327}
\end{figure*}

\begin{figure*}  
\centering{ 
\includegraphics[width=2.\columnwidth]{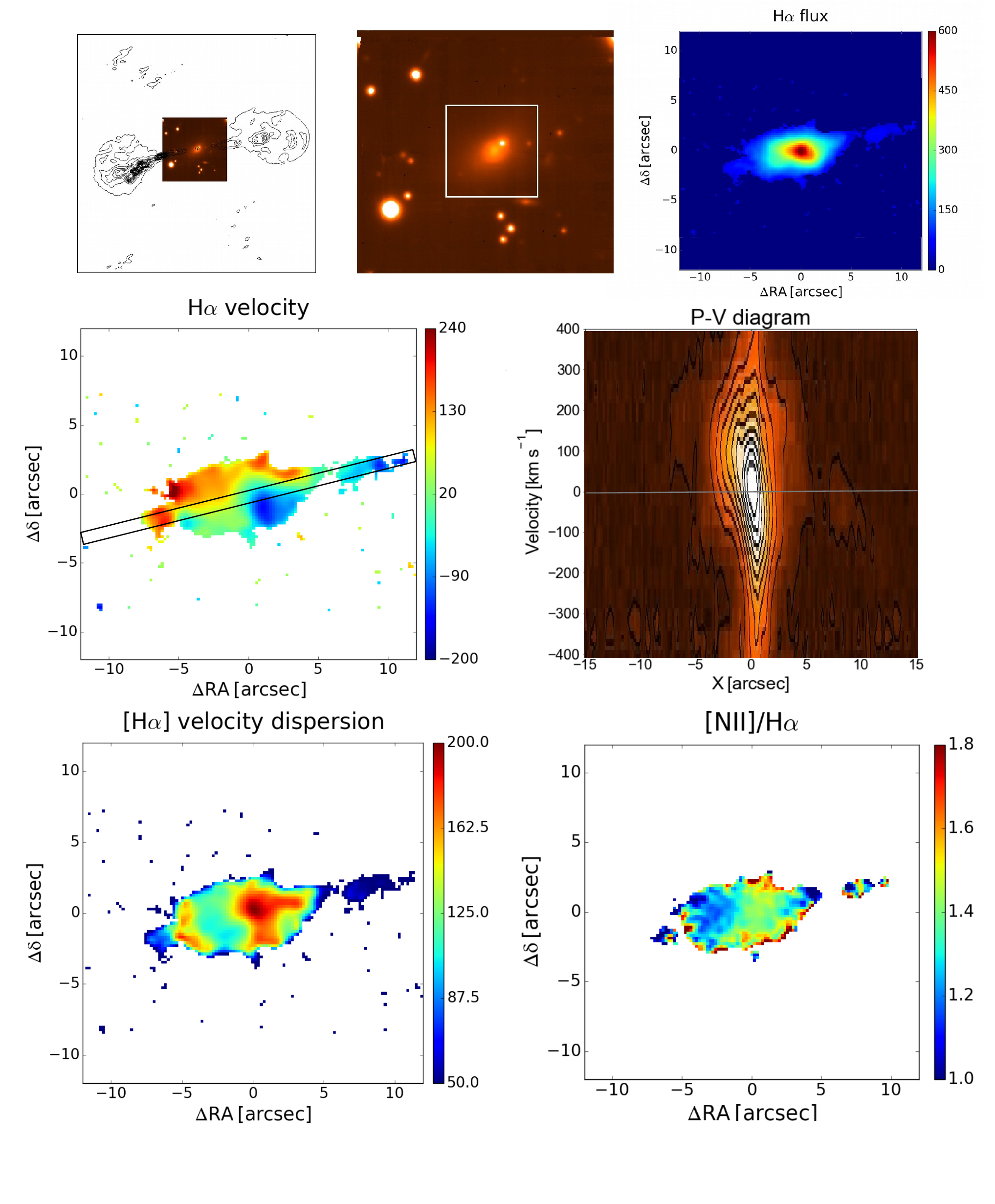}
\caption{3C~348, FR~I/ELEG, 1$\arcsec$ = 2.71 kpc. 
Top left: radio contours overlaid onto the Muse optical continuum image .  Top center: Muse optical continuum image . Top right: \ha\ 
emission line image extracted from the white square in the top center panel.  Middle: velocity field from the \ha\ line and PV diagram (the synthetic slit  is centered on the nucleus, has a width of 10 pixels and it is oriented at an angle of 14$^\circ$ from the X axis). Bottom: \ha\ velocity dispersion and \nii/\ha\ ratio.}}
\label{3C348}
\end{figure*}  

\begin{figure*}  
\centering{ 
\includegraphics[width=2.\columnwidth]{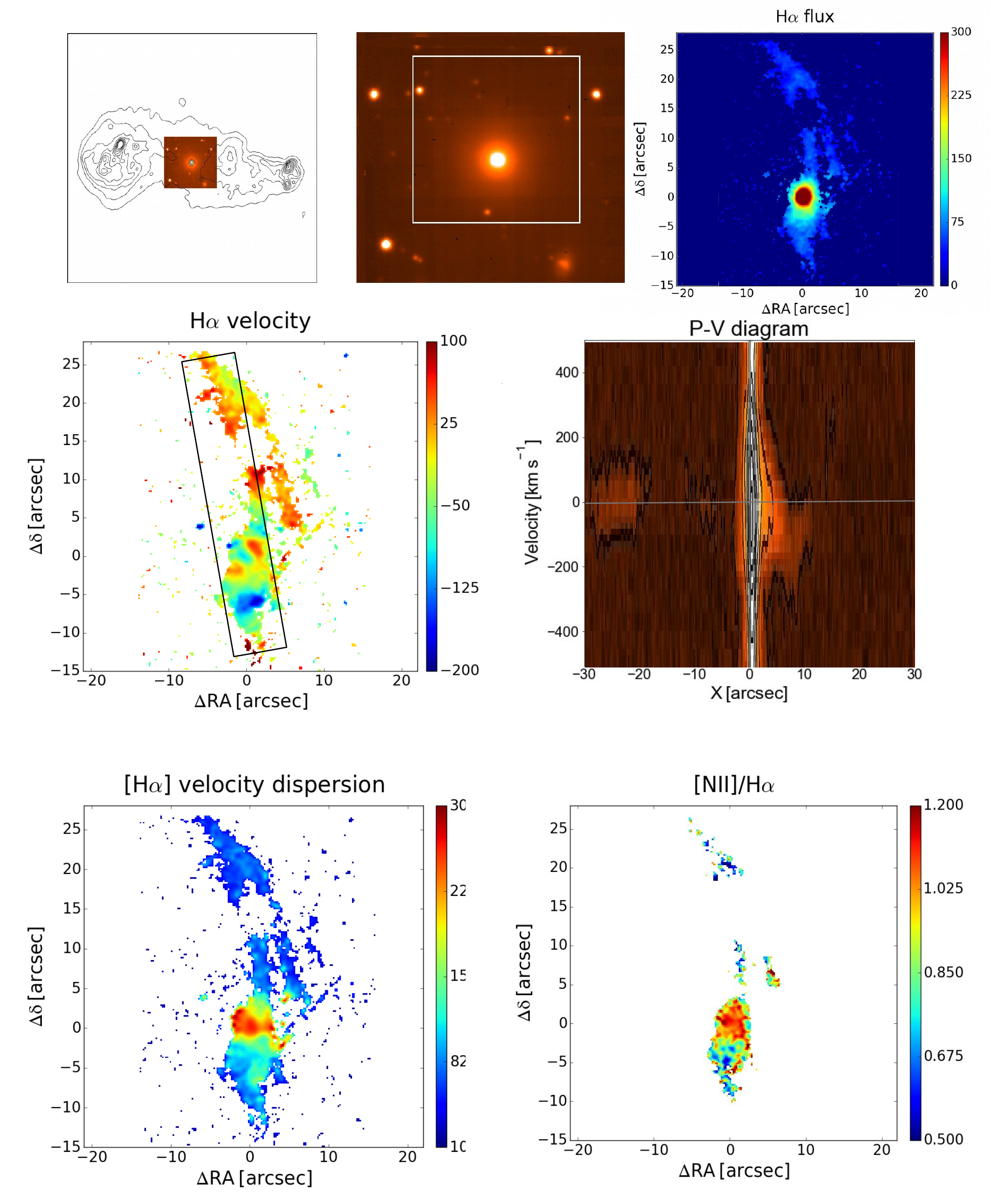}
\caption{3C~353, FR~II/LEG, 1$\arcsec$ = 0.60 kpc. Top left: radio contours overlaid onto the Muse optical continuum image .  Top center: Muse optical continuum image . Top right: \ha\ 
emission line image extracted from the white square in the top center panel.  Middle: velocity field from the \ha\ line and PV diagram (the synthetic slit  is centered on the nucleus, has a width of 30 pixels and it is oriented at an angle of -80$^\circ$ from the X axis). Bottom: \ha\ velocity dispersion and \nii/\ha\ ratio.}}
\label{3C353}
\end{figure*}  

\begin{figure*}  
\centering{ 
\includegraphics[width=2.\columnwidth]{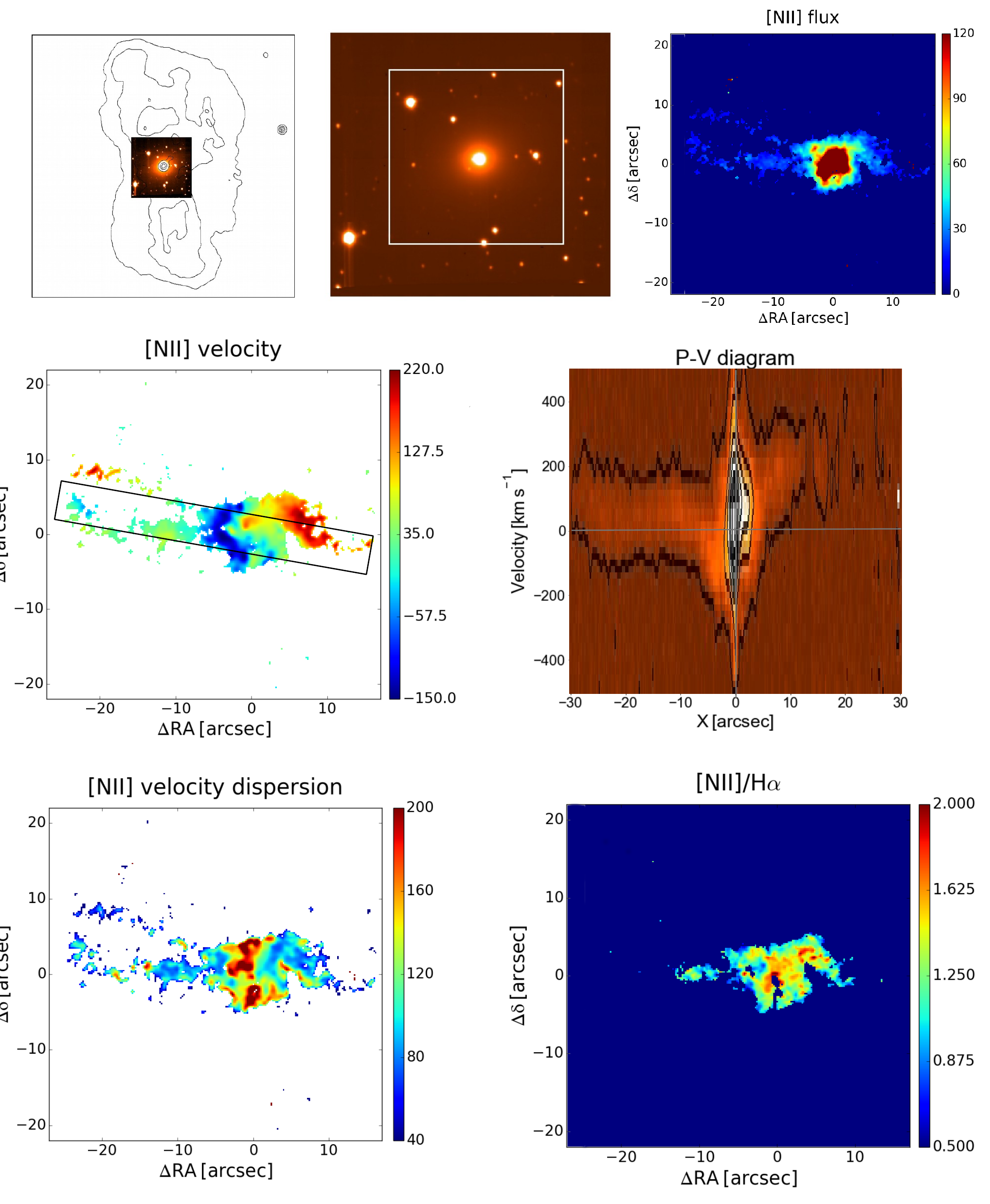}
\caption{3C~386, FR~II, 1$\arcsec$ = 0.35 kpc. Top left: radio contours
  overlaid onto the Muse optical continuum image . Top center: Muse optical continuum image . Top right: [N~II] emission line
  image extracted from the white square in the top center panel. Middle: velocity field from the \nii\ line and position
  velocity diagram extracted from the synthetic slit shown overlaid
  onto the velocity field (the boxy region is centered on the nucleus,
  has a width of 30 pixels and it is oriented at an angle of
  -10$^\circ$ measured from the X axis).  Bottom: \nii\ velocity
dispersion and \nii/\ha\ ratio.}}
\label{3C386}
\end{figure*}  

\section{Observation and data reduction}
\label{sample}

We observed a sample of 20 radio galaxies with MUSE as part of the
MURALES survey. The sample is formed by all the 3C radio-sources
limited to $z<0.3$ and $\delta<20^\circ$, visible during the
April-September semester, i.e., R.A. $<$ 3$^{\rm h}$ and R.A. $>$
15$^{\rm h}$. Their main properties are listed in Tab. \ref{tab1}.
They are in the redshift range $ 0.018 < $z$ < 0.289$, with 11 sources
located at z$<0.1$. Their radio power spans more than two orders of
magnitude, from $\sim 10^{24}$ to $\sim 2 \times 10^{26}$ W Hz$^{-1}$
at 178 MHz. Most of them (15) are FR~II. All optical spectroscopic
classes (LEGs, HEGs, and broad lined objects, BLOs) are represented with an almost
equal share of LEGs (including four FR~II/LEGs) and HEGs/BLOs.  We
compare the redshift and radio power distribution of our sample with
that of the entire population of 114 3C radio galaxies at z$<$0.3
presented by \citet{buttiglione09}.  The mean redshift and radio power
are $z=0.11$ and $\log L_{178}=33.58 \ergs$, respectively, not dissimilar
from the values measured for the entire 3C sub-sample with $z<0.3$
($z=0.13$ and $\log L_{178}= 33.61 \ergs$). The Kolmogorov-Smirnov test
confirms that the two distributions of $z$ and $L_{178}$ are not
statistically distinguishable. Our sub-sample can then be considered
as well representative of the population of powerful, low redshift,
radio galaxies.

The observations were obtained as part of the program ID
099.B-0137(A). Two exposures of 10 minutes each (except for 3C~015 and
3C~348 for which the exposure times were 2$\times$13 and 2$\times$14
minutes, respectively) were obtained with the VLT/MUSE spectrograph
between June 3rd, 2017, and July 22nd, 2017 covering the
  wavelength range 4800-9300 \AA. The median seeing of the
observations is 0\farcs65.  We split the total exposure time on
  source in two sub-exposures, applying a small dithering pattern and
  a 90 degrees rotation between on-object exposures to reject cosmic
  rays.  A negligible cross-pattern remains in the final image, due to
  the observations with the two position angles of 0 and 90 degree.
We used the ESO MUSE pipeline (version 1.6.2) with default
  settings to obtain fully reduced and calibrated data cubes. In
  detail, the pipeline applies the standard reduction procedure,
  i.e. corrects for the bias, the dark, the flat, the vignetting,
  removes cosmic rays and calibrates each exposure in wavelength and
  in flux.  In each pixel a sky background model is subtracted and a
  sigma clipping method is implemented to remove cosmic rays from
  individual exposures.  For the flux calibration, the pipeline uses a
  pre-processed spectro-photometric standard star observation, taken
  during clear or photometric nights, selected to be the closest to
  the expected position of the reference source.
A known issue of the MUSE calibration procedure  is that the 
sky subtraction is currently not optimal and the astrometry in some cases not precise, failing the identification of the stars depending on the observing conditions.We used the position of bright sources in the MUSE field observed by Pan-STARRS survey \citep{chambers16} to
correct for astrometric offset. 
More details about the  reduction 
strategy are described in the MUSE Pipeline User Manual\footnote{ftp://ftp.eso.org/pub/dfs/pipelines/muse/muse-pipeline-manual-2.4.1.pdf}.

We followed the same strategy for the data analysis described in
\citet{balmaverde18}. Summarizing, we resampled the data cube with the
Voronoi adaptive spatial binning \citep{cappellari03}, requiring an
average signal-to-noise ratio on the continuum per wavelength channel
of at least 50. We then used Penalized Pixel-Fitting code
\citep{cappellari17} to fit the stellar continuum and absorption
features, which we finally subtracted from each spaxel in the data
cube. Over most of the field-of-view, well outside of the host
galaxies, the continuum emission is actually negligible and the
spectra are completely dominated by emission lines. This procedure,
that nonetheless we applied to all spaxels, often does not have any
effect on the data.

We simultaneously fit all emission lines in the blue (namely
  \hb$\lambda$4863, \oiii$\lambda\lambda$4960,5008) and in the red
  ([O~I]$\lambda\lambda$6302,6366, [N~II]$\lambda\lambda$6550,6685,
  \ha$\lambda$6565, [S~II]$\lambda\lambda$6718,6733) portion of the
  spectrum in the continuum subtracted spectra in each spaxel. We
assumed that all lines in the blue and red portion of the spectra have
the same profile. For the broad lined objects (BLO) we allowed for the
presence of a broad component in the Balmer lines on the nucleus.

A single gaussian component usually reproduces accurately the line
profiles.  However, for some objects, we had to include additional
components in the central regions.

\section{Results}
\label{results}
We here focus on the properties of the ionized gas as probed by
various emission lines. The results of the analysis are presented in
Figs. 1 through 20. For each source we derived a
continuum image (integrating the rest-frame line-free region between
5800 and 6250 \AA) on which we superposed the radio contours from
radio maps, retrieved from the NRAO VLA Archive Survey. In many cases
the size of the radio source exceeds the whole MUSE field of view and,
in most objects, the size of the region where emission lines are
detected. For this reason we did not overlay the radio contours onto
the emission line images. We also produced images in the different
emission lines and obtained the correspondent gas velocity field. 
  The velocity map are referred to the recession velocity derived from
  redshift. When necessary, we add a velocity shift to set the
  velocity in the nuclear region equal to zero. In the figures we
show the intensity of the brightest line, usually the \nii\ or
\oiii. When the line emission is resolved we present the 2D gas
velocity maps and the position-velocity diagram extracted along a
synthetic long-slit aperture aligned with interesting line
structures. This is possible for all sources except 3C~015, 3C~029,
and 3C~456.  All maps are centered at the position of the peak in
  the optical continuum images obtained from the MUSE datacube over a
  wavelength range free of emission lines. We show only spaxels with a
  signal to noise ratio (S/N) larger than 2. North is up and East is left; angles are all
  measured counterclockwise.  In the
following we describe the results obtained for the individual sources.

\subsection{Notes on the individual sources}
\label{notes}

We here present the main results for the individual sources. In
  the Figures from 1 to 20 we show the continuum image with superposed
  the radio contours, the image of the brightest emission lines and
  the corresponding velocity field and velocity dispersion. We also
  produced a position-velocity diagram along the direction of largest
  line extent for all well resolved line emission regions as well as a
  ratio image between the brightest line ([O~III] or [N~II]) and
  \ha. In the Appendix we present the nuclear spectra of all sources
  extracted from a single pixel at the continuum peak and the spectra
  of the 14 sources for which we explore the ionization properties of
  the extended emission line regions (see Fig. \ref{spettri}).

\noindent
{\bf 3C~015:} FR~I/LEG, 1$\arcsec$ = 1.40 kpc. The emission lines are
only slightly extended and confined within the central
$\sim$3\arcsec\ ($\sim 4$ kpc). The compactness of this source
prevents us from producing a well resolved gas velocity field. The
velocity dispersion is larger along the radio axis, reaching
$\gtrsim$ 800 \kms.

\smallskip
\noindent
{\bf 3C~017:} FR~II/BLO, 1$\arcsec$ = 3.58 kpc. Beside the bright
central nuclear component, characterized by the presence of broad
Balmer lines, the ionized gas extends in the NW direction out to
$\sim$ 15 kpc. The gas velocity increases moving to the West up to
$\sim 100$ \kms, but when turning toward the North it decreases to
blueshifted velocities, down to $\sim -200$ \kms. In addition, a
blueshifted compact emission line knot is found $\sim 20$kpc to the
North. The velocity dispersion is, except on the nucleus, rather small. The
\oiii/\ha\ line ratio in the west region is enhanced by a factor
$\sim$ 3 compared to the nucleus.

\smallskip
\noindent {\bf 3C~018:} FR~II/BLO, 1$\arcsec$ = 3.17 kpc. Diffuse line
emission, elongated in the EW direction, surrounds the broad lined
nucleus out to $\sim$15 kpc. The lines are redshifted on both sides of
the nucleus, by $\sim 200$\kms\ on the West and by $\sim 100$ \kms\ on
the East where they are slightly blue-shifted in the perpendicular N-S
direction. The velocity dispersion is always much larger than the instrumental
width, with typical values of $\sim$ 200 \kms. The \oiii/\ha\ ratio
decreases the nucleus toward the extended regions, from $\sim$4 to
$\sim$0.5.

\smallskip
\noindent {\bf 3C~029:} , FR~I/LEG, 1$\arcsec$ = 0.89 kpc. Similarly
to 3C~015, the line emission is only marginally extended. There is a
hint of rotation along a line of nodes oriented at $\sim$ 120$^\circ$.

\smallskip
\noindent {\bf 3C~033:} FR~II/HEG, 1$\arcsec$ = 1.17 kpc. The MUSE
field-of-view covers only $\sim$ 1/4 of the extension of the radio
source. The ionized gas has an elliptical shape, extending out to
$\sim 8\arcsec (\sim 9$ kpc) on both sides of the nucleus, and
elongated along PA$\sim$ 50$^\circ$, then twisting along PA$\sim$
70$^\circ$.  The velocity field is broadly dominated by ordered rotation, with
a line of nodes initially at PA $\sim$ 75$^\circ$, then twisting at
smaller angles. Outside the central regions of high velocity gradient,
the most distant gas shows a constant rotation of $\sim$ 300 \kms. The
velocity dispersion is enhanced along a linear region parallel to the
line of nodes. . The \oiii/\ha\ ratio is smaller in the west and east
regions, and larger in a region approximately aligned with the radio
axis, suggestive of a ionization cone.

\smallskip
\noindent {\bf 3C~040:} FR~I/LEG, 1$\arcsec$ = 0.37 kpc. This source
shows a FR~I morphology, with diffuse radio plumes (not fully shown in
Fig. 6) extending by more than $\sim 20\arcmin$. Emission lines are
detected in a compact but resolved region extending by $\sim 3\arcsec$
($\sim 1$ kpc). Gas rotation is detected, around PA $\sim$ 75$^\circ$,
with an amplitude of $\sim$ 300 \kms.

\smallskip
\noindent {\bf 3C~063:} FR~II/HEG, 1$\arcsec$ = 2.99 kpc. The size of
the emission line region is similar to that of this double-lobed radio
source. On the east side the gas extends along PA$\sim 80^\circ$ for
$\sim 8\arcsec (\sim 25$ kpc) with a small velocity gradient. On the
opposite side, emission lines extend initially toward the South-West,
out to $\sim 10\arcsec$, well aligned with the radio axis.  At the
location of the southern hot spot, the gas sharply bend toward the
North, forming an arc-like structure, located at a distance of $\sim$
33 kpc from the nucleus, wrapping around the southern lobe. This
feature is dominated by a series of compact knots, surrounding the
outer edge of the radio lobe. The gas velocity initially has a
positive gradient, reaching $\sim$ 200 \kms, out to $\sim 6\arcsec$,
where it reverses. On the arc-like emission line feature the velocity
smoothly decreases down to $\sim$ -400 \kms.  The velocity dispersion is always
much larger than the instrumental width, with typical values of $\sim$
200 \kms\ and above. The \oiii/\ha\ ratio decreases from the nucleus
toward the more extended regions.

\smallskip
\noindent {\bf 3C~318.1:} a very peculiar source from the point of
view of its radio properties, a likely relic source, with an extremely
steep spectrum ($\alpha_{235 {\rm MHz}}^{1.28 {\rm GHz}}$=2.42)
\citep{giacintucci07}.  The line emission, beside the bright region
cospatial with the host, shows two linear plumes emerging from the
center and extending toward the South, at slightly different angles,
for $\sim$25 kpc. One of them terminates with a bright spot, not
associated with any continuum structure. The plumes show an almost
constant velocity along their whole length, at slightly different
velocities: the plume pointing to South is blueshifted by $\sim$ 230
\kms, while that to the South-West has a larger blueshift, $\sim$ 280
\kms. The line dispersion is much larger on the host than on the
plumes, where it is generally consistent with the MUSE spectral
resolution.

\smallskip
\noindent {\bf 3C~327:} FR~II/HEG, 1$\arcsec$ = 1.94 kpc. The line
emission extends in the NE-SW direction for $\sim 18$ kpc. The
velocity field in the central regions is apparently ordered, with a line
of node inclined with respect to the geometrical axis by $\sim
40^\circ$. The gas velocity at radii larger than $\sim 3$ kpc shows a
full amplitude of $\sim$ 600 \kms.  The velocity dispersion is
enhanced along the line of nodes, likely due to the unresolved
velocity gradients in this region. The \oiii/\ha\ ratio is smaller in
the north and south regions, and larger in a region approximately
aligned with the radio axis, suggestive of a ionization cone.

\smallskip
\noindent {\bf 3C~348:} FR~I/LEG, 1$\arcsec$ = 2.71 kpc. A close
companion, at the same redshift, is seen at $\sim 9$ kpc to the
NW. This is the only FR~I source showing substantially extended
emission line, reaching a distance of $\sim$ 35 kpc on the west
side. The gas is mainly in rotation in the central regions, but the
kinematic axis (at $PA \sim 20^\circ$) is almost perpendicular to the
geometrical one. Following the large scale gas filaments, the gas
velocity increases steadily toward larger radii, reaching an amplitude
of $\sim$ 400 \kms. The gas velocity dispersion in highly enhanced at
the nucleus and in the region cospatial with the west radio jet.

\smallskip
\noindent {\bf 3C~353:} FR~II/LEG, 1$\arcsec$ = 0.60 kpc. In the
central region the gas is in ordered rotation with a line of node
oriented at $PA \sim 30^\circ$. An S-shaped filament extends for $\sim
22$ kpc in the north-south direction. Its velocity field is rather
complex: on the north side the gas is initially blueshifted by $\sim
100$ \kms, but its velocity then steadily increases at larger radii
before falling back to the systemic velocity for $r> 10$ kpc.  In the
central $\sim$ 3 kpc both the velocity dispersion and the
\nii/\ha\ ratio are larger than in the regions at larger radii.

\smallskip
\noindent {\bf 3C~386:} FR~II, 1$\arcsec$ = 0.35 kpc. Line emission is
detected out to the edges of the MUSE field of view, covering a
distance of at least 16 kpc. Beside the central source, it forms
linear diffuse structures in the E-W direction, perpendicular to the
radio lobes. Gas rotation around PA$\sim$ 30$^\circ$ is seen in the
central regions. On the west side the velocity amplitude increases
steadily reaching $\sim$ 300 \kms. On the opposite side the line is
also generally redshifted, but reaching lower velocities, with the two
filaments showing different velocities.  Lines are broader along the
line of nodes which coincides with the radio axis. Our data confirm
the presence of a star superposed onto the nucleus
\citep{lynds71,buttiglione09}.

\smallskip
\noindent {\bf 3C~403:} FR~II/HEG, 1$\arcsec$ = 1.15 kpc. The ionized
gas extends over $\sim$ 45 kpc forming a diffuse structure in the NE -
SW direction. Most of the gas is {apparently} in very regular rotation with a line
of nodes oriented at $PA \sim -30^\circ$ with a large velocity
gradient in the central $\sim 3$ kpc, reaching an amplitude of $\pm
300$ \kms, followed by a smooth decrease at larger radii. High
velocity blueshifted gas, reaching $\sim 800$\kms is seen on the
nucleus.  Similarly to what is seen in 3C~033, lines are broader in a
region almost aligned with the line of nodes. The \oiii/\ha\ ratio is
larger in the off-nuclear regions.

\smallskip
\noindent {\bf 3C~403.1:} FR~II/LEG, 1$\arcsec$ = 1.07 kpc. The line
morphology in this source is particularly complex. There is a central
region extending $\sim 9$ kpc around the host galaxy and apparently showing well
ordered rotation. However, the ionized gas also forms a series of
elongated structures, mainly on the E and SE direction, with compact
knots joined by more diffuse emission. There are several galaxies in
the MUSE field of view, but only a few emission line knots are
associated with them. These extended structures, reaching a distance
of $\sim$35 kpc, are all found at very similar velocities, being
redshifted by $\sim 150 - 200$ \kms\ with respect to the host
velocity. Overall, the gas structure is reminiscent of the gas bubble
seen in 3C~317. In 3C 403.1, however, there is no apparent
correspondence with the location of the radio emission which extends
to much larger radii, $\sim$ 230 kpc, although the available radio
image is at very low spatial resolution ($\sim 45\arcsec$).

\smallskip
\noindent {\bf 3C~424:} FR~II/LEG, 1$\arcsec$ = 2.29 kpc. The emission line
structure is dominated by a bright linear feature in the EW direction
which shows an overall blueshift of $\sim$ 100 \kms. More diffuse
emission extends along a perpendicular axis out to $\sim$ 30 kpc with
an overall redshift of $\sim 150$ \kms\ on the SE side but reaching
$\sim 450$ \kms\ on the NW side where we also find the regions of
highest velocity dispersion, up to $\sim 450$ \kms. Generally, the
nuclear regions have a higher \nii/\ha\ ratio that the rest of
the emission line nebula.

\smallskip
\noindent {\bf 3C~442:} FR~II/LEG, 1$\arcsec$ = 0.53 kpc. Two compact knots,
separated by $\sim 0.7$ kpc dominate the emission line structure. They
show a velocity offset of $\sim 130$\kms\ but their spectra are very
similar from the point of view of the line ratios and this does not
suggest, unlike the case of 3C~459 discussed below, the presence of a
dual AGN. At large radii two tongues of ionized gas extends in the SW
direction for $\sim$ 4 kpc, a feature reminiscent of the edges of the
cavity seen in the 3C~317 \citep{balmaverde18}, but the resolution of
the available radio images does not allow us to perform a detail
comparison between these structures. Along the same SW direction
diffuse gas is detected out to a radius of $\sim$ 14 kpc, showing an
almost constant velocity of $\sim$150 \kms. On the SW side there is
another diffuse structure, apparently unrelated to the filaments
pointing more to the South based on both its location and much larger
blueshift (up to $\sim 400$ \kms) not associated with any galaxy in
the field.

\smallskip
\noindent {\bf 3C~445:} FR~II/BLO, 1$\arcsec$ = 1.09 kpc. The emission line
region in this source has a triangular shape, elongated in the SW
direction out to a radius of $\sim$ 20 kpc. The gas velocity is
roughly constant, with a redshift of $\sim$ 150 \kms. The
\oiii/\ha\ ratio increases from the nucleus to the outer regions.

\smallskip
\noindent {\bf 3C~456:} FR~II/HEG, 1$\arcsec$ = 3.74 kpc. The emission
line region in this source is compact, confined within $\sim$10
kpc. Nonetheless, it clearly shows rotation around PA $\sim$45$^\circ$
and large velocity dispersion up to $\sim$ 350 \kms.

\smallskip
\noindent {\bf 3C~458:} FR~II/HEG, 1$\arcsec$ = 4.38 kpc. Ionized gas
is detected at a distance of more than 100 kpc. It is located in
various clumps forming various elongated structures, in a general
NE-SW direction. The gas velocity field is also rather complex:
although there is a tendency for the gas located in the NE quadrant to
be generally redshifted (and to be blue-shifted in the SW one) there
are changes of speed occurring on small scale.

\smallskip
\noindent {\bf 3C~459:} FR~II/BLO, 1$\arcsec$ = 3.58 kpc. We presented
the results obtained for this source in \citet{balmaverde18}. We
detected diffuse nuclear emission and a filamentary ionized gas
structure forming a one sided triangular-shaped region extending out
to $\sim$80 kpc. The central emission line region is dominated by two
compact knots of similar 
surface brightness separated by 1$\farcs$2 (5.3 kpc). Based
on the dramatic differences in velocity, line widths, and line ratios,
we argued that we are observing a dual AGN system, formed by a
radio-loud AGN and a type 2 QSO companion.

\begin{figure*}  
\centering{ 
\includegraphics[width=2.\columnwidth]{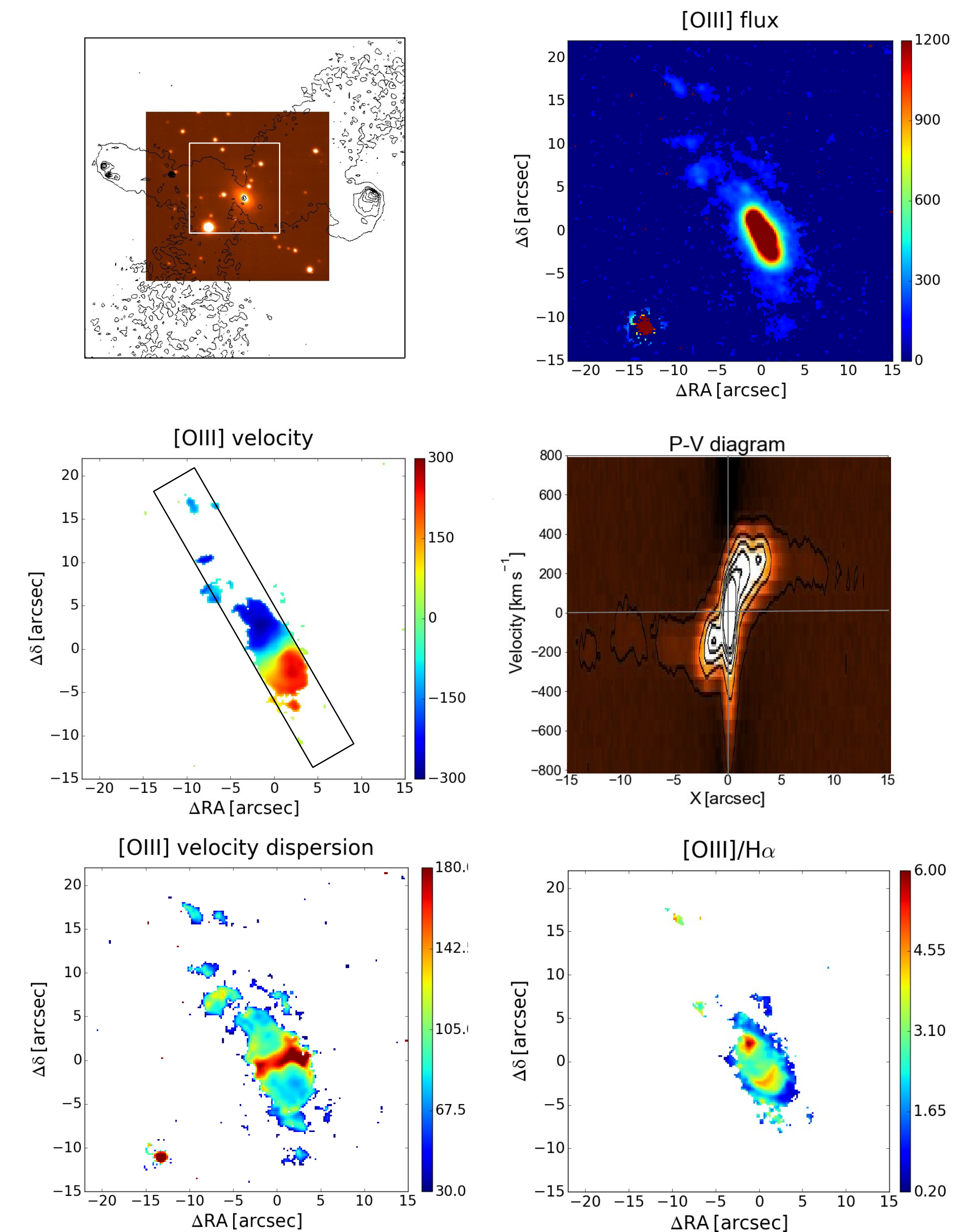}
\caption{3C~403, FR~II/HEG, 1$\arcsec$ = 1.15 kpc.Top left: radio contours overlaid onto the Muse optical continuum image . Top right: \oiii\
emission line image extracted from the white square in the top left panel.  Middle: velocity field from the \oiii\ line and PV diagram (the synthetic slit  is centered on the nucleus, has a width of 30 pixels and it is oriented at an angle of -60$^\circ$ from the X axis). Bottom: \oiii\ velocity dispersion and \oiii/\ha\ ratio.}}
\label{3C403}
\end{figure*}

\begin{figure*}  
\centering{ 
\includegraphics[width=2.\columnwidth]{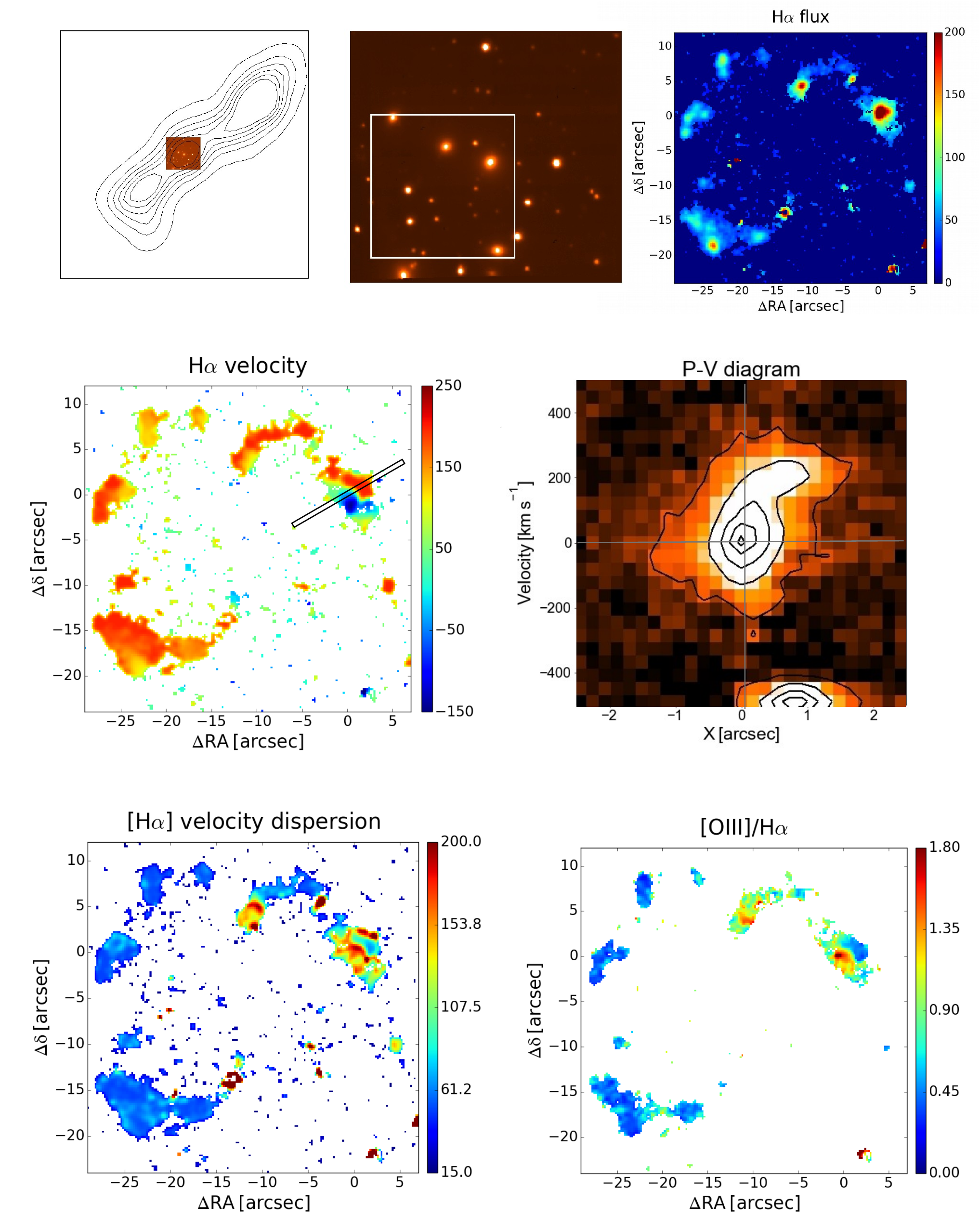}
\caption{3C~403.1, FR~II/LEG, 1$\arcsec$ = 1.07 kpc. Top left: radio contours overlaid onto the Muse optical continuum image .  Top center: Muse optical continuum image . Top right: \ha\ 
emission line image extracted from the white square in the top center panel.  Middle: velocity field from the \ha\ line and PV diagram (the synthetic slit  is centered on the nucleus, has a width of 5 pixels and it is oriented at an angle of 60$^\circ$ from the X axis). Bottom: \ha\ velocity dispersion and \oiii/\ha\ ratio.}}
\label{3C403.1}
\end{figure*}  

\begin{figure*}  
\centering{ 
\includegraphics[width=2.\columnwidth]{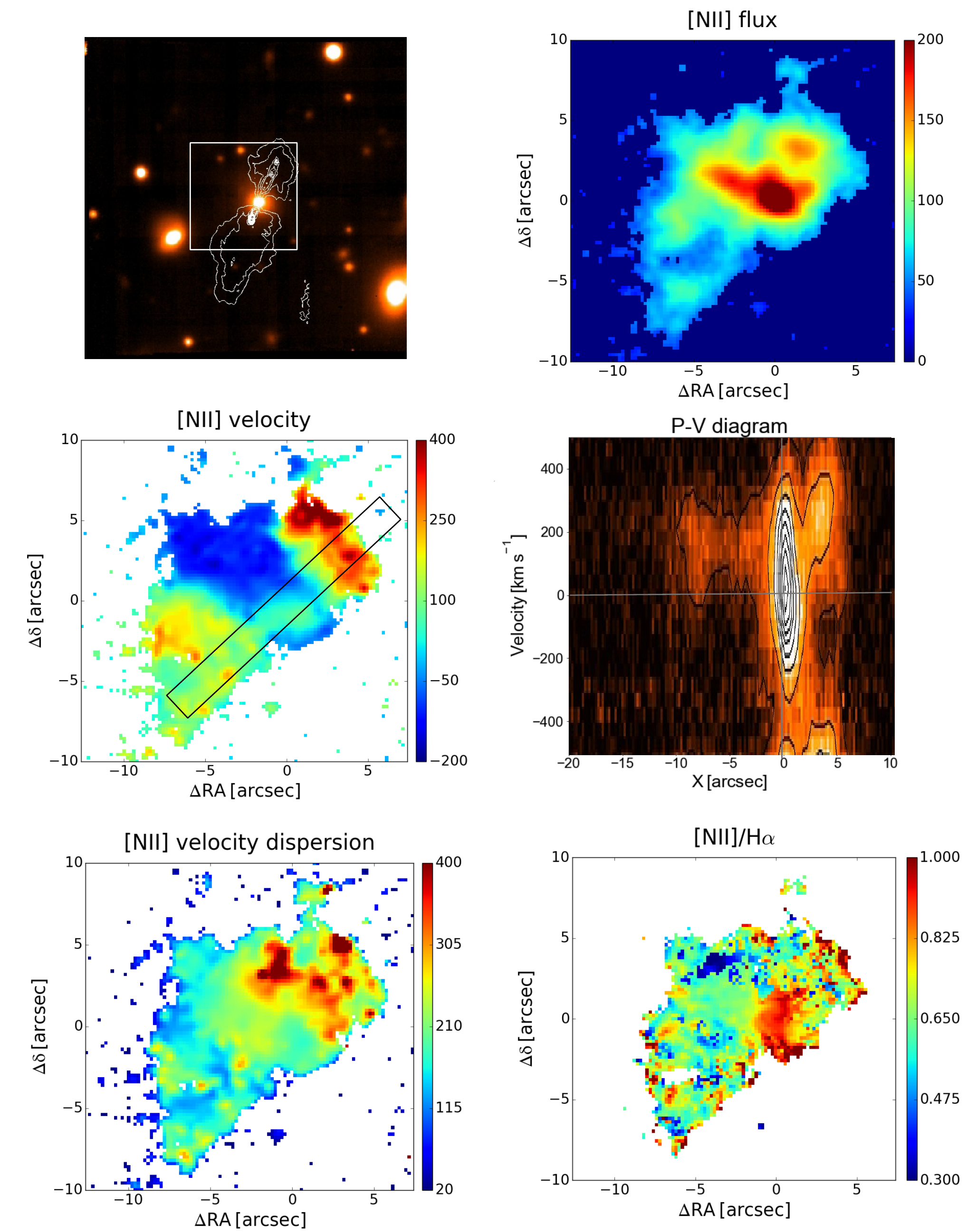}
\caption{3C~424, FR~II/LEG, 1$\arcsec$ = 2.29 kpc. Top left: radio contours overlaid onto the Muse optical continuum image . Top right: \nii\ 
emission line image extracted from the white square in the top left panel.  Middle: velocity field from the \nii\ line and PV diagram (the synthetic slit  is centered on the nucleus, has a width of 10 pixels and it is oriented at an angle of 47$^\circ$ from the X axis). Bottom: \nii\ velocity dispersion and \nii/\ha\ ratio.}}
\label{3C424}
\end{figure*}  

\begin{figure*}  
\centering{ 
\includegraphics[width=2.\columnwidth]{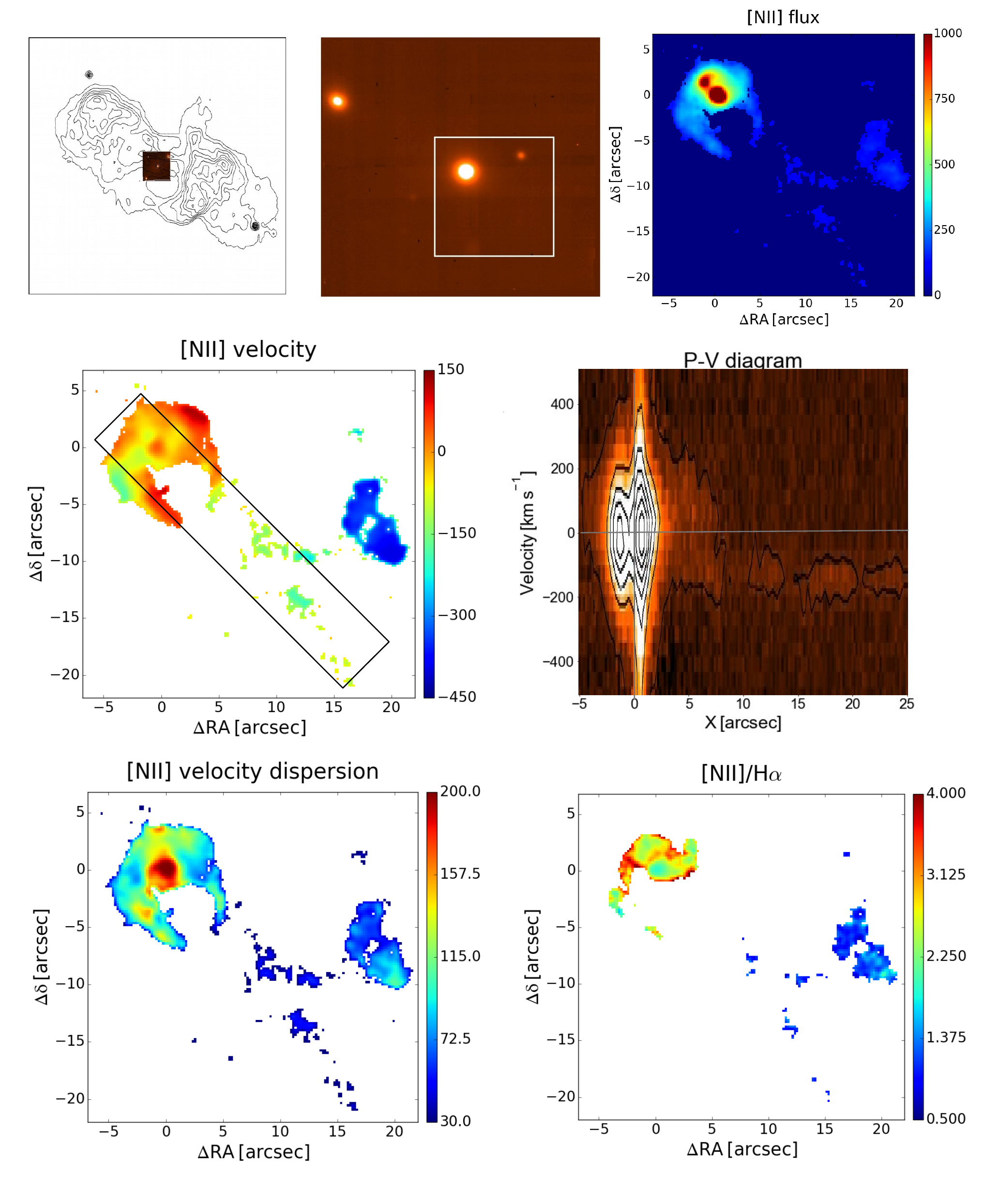}
\caption{3C~442, FR~II/LEG, 1$\arcsec$ = 0.53 kpc. Top left: radio contours overlaid onto the Muse optical continuum image .  Top center: Muse optical continuum image . Top right: \nii\ 
emission line image extracted from the white square in the top center panel.  Middle: velocity field from the \nii\ line and PV diagram (the synthetic slit  is centered on the nucleus, has a width of 30 pixels and it is oriented at an angle of -60$^\circ$ from the X axis). Bottom: \nii\ velocity dispersion and \nii/\ha\ ratio.}}
\label{3C442}
\end{figure*}  

\begin{figure*}  
\centering{ 
\includegraphics[width=2.\columnwidth]{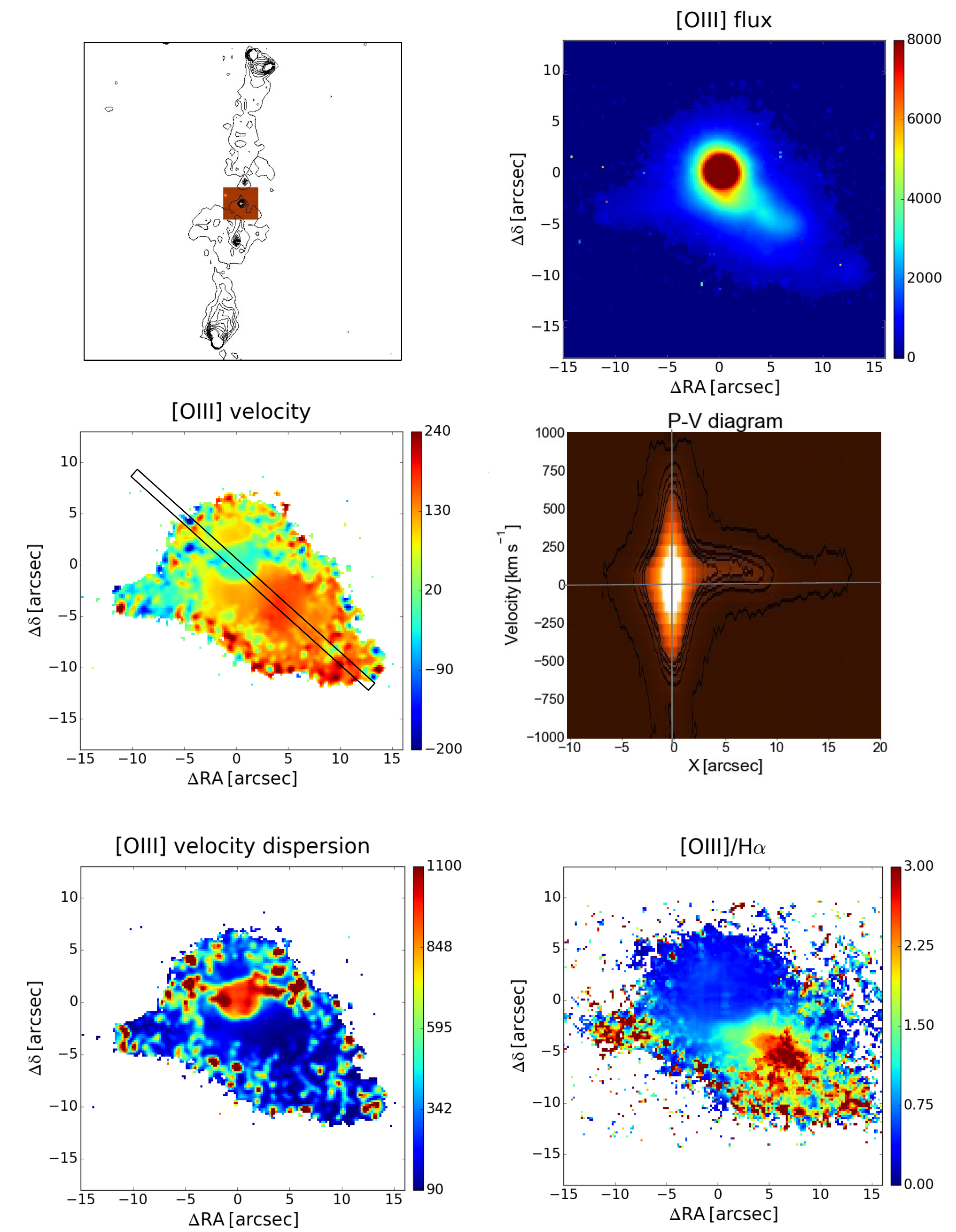}
\caption{3C~445, FR~II/BLO, 1$\arcsec$ = 1.09 kpc. Top left: radio contours overlaid onto the Muse optical continuum image .  Top right:  \oiii\ 
emission line image extracted from the white square in the top left panel.  Middle: velocity field from the \oiii\ line and PV diagram (the synthetic slit  is centered on the nucleus, has a width of 10 pixels and it is oriented at an angle of -42$^\circ$ from the X axis). Bottom: \oiii\ velocity dispersion and \oiii/\ha\ ratio. }}
\label{3C445}
\end{figure*}  

\begin{figure*}  
\centering{ 
\includegraphics[width=2.\columnwidth]{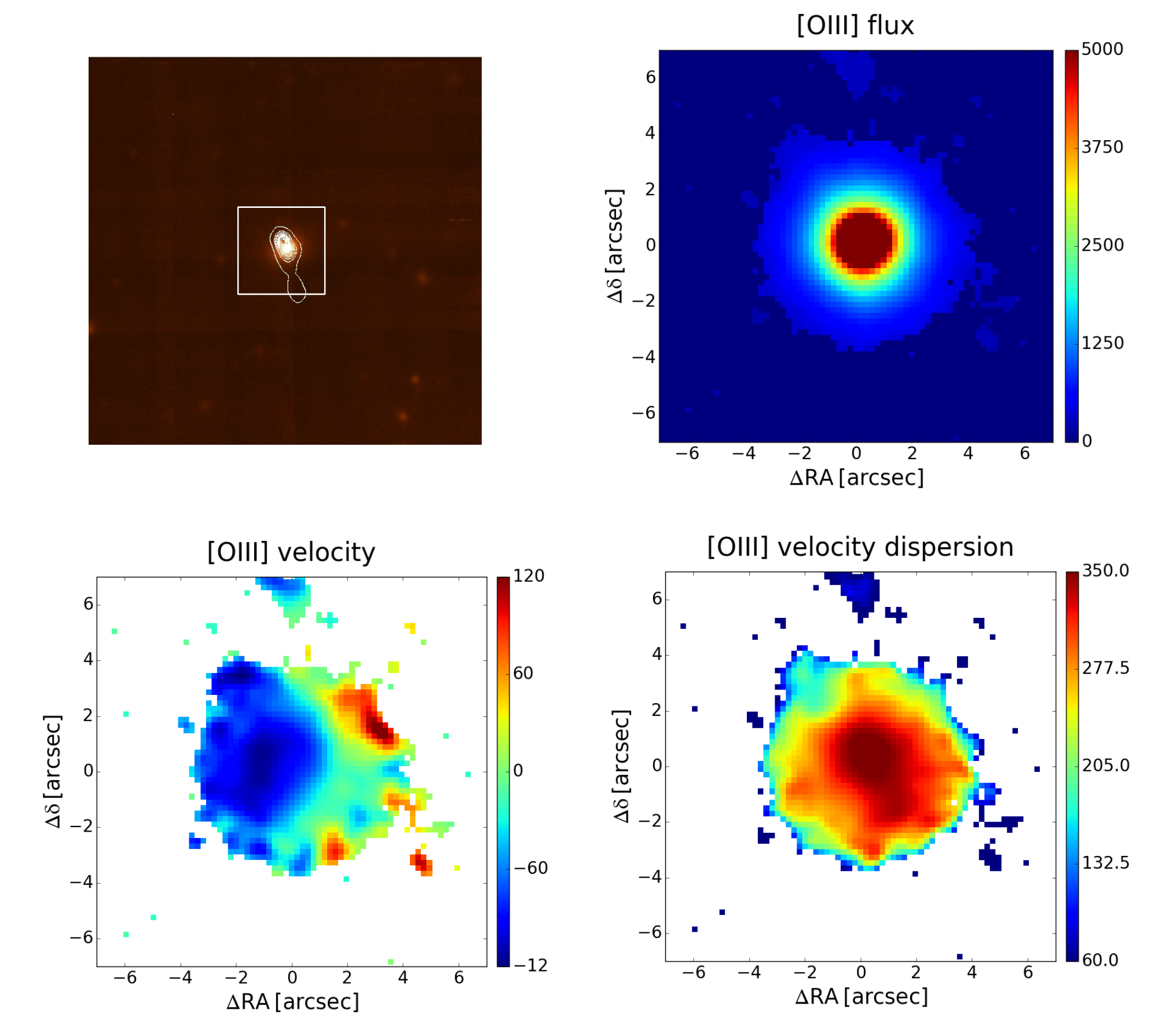}
\caption{3C~456, FR~II/HEG, 1$\arcsec$ = 3.74 kpc. Top left: radio contours
  overlaid onto the Muse optical continuum image . Top right: [O~III]
  emission line image extracted from the white square in the top left panel. Bottom: velocity field from the \oiii\ line and
  velocity dispersion. }}
\label{3C456}
\end{figure*}  

\begin{figure*}  
\centering{ 
\includegraphics[width=2.\columnwidth]{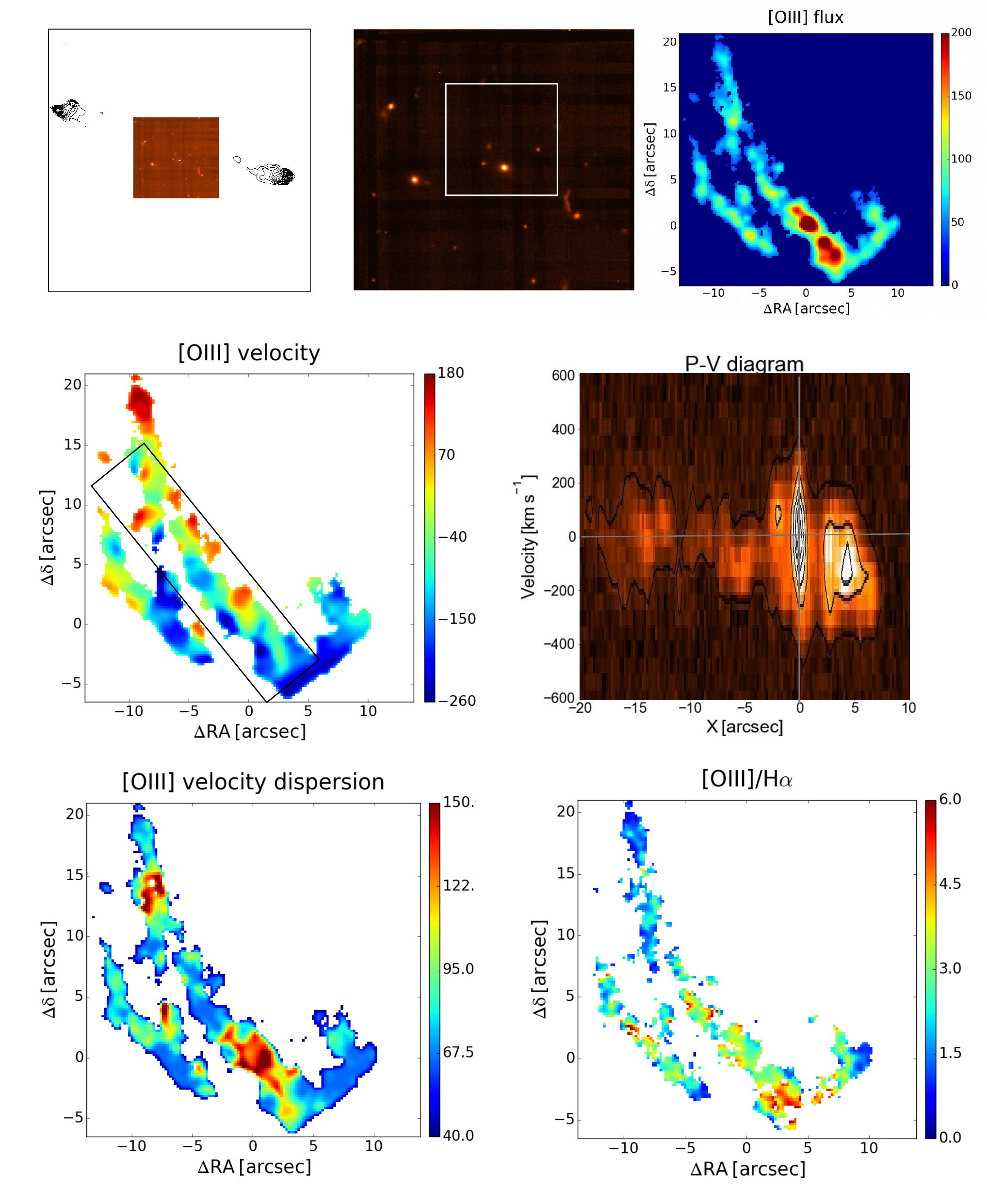}
\caption{3C~458, FR~II/HEG, 1$\arcsec$ = 4.38 kpc. Top left: radio contours overlaid onto the Muse optical continuum image .  Top center: Muse optical continuum image . Top right: \oiii\
emission line image extracted from the white square in the top center panel.  Middle: velocity field from the \oiii\ line and PV diagram (the synthetic slit  is centered on the nucleus, has a width of 30 pixels and it is oriented at an angle of -51$^\circ$ from the X axis). Bottom: \oiii\ velocity dispersion and \oiii/\ha\ ratio.}}
\label{3C458}
\end{figure*}  

\begin{figure*}  
\centering{ 
\includegraphics[width=2.\columnwidth]{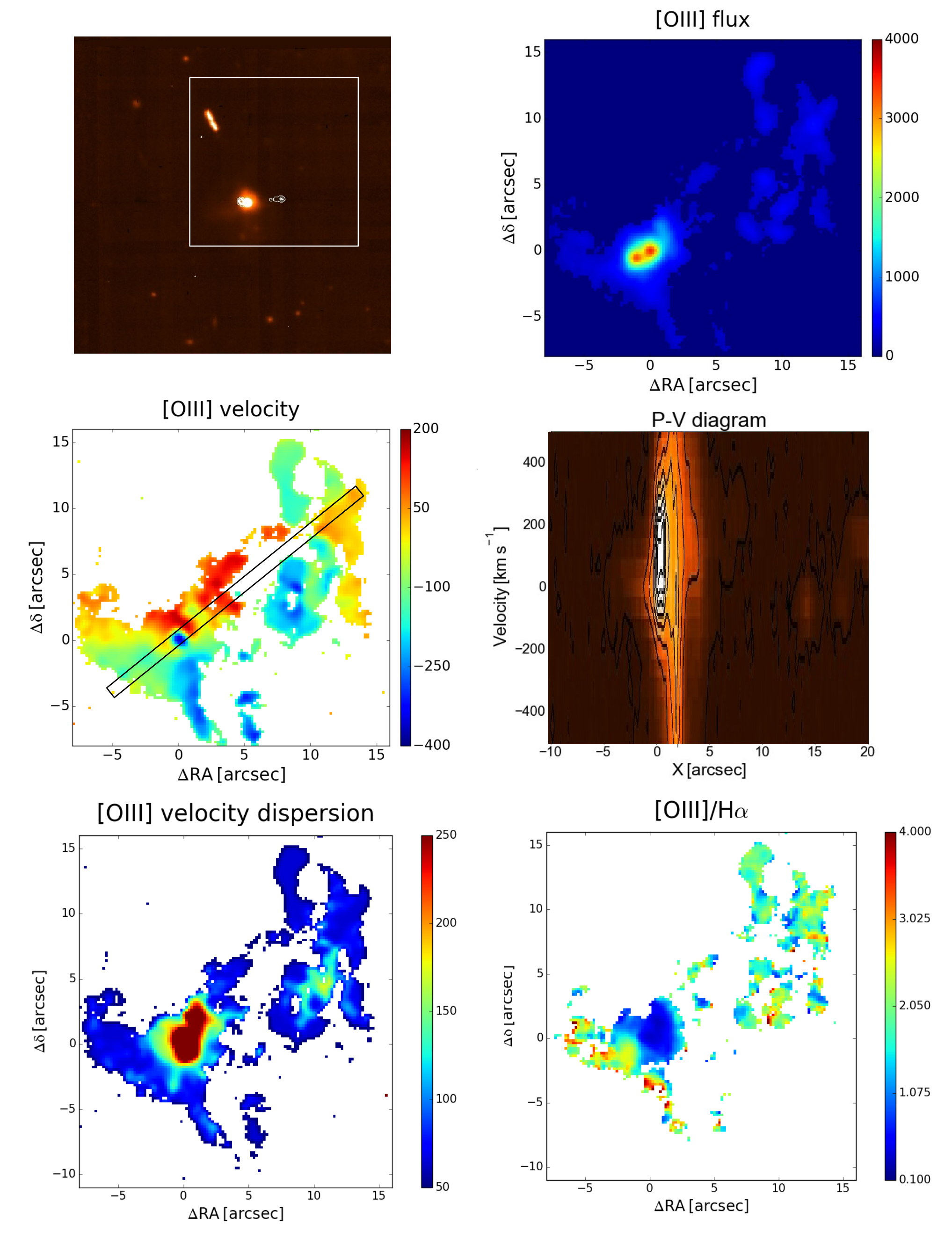}
\caption{3C~459, FR~II/BLO, 1$\arcsec$ = 3.58 kpc. Top left: radio contours overlaid onto the Muse optical continuum image .  Top right: \oiii\
emission line image extracted from the white square in the top left panel.  Middle: velocity field from the \oiii\ line and PV diagram (the synthetic slit  is centered on the nucleus, has a width of 10 pixels and it is oriented at an angle of 35$^\circ$ from the X axis). Bottom: \oiii\ velocity dispersion and \oiii/\ha\ ratio. }}
\label{3C459}
\end{figure*}  

\section{Ionization properties of the extended emission line regions}

\begin{table*}
\caption{Synthetic aperture for the off-nuclear spectra and diagnostic line ratios}
\begin{tabular}{ l r r  r r | r c c r | r  r r r}
\hline
Name (type) & Size & \multicolumn{3}{c}{Distance ($\arcsec$)}  (kpc)  &
[O~III]/H$\beta$ & [N~II]/H$\alpha$ & [S~II]/H$\alpha$ & [O~I]/H$\alpha$ & Type\\
\hline
3C~017   (BLO) & 1.2 &  2.4  W &  0.0 N & 10.5 &    4.74 & 0.74 & 0.64 & 0.31 & HEG  \\
3C~018   (BLO) & 2.0 &  3.4  W &  0.6 N & 13.0 &   16.72 & 1.11 & 0.60 & 0.13 & HEG  \\ 
3C~033   (HEG) & 2.0 &  4.0 ~E &  3.6 N &  6.3 &   14.54 & 0.34 & 0.26 & 0.12 & HEG  \\ 
3C~063   (HEG) & 4.0 &  8.6  W &  1.4 S & 30.1 &    1.18 & 0.30 & 0.52 & 0.22 & Peculiar \\ 
3C~318.1 (---) & 1.6 &  0.4 ~E & 13.0 S & 11.7 &    0.23 & 1.24 & 0.43 & 0.16 & Peculiar \\  
3C~327   (HEG) & 2.0 &  0.4 ~E &  3.8 N &  8.1 &   11.64 & 1.33 & 0.75 &      & HEG  \\  
3C~353   (LEG) & 4.0 &  0.8 ~E & 19.8 N & 12.7 & $<$1.90 & 0.66 & 0.69 &      & LEG  \\ 
3C~386   (---) & 4.0 & 11.4 ~E &  0.0 N &  4.2 &    0.66 & 1.01 & 0.28 &      & Peculiar \\  
3C~403   (HEG) & 4.0 &  6.8 ~E &  6.6 N & 11.3 & $>$9.50 & 1.15 & 0.72 &      & HEG  \\   
3C~424   (LEG) & 4.0 &  5.8 ~E &  6.8 S &  2.4 &    0.77 & 0.68 & 0.58 & 0.18 & Peculiar \\  
3C~442   (LEG) & 1.2 &  0.4  W &  4.6 S &  3.0 & $>$0.52 & 3.00 & 1.74 & 0.40 & LEG  \\  
3C~445   (BLO) & 4.0 & 11.8  W &  9.8 S & 17.0 &    8.20 & 0.15 & 0.35 & 0.08 & HEG  \\  
3C~458   (HEG) & 2.0 &  7.8 ~E & 11.6 N & 80.2 &    5.60 & 0.58 & 0.99 & 0.53 & LEG  \\   
3C~459   (BLO) & 1.2 & 12.4  W &  9.8 N & 69.0 &    5.50 & 0.18 & 0.68 &      & HEG  \\   
  \hline
\end{tabular}
\label{tab2}

\smallskip
\small{Column description: (1) source name; (2) region size (arcseconds); (3)
  distance from the nucleus in arcseconds and (4) kpc, (5,6, and 7) diagnostic
  line ratios, (8) spectroscopic type}
\end{table*}

\begin{figure*}  
\centering{ 
\includegraphics[width=4.5cm]{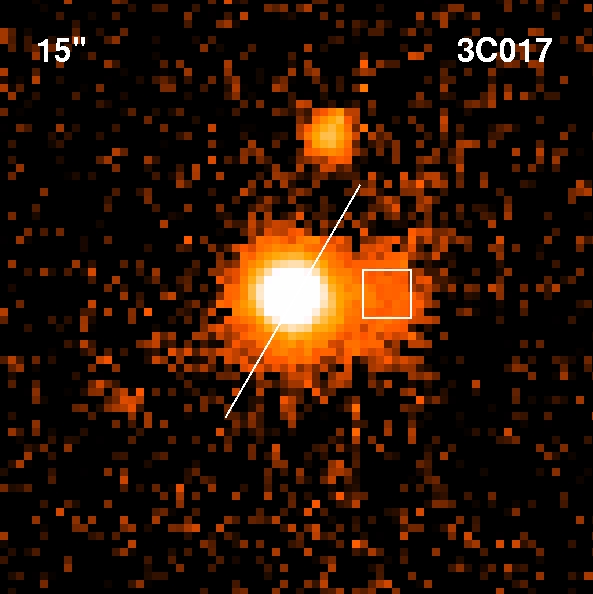}
\includegraphics[width=4.5cm]{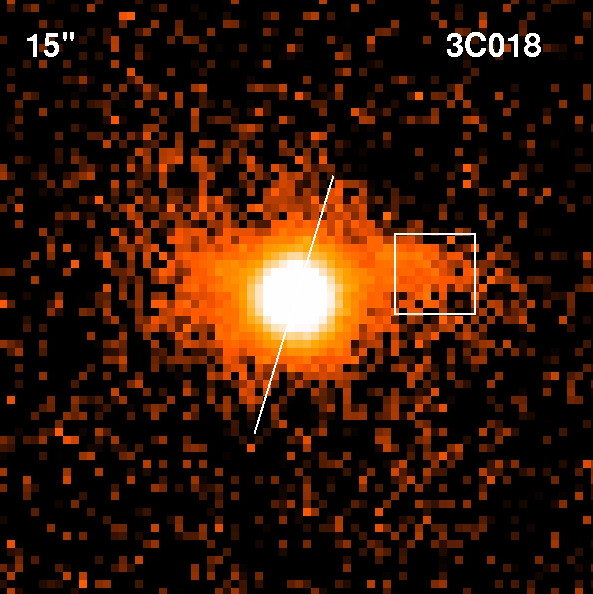}
\includegraphics[width=4.5cm]{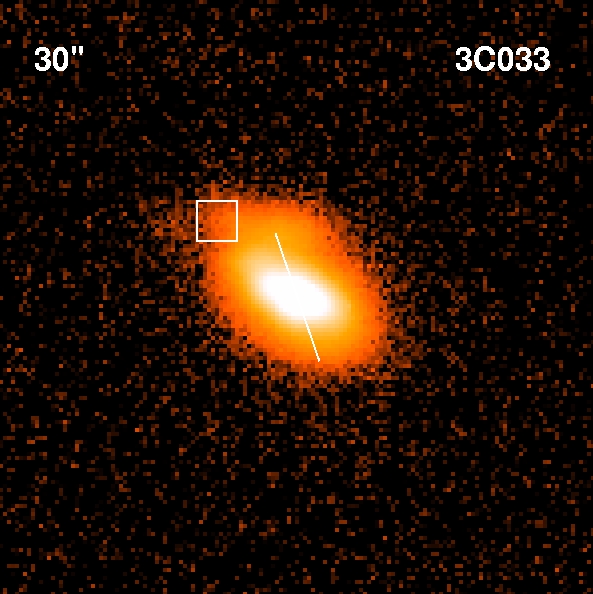}
\includegraphics[width=4.5cm]{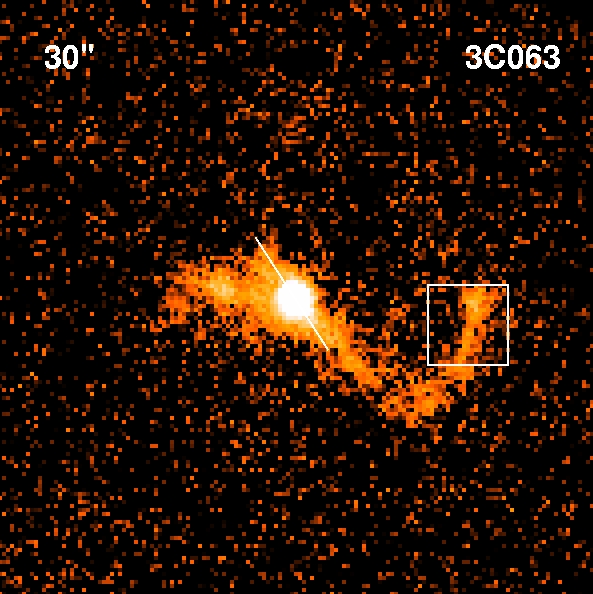}

\smallskip

\includegraphics[width=4.5cm,height=4.51cm]{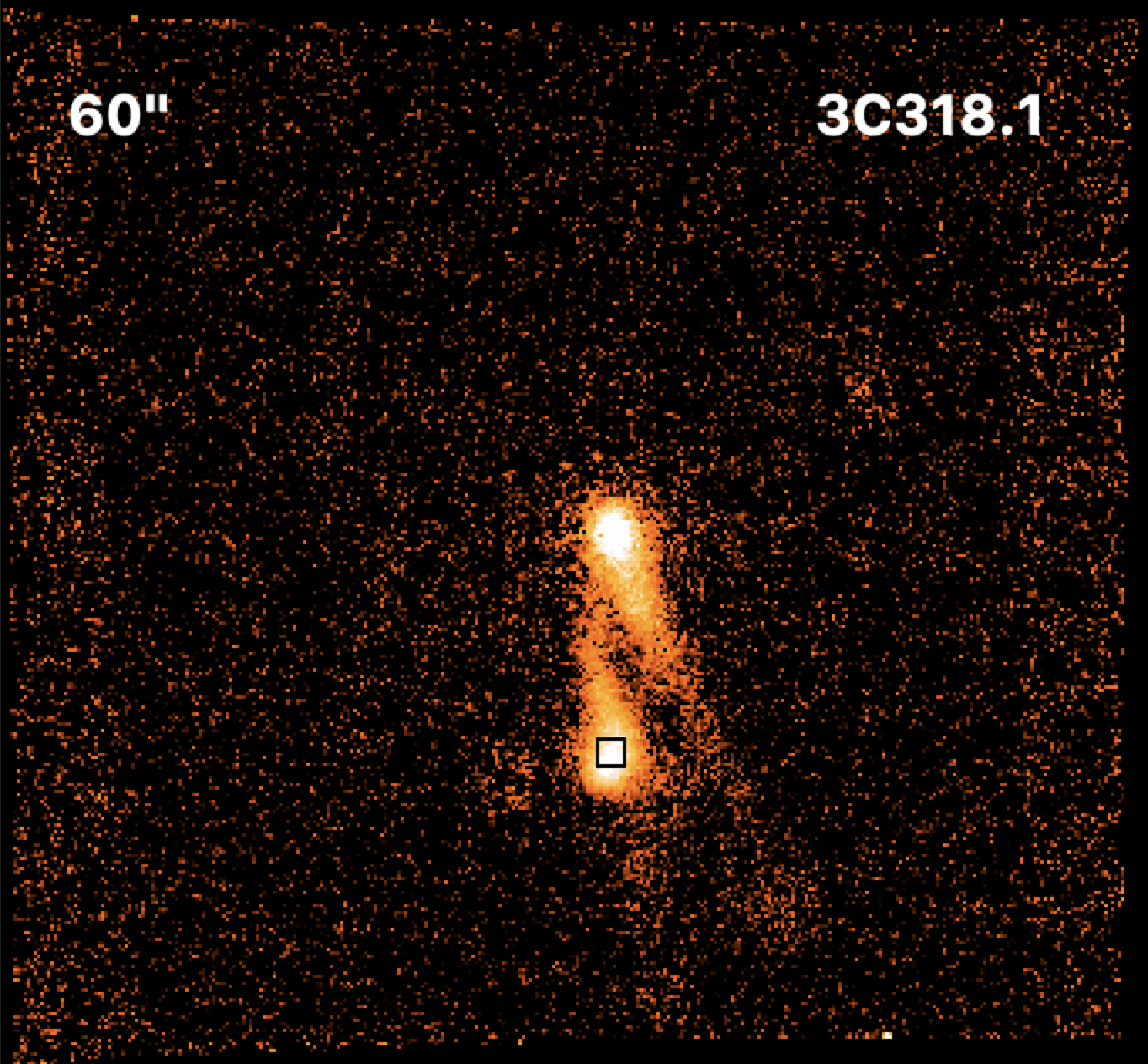}
\includegraphics[width=4.5cm]{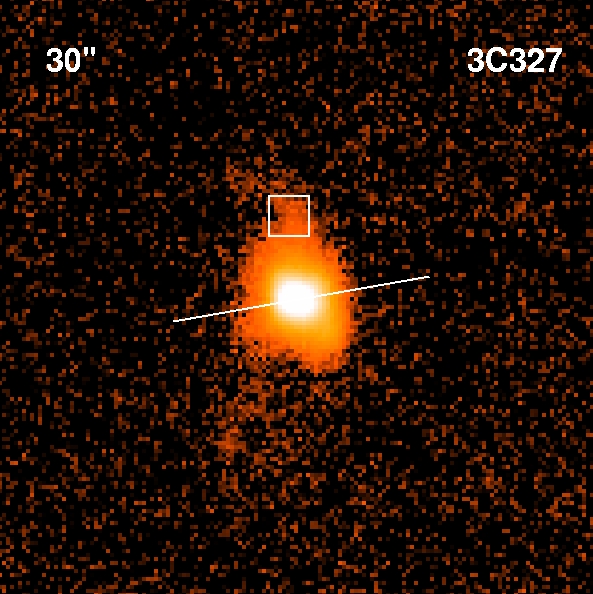}
\includegraphics[width=4.5cm]{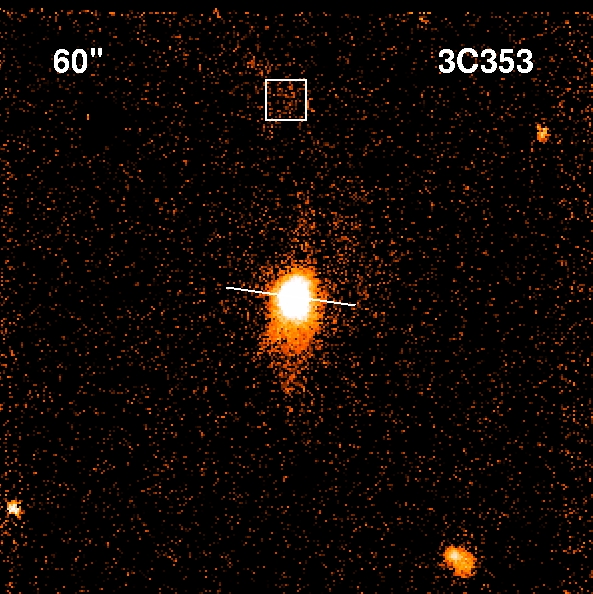}
\includegraphics[width=4.5cm]{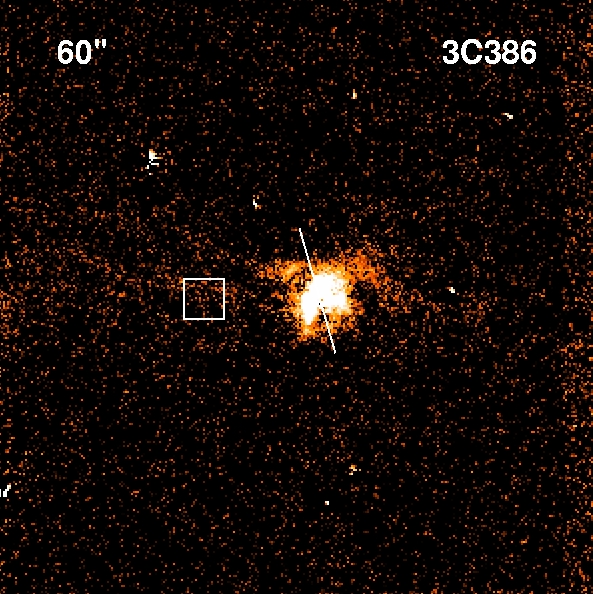}

\smallskip
\includegraphics[width=4.5cm]{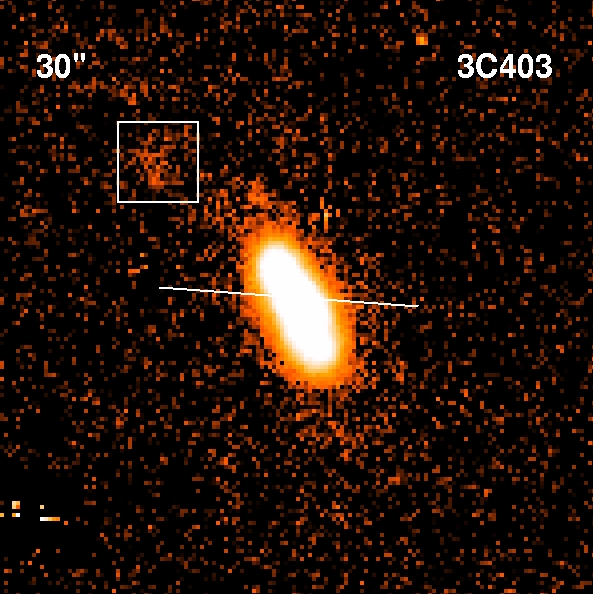}
\includegraphics[width=4.5cm]{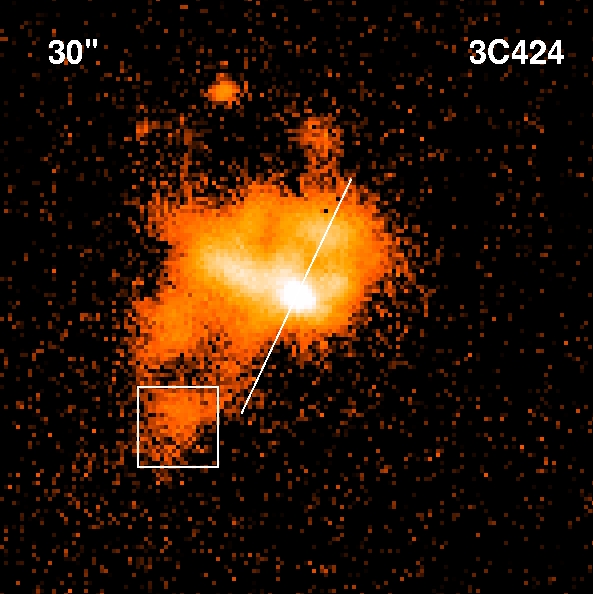}
\includegraphics[width=4.5cm]{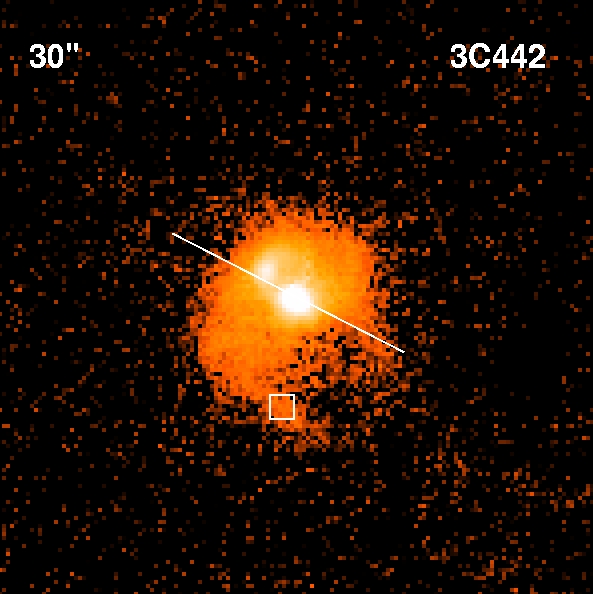}
\includegraphics[width=4.5cm]{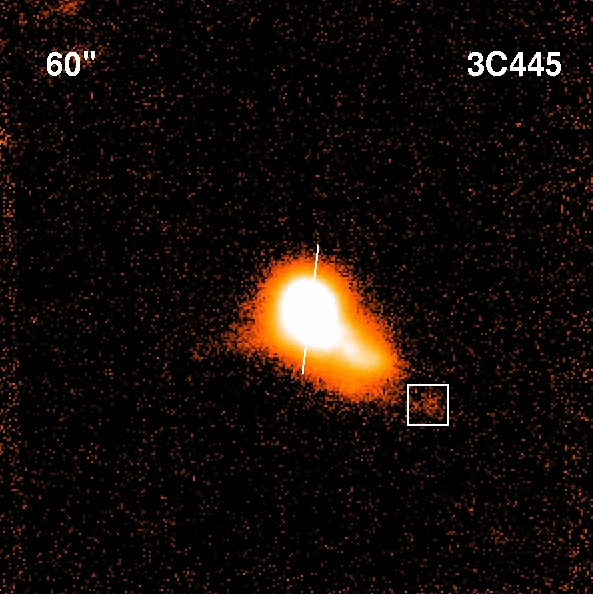}

\smallskip
\includegraphics[width=4.5cm]{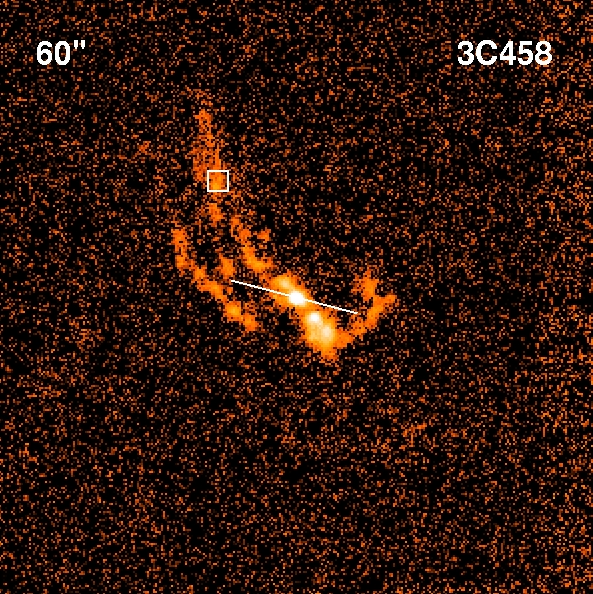}
\includegraphics[width=4.5cm]{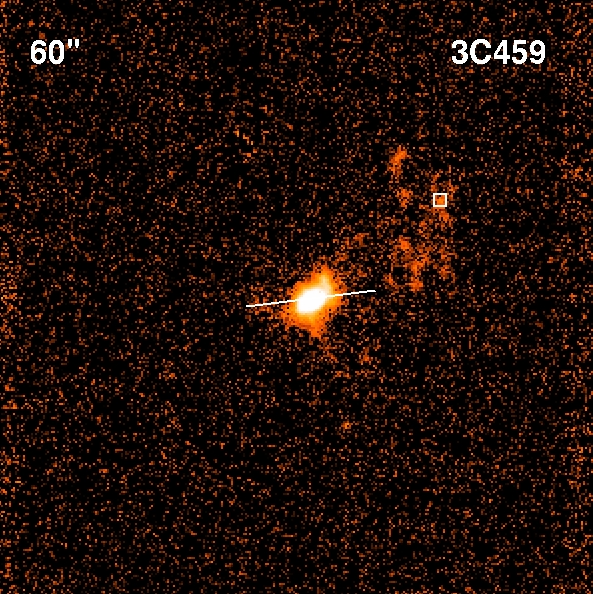}
\caption{Emission line images (in logarithmic scale) of the 14 radio
  galaxies observed in MURALES with extended emission lines. The
  fields of view are indicated in the upper left corner of each
  image. The white segments are parallel to the radio axes, the with
  boxes mark the synthetic aperture from which we extracted the
  off-nuclear spectra.}}
\label{franco}
\end{figure*}   

\begin{figure*}  
\centering{ 
\label{diagfig}
\includegraphics[width=18.5cm]{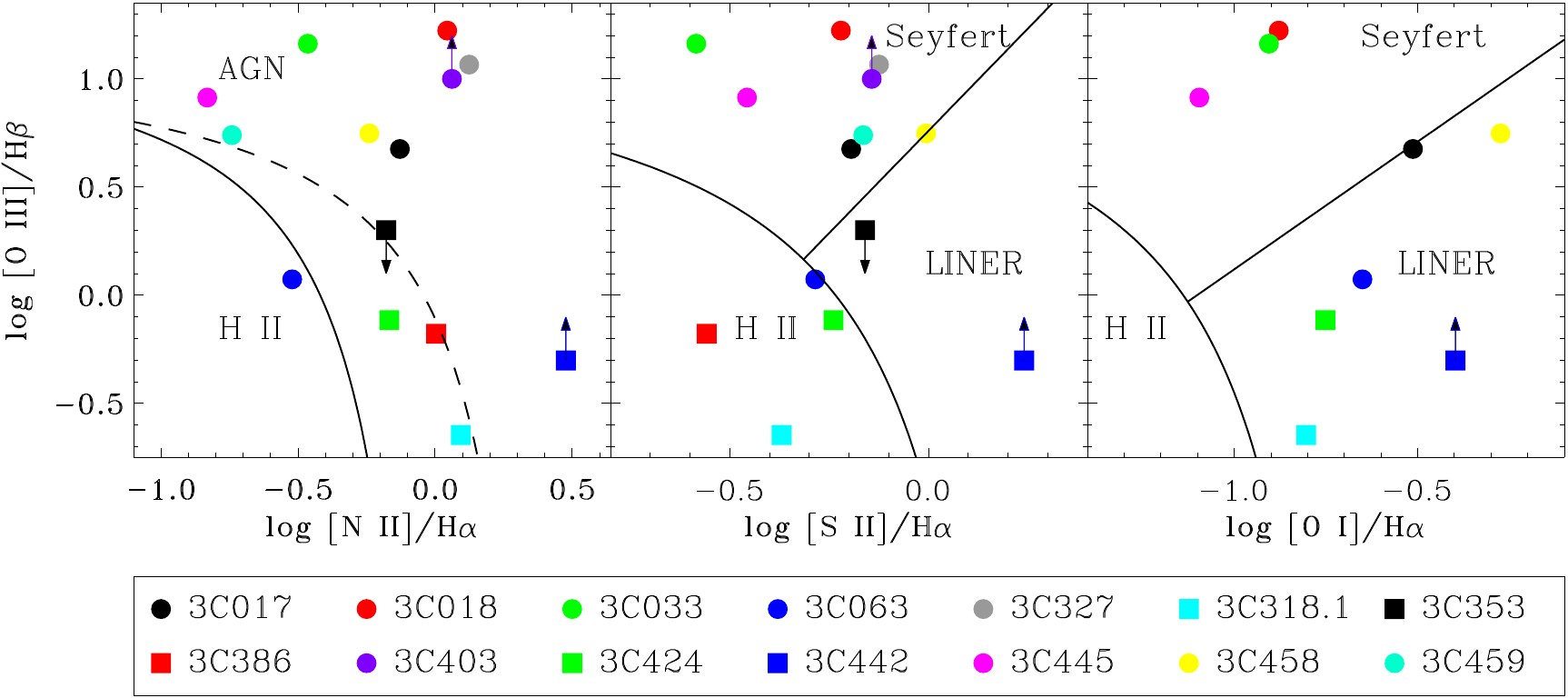}}
\caption{Location of the 13 FR~II radio galaxies with extended line
  emission in the spectroscopic diagnostic diagrams. Circles (squares)
  correspond to source classified as HEGs (LEGs or unclassified) based
  on their nuclear emission line ratios \citep{buttiglione10}. The
  solid lines separate star-forming galaxies, LINERs, and Seyferts
  \citep{kewley06}. }
\end{figure*}   

The integral field spectroscopic data allow us to explore the
ionization properties by comparing the emission line strength in
different spaxels (e.g., \citealt{singh13}), but this approach
requires to have detected at least four emission lines in each
spaxels. To increase the S/N, we extracted spectra from
synthetic apertures as far as possible from the nuclei, but still in
regions of sufficient signal to derive useful emission line
measurements. We discarded the sources which are too compact to obtain
off-nuclear measurements. The selected areas are marked in
Fig. \ref{franco} and listed in Tab. \ref{tab2}. They are located at a
median distance of $\sim$13 kpc from the nucleus, but reach distances
of $\sim$80 kpc. In Tab. \ref{tab2} we list the line ratios for each
source and derive a spectroscopic classification based on the location
of each source in the diagnostic diagrams defined by \citep{kewley06},
see Fig. \ref{diagfig}.

Seven sources fall into the region populated by Seyfert/HEGs: they are
all objects classified as HEGs also based on the nuclear line ratios
\citep{buttiglione10}, see Tab. \ref{tab2}. Similarly, we find the
same classification for the nucleus and for the extended region in two
LEGs, namely 3C~353 and 3C~442. However, there is one HEGs, 3C~458, in
which the emission line ratios measured on the extended emission
correspond to a different spectroscopic type, for which we derive a
LEG classification. In this case the extended emission line region is
located at very large distance from the nucleus (80 kpc). The
transition from high to low ionization appears to occur gradually,
based on the decrease of the \oiii/\ha\ ratio with distance (see
Fig. 18).

Finally, there are four sources with a peculiar behavior, namely 3C~63
(a HEG), 3C~318.1 and 3C~386 (both of uncertain spectral type), and
3C~424 (a LEG): they are located in different regions of the
diagnostic diagrams depending on the panel considered, an indication
that we might not be observing regions in which photoionization from
an active nucleus or from young stars is not the dominant
process. This is reminiscent of the results found with MUSE
observations by \citet{balmaverde18} for 3C~317: we argued that
ionization due to slow shocks \citep{dopita95} or collisional heating
from cosmic rays \citep{ferland08,ferland09,fabian11} might be
important. The general small sigma of the emission lines in these
regions, $\sim 50 - 100$ \kms, argues against the importance of
shocks, thus favoring ionization from energetic particles.

 \section{Discussion}
\label{discussion}

The MUSE data reveal a relationship between the radio morphology and
the warm ionized gas structures.  Objects belonging to the FR~I and
FR~II classes behave differently: we did not detect extended line
emission in three out of four FR~I despite their lower redshift
(ranging 0.018 to 0.073) with respect to FR~II. The only exception is
the FR~I, 3C~348.  In all the FR~II observed (with the only exception
of 3C~456) we observe ionized gas extending up to 20 effective radii
of the host. In most cases these structures exhibit a filamentary
shape, reminiscent of tidal tails (3C~327, 3C~348, 3C~353, 3C~386,
3C~403.1, 3C~442, 3C~458, and 3C~459). In some cases these filaments
seem to connect the radio galaxy with a galaxy at the same redshift
(e.g., 3C~403.1, 3C~018, and 3C~424), suggesting that FR~II radio
galaxies often inhabit a dynamic environment, as seen also by high
redshift studies \citep{chiaberge18}. In one case (3C~063) we observe
ionized gas emission around the expanding radio lobe, similarly to
3C~317 \citep{balmaverde18}. No apparent difference emerges between
the FR~II of HEG or LEG spectral class.

Overall, we observe extended line emission structures in FR~II are
preferentially (but not exactly) oriented perpendicularly to the radio
jets. The tendency for the radio ejection to occur along the rotation
axis of gas/dust disks has been strongly debated in past, resulting in
contrasting results. \citet{heckman85} found that in radio galaxies
with radio emission larger than $>$100 kpc, the radio jet and the gas
rotational axis are typically aligned to within a few tens of degree.
We defer the detailed analysis of the velocity fields and of
the relationship between radio and line emission to a forthcoming
paper.

We compare our results with the study of extended optical
emission-lines gas in low redshift radio galaxies, presented by
\citet{baum88,baum89}, in which they presented narrow band images of a
representative sample of powerful radio galaxies. The spatial
resolution of the images was $\sim$1.5-2$\arcsec$, down to a 
surface brightness
limit in the range 10$^{-17}$ - 10$^{-16}$ erg s$^{-1}$ cm$^{-2}$
arcsec$^{-2}$. They commonly detected extended optical-line-emitting
gas and in some cases filaments of ionized gas on scales of 40-100
kpc, departing from the host galaxy. Seven MURALES objects are in
common with their sample: the comparison of the contour map with MUSE
images shows that, although with lower spatial resolution and
sensitivity, their narrow band filter images already captures the
overall morphology of the extended gaseous structures. For example,
they observed the S-shaped regions of line emission in 3C~063 and
around the cavity in 3C~317.

The HST images of emission lines present a rather different picture
with respect to the MUSE data, in particular both \citet{privon08} and
\citet{baldi19} found a general alignment between the line and radio
axis. This is likely due to the fact that HST is able to resolve the
innermost regions of the narrow line regions while it is not sensitive
to the large scale low brightness filaments. 

In many galaxies the gas shows ordered motion, usually indicative of
rotation on both the galactic and the larger scale. However, several
sources depart from this description, e.g. 3C~063, 3C~424, and 3C~458,
while in other objects the gas velocity field is highly complex.

\citet{baum90} obtained long-slit optical spectra along the extended
emission line features (see also \citealt{heckman85}). Inspecting the
velocity curves along different position angles, they classified the
emission line nebulae into rotators, calm rotators or violent
rotators. Their kinematical classification corresponds generally to
the morphological separation into FR~I and FR~II radio galaxies:
rotators or violent rotators are nearly always associated with FR~II
HEGs, while instead the calm non rotators are preferentially of FR~I
or FR~II LEGs type. Our MUSE data reveal that objects of these latter
two classes (e.g. 3C~353) actually show rotational motions that was
not spatially resolved in their data. This applies also to at least
two FR~Is of our sample (namely, 3C~040 and 3C~348). We thus believe
the objects belonging to the ``calm non rotators'' show well defined
rotation with data of adequate spatial resolution, like in this work.

The maps of velocity dispersion often show quasi-linear regions of
high velocity dispersion: the best examples are 3C~033 and 3C~403 (but
a similar feature is seen also in 3C~327, 3C~348, 3C~353, and 3C~386).
In 3C~033 the velocity dispersion reaches $\sim 200$ \kms, to be
compared with a value of $\sim$ 70 \kms\ across the remaining of the
emission line region, similar values are seen also in 3C~403. These
features do not show a preferential relation with the radio axis,
while they are generally perpendicular to the axis along which the
emission line region is elongated. The origin of this effect is
unclear.  The high velocity dispersion could be due to interstellar
material shocked by the expanding radio jet (\citealt{couto17}) or to
higher turbulence due to the presence of strong backflow toward the
inner regions produced by the expansion of the radio cavity
(e.g. \citealt{cielo17,antonuccio10}).

From the point of view of the line ratios we find a variety of
behaviors, with sources in which the \oiii/\ha\ ratio decreases with
radius (e.g., 3C~018 and 3C~458) but also sources in which it
increases (e.g., 3C~017 and 3C~445). In 3C~033, 3C~327, and 3C~403,
the map of the emission line ratios show the pattern reminiscent of a
ionization cone, in which the [O~III]/\ha\ flux ratio map shows a
biconical shape.

Our analysis of the large scale emission line regions shows that
generally the spectroscopic type derived from the line ratios is
consistent with the spectral classification based on the nuclear
properties. In other sources, the emission line ratios are
inconsistent with those produced by photo-ionization. Given the low
velocities observed, leaving the possibility of cosmic rays heating or
ionization due to slow shocks. This suggests that at very large
distances the geometric dilution of the nuclear radiation field allows
other ionization mechanism to prevail.

The spectra obtained also revealed significant departures from single
gaussian profiles for the emission lines, mostly in the nuclear
regions (see, e.g., the high velocity component visible in the P-V
diagram of 3C~403 reaching $\sim$800 \kms), that usually show
asymmetric profiles with prominent high velocity wings. In a
forthcoming paper, we will analyze in detail the line shapes, looking
for the signature of outflows. In particular we will investigate the
role of relativistic collimated jets in dragging outflows.

It has been suggested that LEGs and HEGs are related to a different
accretion process, hot versus cold gas
\citep{hardcastle07,buttiglione10,baum95}. We are probably witnessing
a somewhat different situation: while we confirm the general paucity
of ionized gas in FR~Is, it is somewhat surprising that the FR~II/LEGs
show similar ionized gas structures to those of FR~IIs/HEGs and
BLOs. In fact LEGs are thought to be powered by hot accretion but we
detect a large reservoir of warm, emitting line gas, similar to those
found in HEGs.  This seems to indicate that FR~IIs of LEG and HEG type
both inhabit regions characterized by a rich gaseous environment on
kpc scale.  However, the extended ionized tails and filaments observed
by MUSE are only the portion of the gas reservoir that is ionized by
AGN photoionization or shocks. The ionized gas structures revealed by
MUSE are likely to be only the tip of the iceberg of a larger amount
of colder (atomic and molecular) gas. Observations of the H~I and CO
emission lines are needed to obtain a robust estimate of the entire
gas content in these sources.

From the spatially resolved diagnostic diagram, we might envisage that
the spectral type of a FR~II varies with time, responding to changes
of the accretion rate on a timescale much shorter than its
lifetime. The measurements of the line ratios on large scale and the
general consistency (with the notable exception of 3C~458) with the
nuclear classification (see Tab.\ref{tab2}) indicates that this does
not generally occur on a timescale shorter than the light travel time
from the nucleus to where these features are located, i.e., $\sim
5\times10^4$ years.

\section{Summary and conclusions}

We presented the first results of the MURALES project (MUse RAdio Loud
Emission lines Snapshot) obtained from VLT/MUSE optical integral field
spectroscopic observations of 20 3C sources with z $<$ 0.3 observed in
Period 99, i.e., between June and July 2017. All classes of radio
morphology (FR~I and FR~II) and optical spectroscopic classification
are represented with an almost equal share of LEGs (including four
FR~II/LEGs) and FR~IIs HEGs and BLOs.  The distribution of redshift
and radio power are not statistically distinguishable from the entire
population of 114 3C radio galaxies at z$<$0.3. Our sub-sample is
therefore well representative of the population of powerful, low
redshift, radio galaxies. 

In this present paper we focused on the properties of the ionized gas.
One of the most interesting result which emerges is the detection of
emission line regions extending up to $\sim 100$ kpc. This is made
possible by the unprecedented depth of the line emission images,
reaching brightness levels, as low as a few $10^{-17}$ erg s$^{-1}$
cm$^{-2}$ arcsec$^{-2}$, an order of magnitude deeper that the
previous studies of emission lines in radio galaxies, despite the
rather short exposure time of 20$^\prime$.

We detected emission lines in all sources. The emission line region is
dominated by a compact component on a scale of a few kpc, likely the
classical narrow line region. Large scale ($\sim 5 - 100$ kpc) ionized
gas is seen in all but one (3C~456) of the 15 FR~II radio galaxies
observed. Usually these structures appear as elongated filaments. Only
one FR~I (3C~348) shows extended emission lines. In some cases the
MUSE field of view (1$\times$1 arcmin square) covers the whole radio
structure (e.g., 3C~017 and 3C~063), but in most cases the radio
emission extends well beyond the portion of the sky covered by MUSE.

We found that the line emission structures are preferentially (but not
exactly) oriented perpendicularly to the radio jets. Conversely, large
scale gas structures aligned with the jets or surrounding the radio
lobes (such as those found in 3C~317) are found only in one source,
namely 3C~063. We are most likely observing gas structure associated
with the secular fueling of the central SMBH. We defer the detailed
analysis of the spatial relation between radio and line emission to a
forthcoming paper.

The interaction between the AGN and the external medium is probably
confined within the innermost regions, as indicated by the HST images,
where most of the ionized gas is located. These regions are not
properly spatially resolved by the MUSE emission line images. An
analysis of the nuclear line profiles will be used elsewhere to study
AGN driven outflows.

We presented also maps of emission line ratios
(\oiii/\ha\ or \nii/\ha), to identify regions characterized by
different ionization states. Large spatial variations in line ratios
are observed in most galaxies: moving outward from the nucleus we
found sources with both increasing or decreasing gas excitation.

When possible, we compared the location of the nuclear and extended
line emission regions into the spectroscopic diagnostic diagrams for
13 FR~IIs: seven (two) sources are classified as HEGs and BLOs (LEGs)
in the nuclear regions similarly show a high (low) excitation spectrum
at larger distances. Conversely, in the HEG 3C~458, the spectrum at
$\sim$ 80 kpc from the nucleus leads to a LEG classification. In the
last three sources (3C~63, 3C~386, and 3C~424) we are probably
observing regions in which collisional heating from cosmic rays,
rather than photoionization, is the dominant process.

For most sources we were able to produce the velocity and velocity
dispersion 2D maps and a position-velocity diagram extracting the
velocity profile from a synthetic long-slit aperture aligned with
extended or interesting line structures. In most objects, the central
gas is in ordered rotation, but it is highly complex in a significant
fraction of our sample. On larger scales, we usually observed regular
velocity fields, with small gradients.

The MUSE images confirm the general paucity of ionized gas in FR~Is,
while FR~II/LEGs show ionized gas structures morphologically similar
to those of FR~IIs/HEGs and BLOs. This could indicate that FR~IIs of
both LEG and HEG type apparently inhabit regions characterized by a
similar content of gas and, possibly, similar triggering/feeding
mechanism. This would challenge previous suggestions that LEGs are
powered by hot accretion (and HEGs by cold gas) while we detected a
large reservoir of gas in both classes. To tackle the long standing
issue of the triggering and of the origin of the fueling material in
radio galaxies, we need to complement the information derived on the
ionized component of the interstellar medium with observations able to
trace also the cold gas component.

\begin{acknowledgements}
We acknowledge the anonymous referee for her/his report.
Based on observations made with ESO Telescopes at the La Silla Paranal
Observatory under programme ID 099.B-0137(A). The National Radio Astronomy
Observatory is a facility of the National Science Foundation operated under
cooperative agreement by Associated Universities, Inc. The radio images were retrieved from the NRAO VLA Archive Survey,
(c) AUI/NRAO, available at http://archive.nrao.edu/nvas/.
B.B. acknowledge
financial contribution from the agreement ASI-INAF I/037/12/0. 
We acknowledge financial contribution from the agreement ASI-INAF n. 2017-14-H.O.
\end{acknowledgements}


\begin{thebibliography}{48}
\expandafter\ifx\csname natexlab\endcsname\relax\def\natexlab#1{#1}\fi

\bibitem[{{Antonuccio-Delogu} \& {Silk}(2010)}]{antonuccio10}
{Antonuccio-Delogu}, V. \& {Silk}, J. 2010, \mnras, 405, 1303

\bibitem[{{Bacon} {et~al.}(2010){Bacon}, {Accardo}, {Adjali}, {Anwand},
  {Bauer}, {Biswas}, {Blaizot}, {Boudon}, {Brau-Nogue}, {Brinchmann},
  {Caillier}, {Capoani}, {Carollo}, {Contini}, {Couderc}, {Daguis{\'e}},
  {Deiries}, {Delabre}, {Dreizler}, {Dubois}, {Dupieux}, {Dupuy}, {Emsellem},
  {Fechner}, {Fleischmann}, {Fran{\c{c}}ois}, {Gallou}, {Gharsa}, {Glindemann},
  {Gojak}, {Guiderdoni}, {Hansali}, {Hahn}, {Jarno}, {Kelz}, {Koehler},
  {Kosmalski}, {Laurent}, {Le Floch}, {Lilly}, {Lizon}, {Loupias}, {Manescau},
  {Monstein}, {Nicklas}, {Olaya}, {Pares}, {Pasquini}, {P{\'e}contal-Rousset},
  {Pell{\'o}}, {Petit}, {Popow}, {Reiss}, {Remillieux}, {Renault}, {Roth},
  {Rupprecht}, {Serre}, {Schaye}, {Soucail}, {Steinmetz}, {Streicher}, {Stuik},
  {Valentin}, {Vernet}, {Weilbacher}, {Wisotzki}, \& {Yerle}}]{bacon10}
{Bacon}, R., {Accardo}, M., {Adjali}, L., {et~al.} 2010, in Society of
  Photo-Optical Instrumentation Engineers (SPIE) Conference Series, Vol. 7735,
  \procspie, 773508

\bibitem[{{Baldi} {et~al.}(2019){Baldi}, {Rodr{\'\i}guez Zaur{\'\i}n},
  {Chiaberge}, {Capetti}, {Sparks}, \& {McHardy}}]{baldi19}
{Baldi}, R.~D., {Rodr{\'\i}guez Zaur{\'\i}n}, J., {Chiaberge}, M., {et~al.}
  2019, \apj, 870, 53

\bibitem[{{Baldwin} {et~al.}(1981){Baldwin}, {Phillips}, \&
  {Terlevich}}]{baldwin81}
{Baldwin}, J.~A., {Phillips}, M.~M., \& {Terlevich}, R. 1981, \pasp, 93, 5

\bibitem[{{Balmaverde} {et~al.}(2018{\natexlab{a}}){Balmaverde}, {Capetti},
  {Marconi}, {Venturi}, {Chiaberge}, {Baldi}, {Baum}, {Gilli}, {Grandi},
  {Meyer}, {Miley}, {O'Dea}, {Sparks}, {Torresi}, \& {Tremblay}}]{balmaverde18}
{Balmaverde}, B., {Capetti}, A., {Marconi}, A., {et~al.} 2018{\natexlab{a}},
  \aap, 619, A83

\bibitem[{{Balmaverde} {et~al.}(2018{\natexlab{b}}){Balmaverde}, {Capetti},
  {Marconi}, \& {Venturi}}]{balmaverde18a}
{Balmaverde}, B., {Capetti}, A., {Marconi}, A.~r., \& {Venturi}, G.
  2018{\natexlab{b}}, \aap, 612, A19

\bibitem[{{Baum} \& {Heckman}(1989)}]{baum89}
{Baum}, S.~A. \& {Heckman}, T. 1989, \apj, 336, 702

\bibitem[{{Baum} {et~al.}(1990){Baum}, {Heckman}, \& {van Breugel}}]{baum90}
{Baum}, S.~A., {Heckman}, T., \& {van Breugel}, W. 1990, \apjs, 74, 389

\bibitem[{{Baum} {et~al.}(1988){Baum}, {Heckman}, {Bridle}, {van Breugel}, \&
  {Miley}}]{baum88}
{Baum}, S.~A., {Heckman}, T.~M., {Bridle}, A., {van Breugel}, W. J.~M., \&
  {Miley}, G.~K. 1988, The Astrophysical Journal Supplement Series, 68, 643

\bibitem[{{Baum} {et~al.}(1992){Baum}, {Heckman}, \& {van Breugel}}]{baum92}
{Baum}, S.~A., {Heckman}, T.~M., \& {van Breugel}, W. 1992, \apj, 389, 208

\bibitem[{{Baum} {et~al.}(1995){Baum}, {Zirbel}, \& {O'Dea}}]{baum95}
{Baum}, S.~A., {Zirbel}, E.~L., \& {O'Dea}, C.~P. 1995, \apj, 451, 88

\bibitem[{{Bennett} {et~al.}(2014){Bennett}, {Larson}, {Weiland}, \&
  {Hinshaw}}]{bennett14}
{Bennett}, C.~L., {Larson}, D., {Weiland}, J.~L., \& {Hinshaw}, G. 2014, \apj,
  794, 135

\bibitem[{{Buttiglione} {et~al.}(2009){Buttiglione}, {Capetti}, {Celotti},
  {Axon}, {Chiaberge}, {Macchetto}, \& {Sparks}}]{buttiglione09}
{Buttiglione}, S., {Capetti}, A., {Celotti}, A., {et~al.} 2009, \aap, 495, 1033

\bibitem[{{Buttiglione} {et~al.}(2010){Buttiglione}, {Capetti}, {Celotti},
  {Axon}, {Chiaberge}, {Macchetto}, \& {Sparks}}]{buttiglione10}
{Buttiglione}, S., {Capetti}, A., {Celotti}, A., {et~al.} 2010, \aap, 509, A6

\bibitem[{{Cappellari}(2017)}]{cappellari17}
{Cappellari}, M. 2017, \mnras, 466, 798

\bibitem[{{Cappellari} \& {Copin}(2003)}]{cappellari03}
{Cappellari}, M. \& {Copin}, Y. 2003, \mnras, 342, 345

\bibitem[{{Carniani} {et~al.}(2016){Carniani}, {Marconi}, {Maiolino},
  {Balmaverde}, {Brusa}, {Cano-D{\'{\i}}az}, {Cicone}, {Comastri}, {Cresci},
  {Fiore}, {Feruglio}, {La Franca}, {Mainieri}, {Mannucci}, {Nagao}, {Netzer},
  {Piconcelli}, {Risaliti}, {Schneider}, \& {Shemmer}}]{carniani16}
{Carniani}, S., {Marconi}, A., {Maiolino}, R., {et~al.} 2016, \aap, 591, A28

\bibitem[{{Chambers} {et~al.}(2016){Chambers}, {Magnier}, {Metcalfe},
  {Flewelling}, {Huber}, {Waters}, {Denneau}, {Draper}, {Farrow}, {Finkbeiner},
  {Holmberg}, {Koppenhoefer}, {Price}, {Rest}, {Saglia}, {Schlafly}, {Smartt},
  {Sweeney}, {Wainscoat}, {Burgett}, {Chastel}, {Grav}, {Heasley}, {Hodapp},
  {Jedicke}, {Kaiser}, {Kudritzki}, {Luppino}, {Lupton}, {Monet}, {Morgan},
  {Onaka}, {Shiao}, {Stubbs}, {Tonry}, {White}, {Ba{\~n}ados}, {Bell},
  {Bender}, {Bernard}, {Boegner}, {Boffi}, {Botticella}, {Calamida},
  {Casertano}, {Chen}, {Chen}, {Cole}, {Deacon}, {Frenk}, {Fitzsimmons},
  {Gezari}, {Gibbs}, {Goessl}, {Goggia}, {Gourgue}, {Goldman}, {Grant},
  {Grebel}, {Hambly}, {Hasinger}, {Heavens}, {Heckman}, {Henderson}, {Henning},
  {Holman}, {Hopp}, {Ip}, {Isani}, {Jackson}, {Keyes}, {Koekemoer}, {Kotak},
  {Le}, {Liska}, {Long}, {Lucey}, {Liu}, {Martin}, {Masci}, {McLean}, {Mindel},
  {Misra}, {Morganson}, {Murphy}, {Obaika}, {Narayan}, {Nieto-Santisteban},
  {Norberg}, {Peacock}, {Pier}, {Postman}, {Primak}, {Rae}, {Rai}, {Riess},
  {Riffeser}, {Rix}, {R{\"o}ser}, {Russel}, {Rutz}, {Schilbach}, {Schultz},
  {Scolnic}, {Strolger}, {Szalay}, {Seitz}, {Small}, {Smith}, {Soderblom},
  {Taylor}, {Thomson}, {Taylor}, {Thakar}, {Thiel}, {Thilker}, {Unger},
  {Urata}, {Valenti}, {Wagner}, {Walder}, {Walter}, {Watters}, {Werner},
  {Wood-Vasey}, \& {Wyse}}]{chambers16}
{Chambers}, K.~C., {Magnier}, E.~A., {Metcalfe}, N., {et~al.} 2016, arXiv
  e-prints, arXiv:1612.05560

\bibitem[{{Chiaberge} {et~al.}(2018){Chiaberge}, {Tremblay}, {Capetti}, \&
  {Norman}}]{chiaberge18}
{Chiaberge}, M., {Tremblay}, G.~R., {Capetti}, A., \& {Norman}, C. 2018, \apj,
  861, 56

\bibitem[{{Cielo} {et~al.}(2017){Cielo}, {Antonuccio-Delogu}, {Silk}, \&
  {Romeo}}]{cielo17}
{Cielo}, S., {Antonuccio-Delogu}, V., {Silk}, J., \& {Romeo}, A.~D. 2017,
  \mnras, 467, 4526

\bibitem[{{Couto} {et~al.}(2017){Couto}, {Storchi-Bergmann}, \&
  {Schnorr-M{\"u}ller}}]{couto17}
{Couto}, G.~S., {Storchi-Bergmann}, T., \& {Schnorr-M{\"u}ller}, A. 2017,
  \mnras, 469, 1573

\bibitem[{{Cresci} \& {Maiolino}(2018)}]{cresci18}
{Cresci}, G. \& {Maiolino}, R. 2018, Nature Astronomy, 2, 179

\bibitem[{{Dopita} \& {Sutherland}(1995)}]{dopita95}
{Dopita}, M.~A. \& {Sutherland}, R.~S. 1995, \apj, 455, 468

\bibitem[{{Fabian}(2012)}]{fabian12}
{Fabian}, A.~C. 2012, Annual Review of Astronomy and Astrophysics, 50, 455

\bibitem[{{Fabian} {et~al.}(2011){Fabian}, {Sanders}, {Williams}, {Lazarian},
  {Ferland}, \& {Johnstone}}]{fabian11}
{Fabian}, A.~C., {Sanders}, J.~S., {Williams}, R.~J.~R., {et~al.} 2011, \mnras,
  417, 172

\bibitem[{{Fanaroff} \& {Riley}(1974)}]{fanaroff74}
{Fanaroff}, B.~L. \& {Riley}, J.~M. 1974, MNRAS, 167, 31P

\bibitem[{{Ferland} {et~al.}(2008){Ferland}, {Fabian}, {Hatch}, {Johnstone},
  {Porter}, {van Hoof}, \& {Williams}}]{ferland08}
{Ferland}, G.~J., {Fabian}, A.~C., {Hatch}, N.~A., {et~al.} 2008, \mnras, 386,
  L72

\bibitem[{{Ferland} {et~al.}(2009){Ferland}, {Fabian}, {Hatch}, {Johnstone},
  {Porter}, {van Hoof}, \& {Williams}}]{ferland09}
{Ferland}, G.~J., {Fabian}, A.~C., {Hatch}, N.~A., {et~al.} 2009, \mnras, 392,
  1475

\bibitem[{{Finlez} {et~al.}(2018){Finlez}, {Nagar}, {Storchi-Bergmann},
  {Schnorr-M{\"u}ller}, {Riffel}, {Lena}, {Mundell}, \& {Elvis}}]{finlez18}
{Finlez}, C., {Nagar}, N.~M., {Storchi-Bergmann}, T., {et~al.} 2018, \mnras,
  479, 3892

\bibitem[{{Gaspari} {et~al.}(2017){Gaspari}, {Temi}, \&
  {Brighenti}}]{gaspari17}
{Gaspari}, M., {Temi}, P., \& {Brighenti}, F. 2017, Monthly Notices of the
  Royal Astronomical Society, 466, 677

\bibitem[{{Giacintucci} {et~al.}(2007){Giacintucci}, {Venturi}, {Murgia},
  {Dallacasa}, {Athreya}, {Bardelli}, {Mazzotta}, \& {Saikia}}]{giacintucci07}
{Giacintucci}, S., {Venturi}, T., {Murgia}, M., {et~al.} 2007, \aap, 476, 99

\bibitem[{{Gitti} {et~al.}(2012){Gitti}, {Brighenti}, \& {McNamara}}]{gitti12}
{Gitti}, M., {Brighenti}, F., \& {McNamara}, B.~R. 2012, Advances in Astronomy,
  2012, 950641

\bibitem[{{Hardcastle} {et~al.}(2007){Hardcastle}, {Evans}, \&
  {Croston}}]{hardcastle07}
{Hardcastle}, M.~J., {Evans}, D.~A., \& {Croston}, J.~H. 2007, \mnras, 376,
  1849

\bibitem[{{Heckman} {et~al.}(1985){Heckman}, {Illingworth}, {Miley}, \& {van
  Breugel}}]{heckman85}
{Heckman}, T.~M., {Illingworth}, G.~D., {Miley}, G.~K., \& {van Breugel},
  W.~J.~M. 1985, \apj, 299, 41

\bibitem[{{Hine} \& {Longair}(1979)}]{hine79}
{Hine}, R.~G. \& {Longair}, M.~S. 1979, MNRAS, 188, 111

\bibitem[{{Kewley} {et~al.}(2006){Kewley}, {Groves}, {Kauffmann}, \&
  {Heckman}}]{kewley06}
{Kewley}, L.~J., {Groves}, B., {Kauffmann}, G., \& {Heckman}, T. 2006, MNRAS,
  372, 961

\bibitem[{{Lynds}(1971)}]{lynds71}
{Lynds}, R. 1971, \apjl, 168, L87+

\bibitem[{{McCarthy} {et~al.}(1995){McCarthy}, {Spinrad}, \& {van
  Breugel}}]{mccarthy95}
{McCarthy}, P.~J., {Spinrad}, H., \& {van Breugel}, W. 1995, The Astrophysical
  Journal Supplement Series, 99, 27

\bibitem[{{McDonald} {et~al.}(2010){McDonald}, {Veilleux}, {Rupke}, \&
  {Mushotzky}}]{mcdonald10}
{McDonald}, M., {Veilleux}, S., {Rupke}, D. S.~N., \& {Mushotzky}, R. 2010,
  \apj, 721, 1262

\bibitem[{{McNamara} \& {Nulsen}(2007)}]{mcnamara07}
{McNamara}, B.~R. \& {Nulsen}, P.~E.~J. 2007, \araa, 45, 117

\bibitem[{{Privon} {et~al.}(2008){Privon}, {O'Dea}, {Baum}, {Axon}, {Kharb},
  {Buchanan}, {Sparks}, \& {Chiaberge}}]{privon08}
{Privon}, G.~C., {O'Dea}, C.~P., {Baum}, S.~A., {et~al.} 2008, The
  Astrophysical Journal Supplement Series, 175, 423

\bibitem[{{Russell} {et~al.}(2019){Russell}, {McNamara}, {Fabian}, {Nulsen},
  {Combes}, {Edge}, {Madar}, {Olivares}, {Salome}, \& {Vantyghem}}]{russsel19}
{Russell}, H.~R., {McNamara}, B.~R., {Fabian}, A.~C., {et~al.} 2019, arXiv
  e-prints, arXiv:1902.09227

\bibitem[{{Singh} {et~al.}(2013){Singh}, {van de Ven}, {Jahnke}, {Lyubenova},
  {Falc{\'o}n-Barroso}, {Alves}, {Cid Fernandes}, {Galbany},
  {Garc{\'\i}a-Benito}, {Husemann}, {Kennicutt}, {Marino}, {M{\'a}rquez},
  {Masegosa}, {Mast}, {Pasquali}, {S{\'a}nchez}, {Walcher}, {Wild}, {Wisotzki},
  \& {Ziegler}}]{singh13}
{Singh}, R., {van de Ven}, G., {Jahnke}, K., {et~al.} 2013, \aap, 558, A43

\bibitem[{{Spinrad} {et~al.}(1985){Spinrad}, {Marr}, {Aguilar}, \&
  {Djorgovski}}]{spinrad85}
{Spinrad}, H., {Marr}, J., {Aguilar}, L., \& {Djorgovski}, S. 1985, \pasp, 97,
  932

\bibitem[{{Tremmel} {et~al.}(2019){Tremmel}, {Quinn}, {Ricarte}, {Babul},
  {Chadayammuri}, {Natarajan}, {Nagai}, {Pontzen}, \& {Volonteri}}]{tremmel19}
{Tremmel}, M., {Quinn}, T.~R., {Ricarte}, A., {et~al.} 2019, Monthly Notices of
  the Royal Astronomical Society, 483, 3336

\bibitem[{{Veilleux} \& {Osterbrock}(1987)}]{veilleux87}
{Veilleux}, S. \& {Osterbrock}, D.~E. 1987, \apjs, 63, 295

\bibitem[{{Voit} {et~al.}(2015){Voit}, {Bryan}, {O'Shea}, \&
  {Donahue}}]{voit15}
{Voit}, G.~M., {Bryan}, G.~L., {O'Shea}, B.~W., \& {Donahue}, M. 2015, The
  Astrophysical Journal, 808, L30

\bibitem[{{Wylezalek} \& {Zakamska}(2016)}]{wylezalek16}
{Wylezalek}, D. \& {Zakamska}, N.~L. 2016, \mnras, 461, 3724

\end{thebibliography}

\begin{appendix}
\section{Nuclear and off-nuclear spectra of selected sources.}

\begin{figure*}  
\centering{ 

\includegraphics[width=9cm]{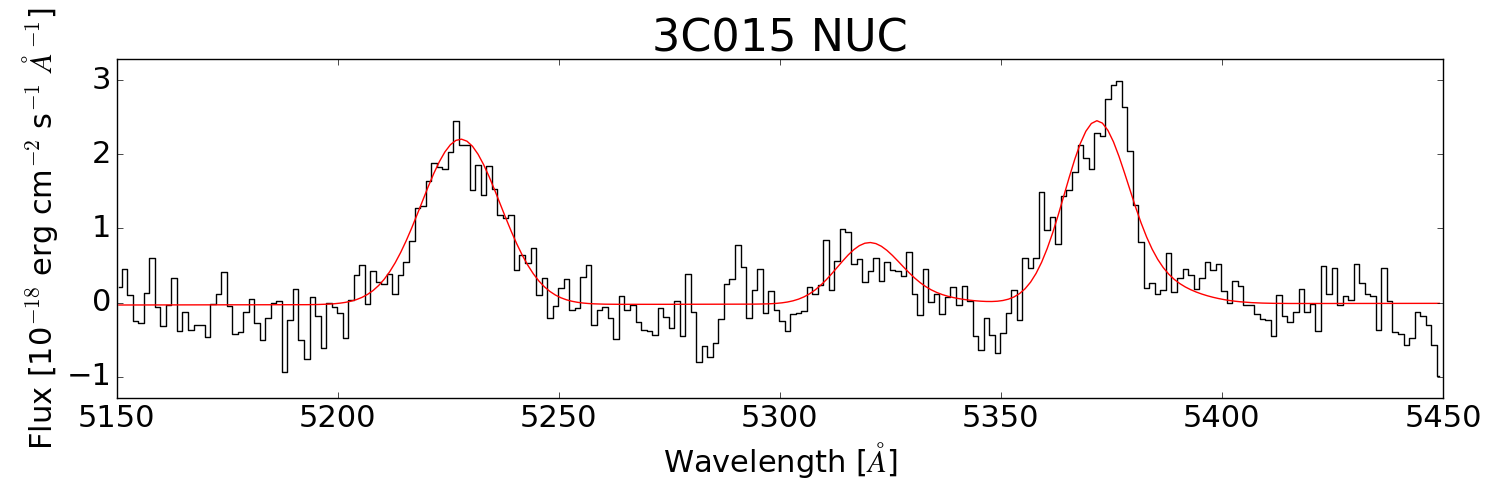}
\includegraphics[width=9cm]{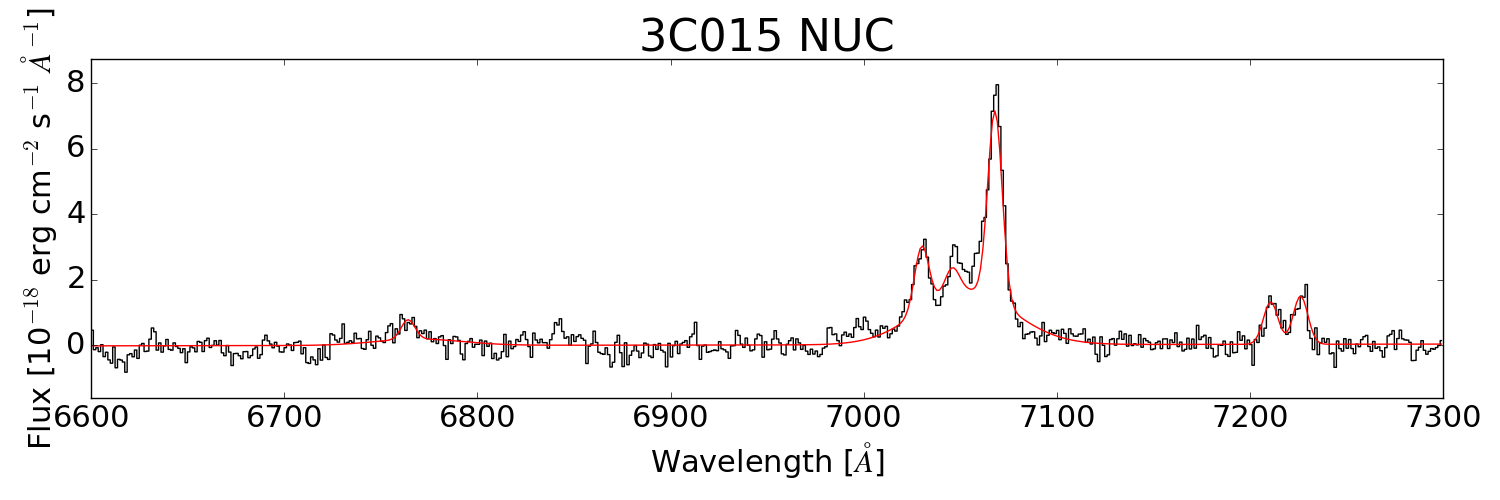}

\includegraphics[width=9cm]{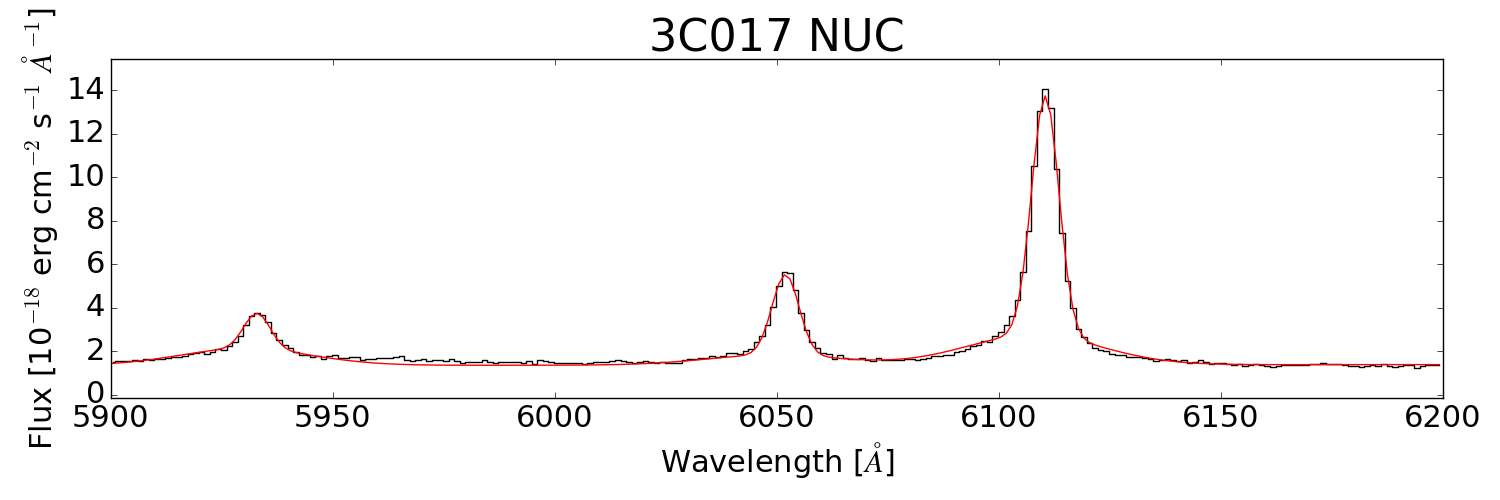}
\includegraphics[width=9cm]{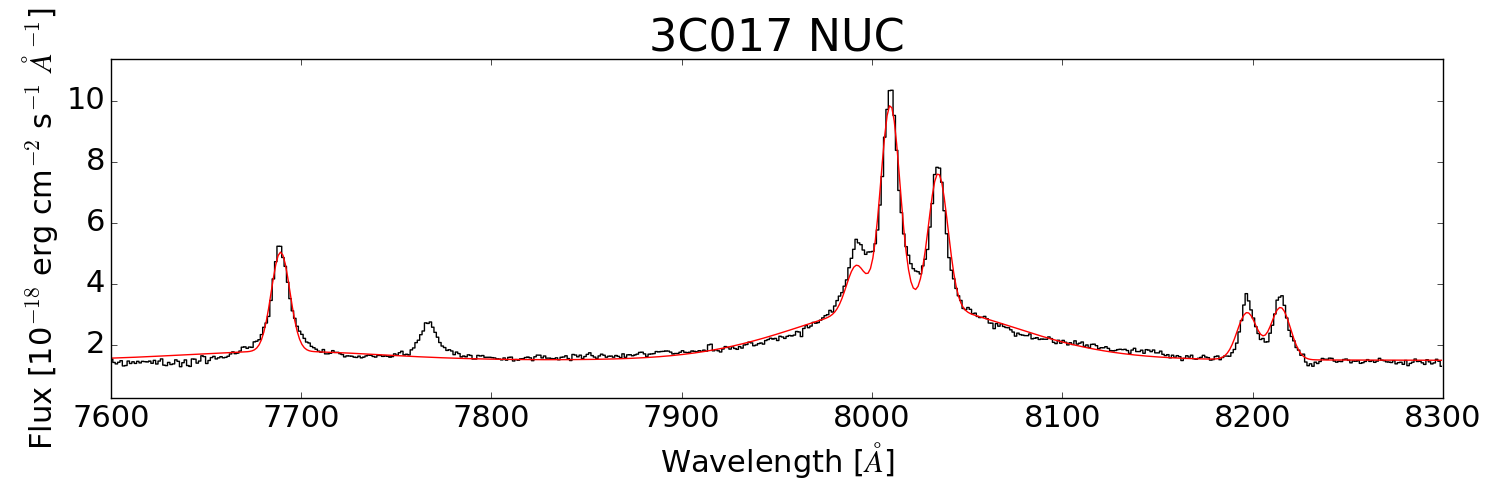}
\includegraphics[width=9cm]{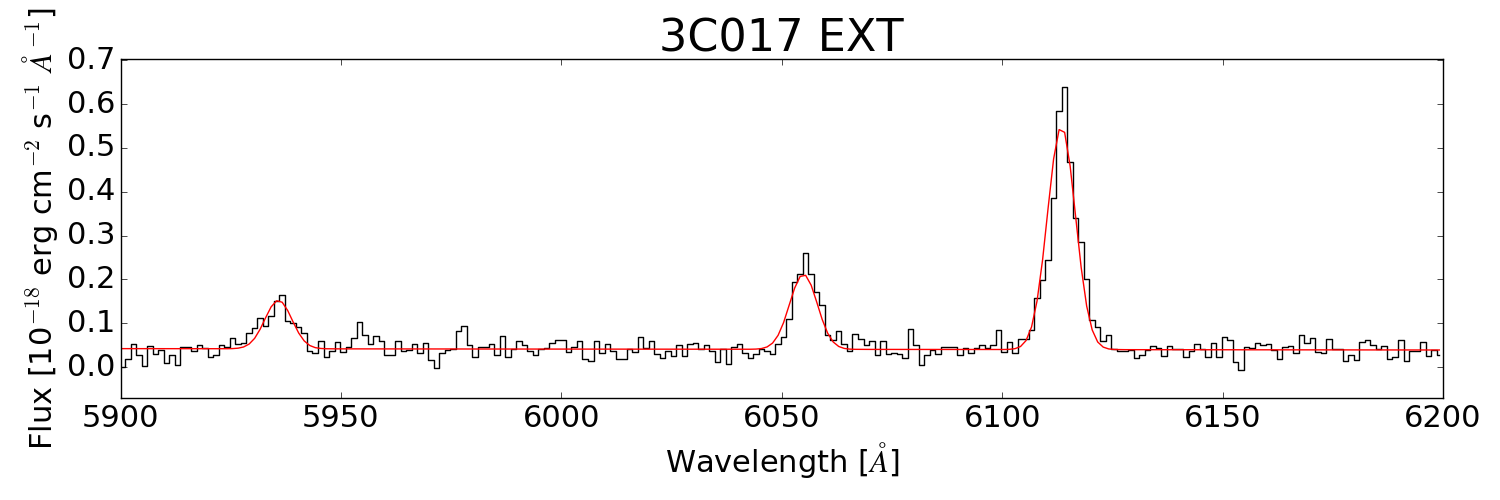}
\includegraphics[width=9cm]{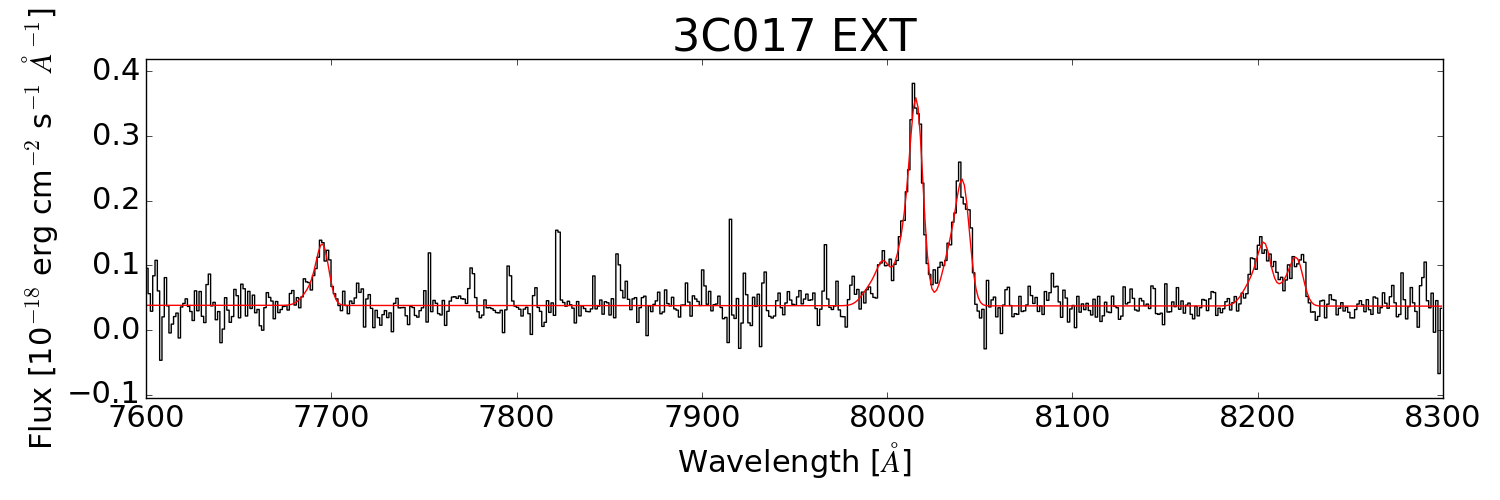}

\includegraphics[width=9cm]{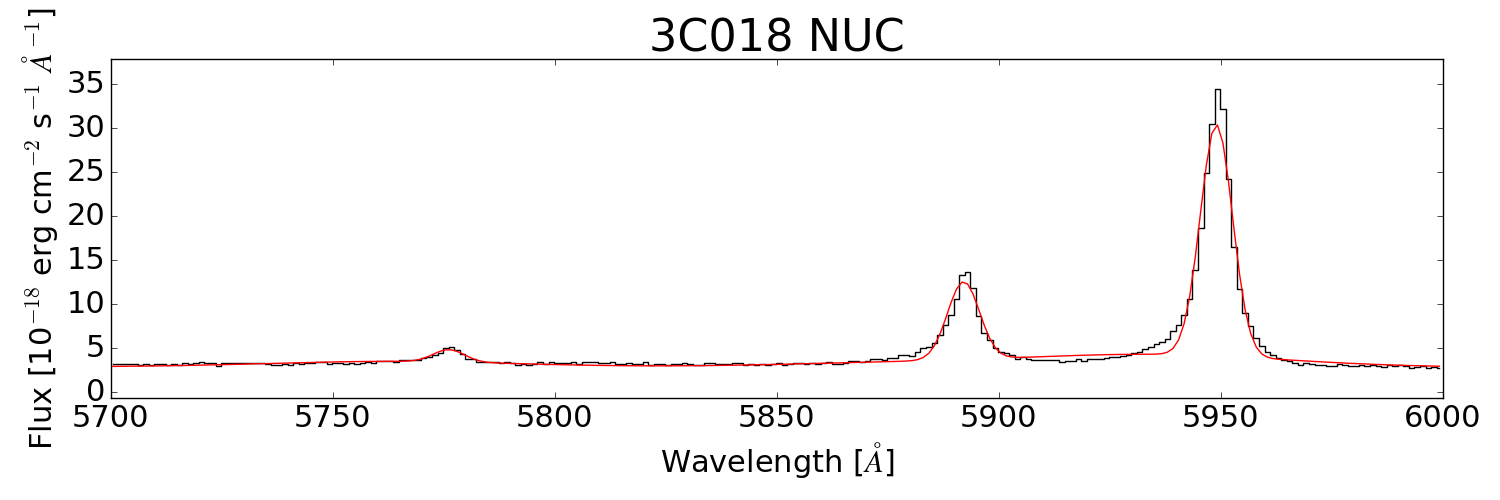}
\includegraphics[width=9cm]{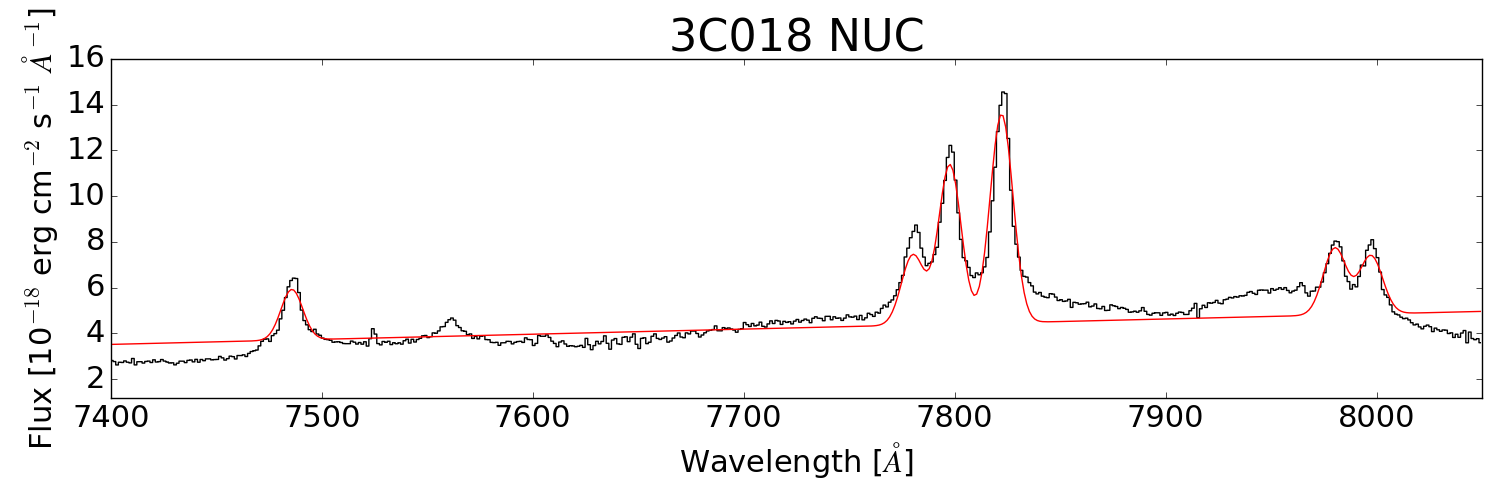}
\includegraphics[width=9cm]{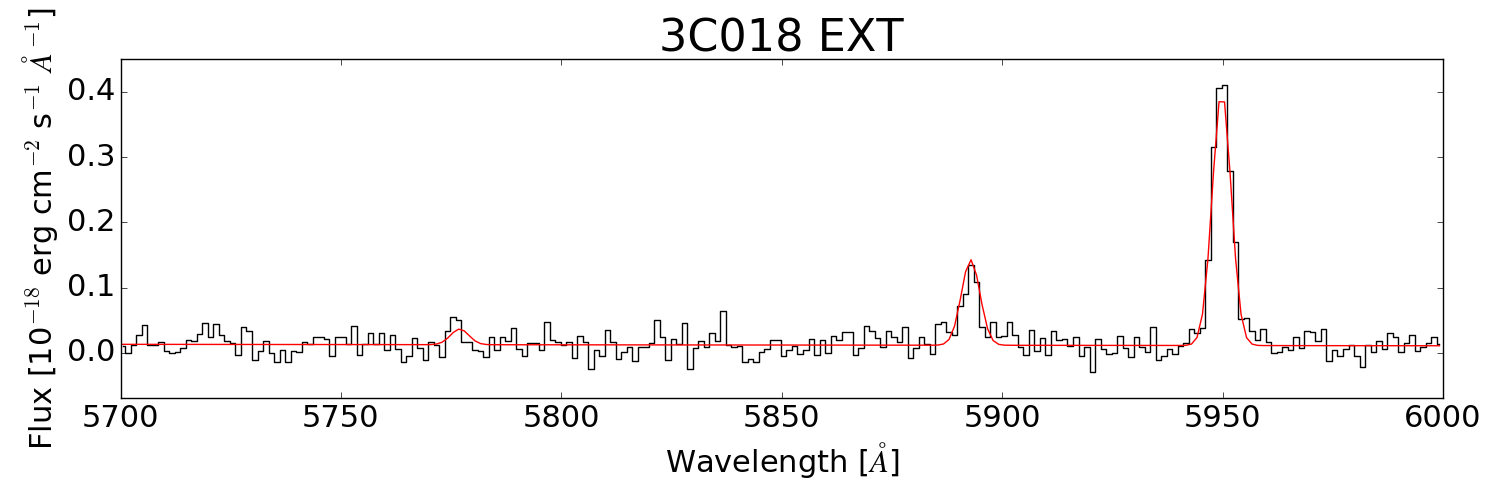}
\includegraphics[width=9cm]{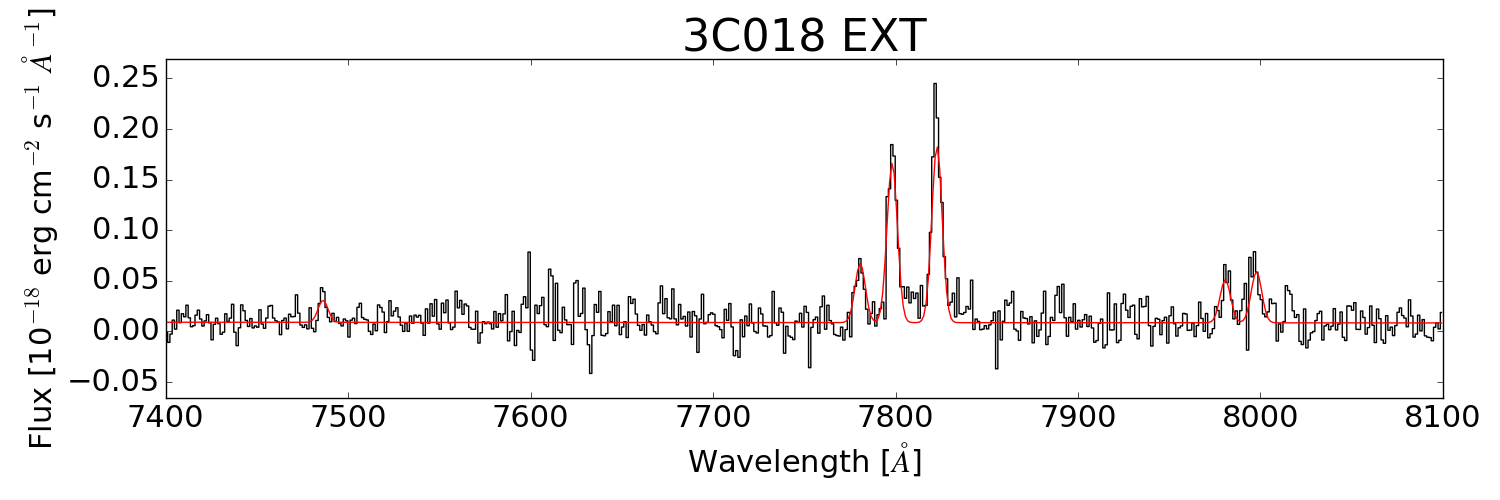}

\includegraphics[width=9cm]{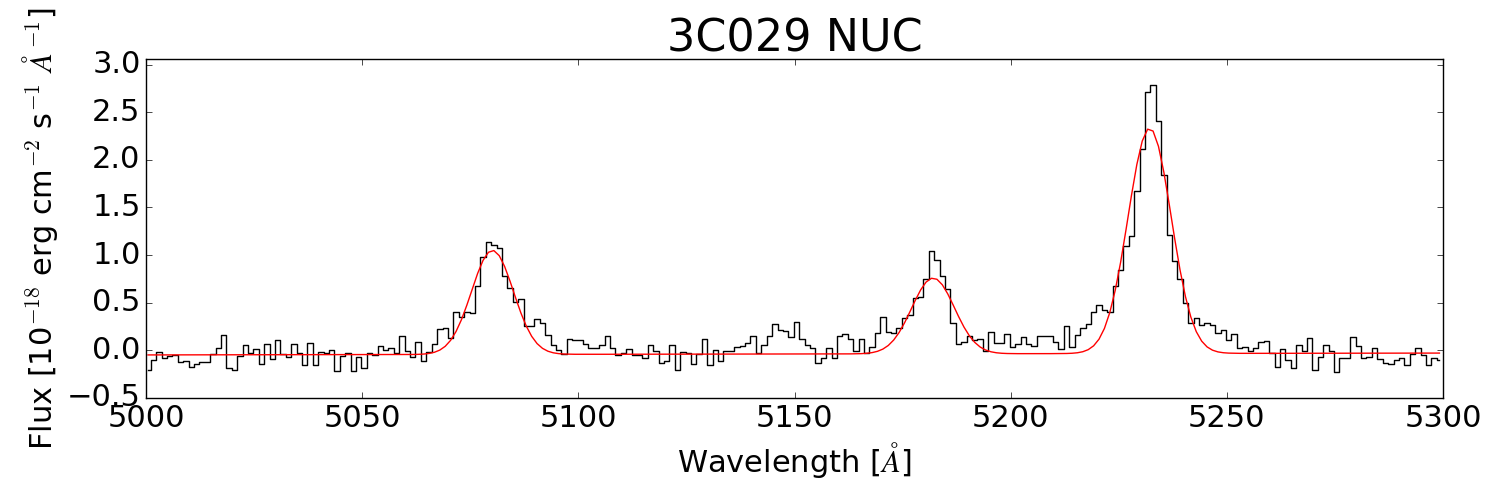}
\includegraphics[width=9cm]{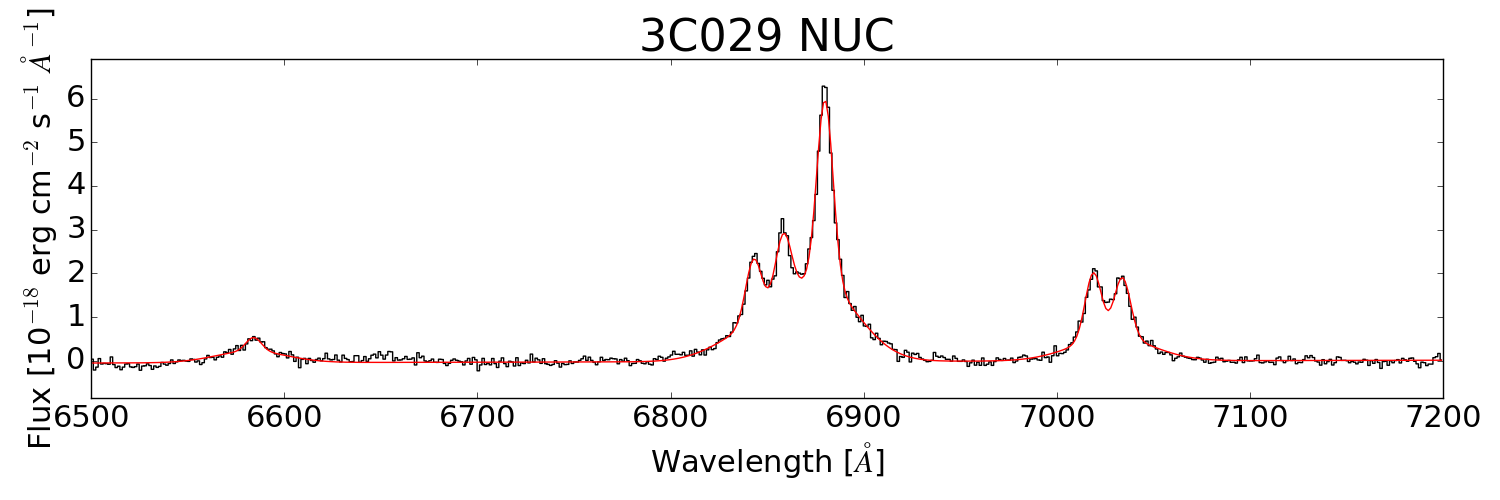}

\includegraphics[width=9cm]{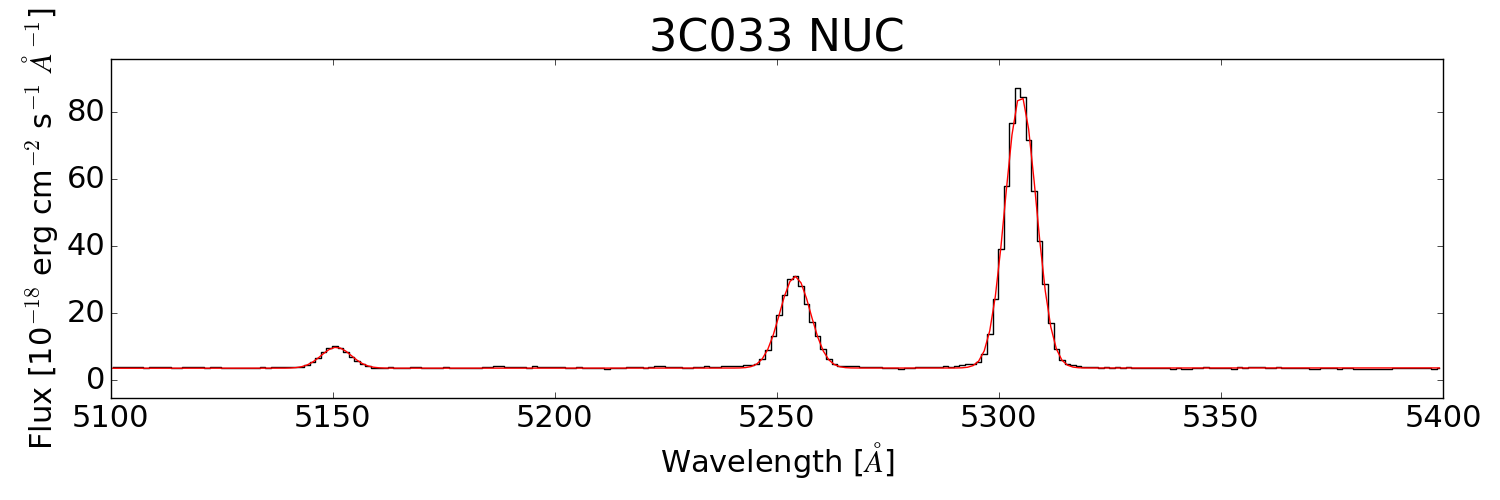}
\includegraphics[width=9cm]{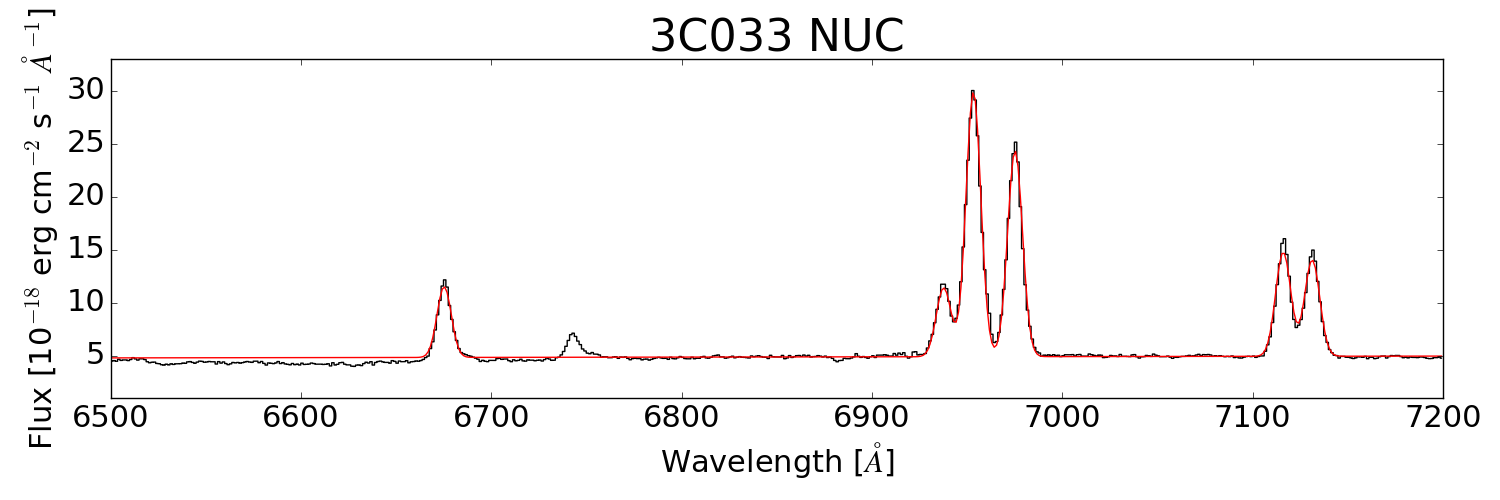}
\includegraphics[width=9cm]{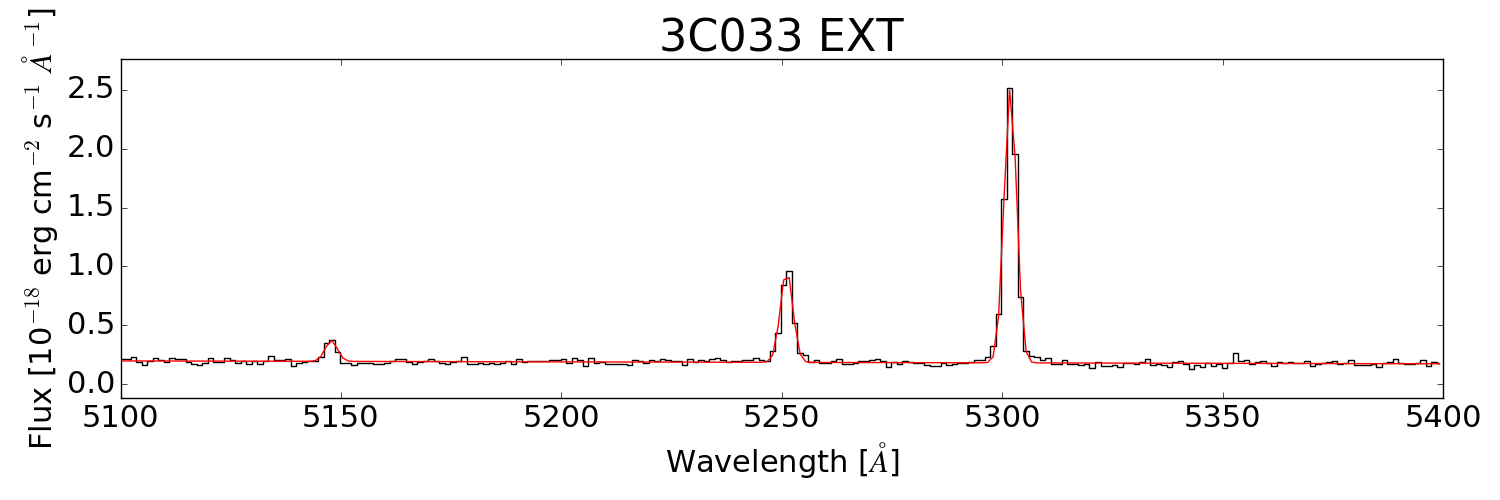}
\includegraphics[width=9cm]{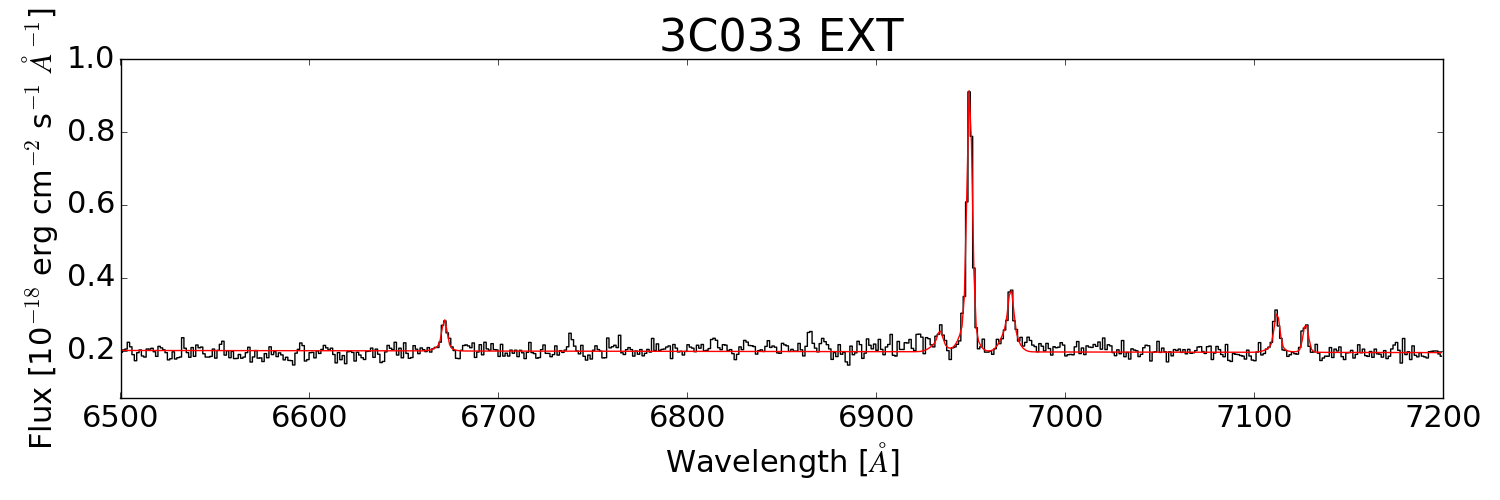}

}
\caption{Blue (left panel) and red (right panel) portion of the
  nuclear spectrum for all sources; for the 14 sources discussed in
  section 4, we also show the off-nuclear spectra extracted from the
  region marked in Fig. \ref{franco}.}
\label{spettri}
\end{figure*}  
 \addtocounter{figure}{-1}
 \begin{figure*}  
\centering{ 
\includegraphics[width=9cm]{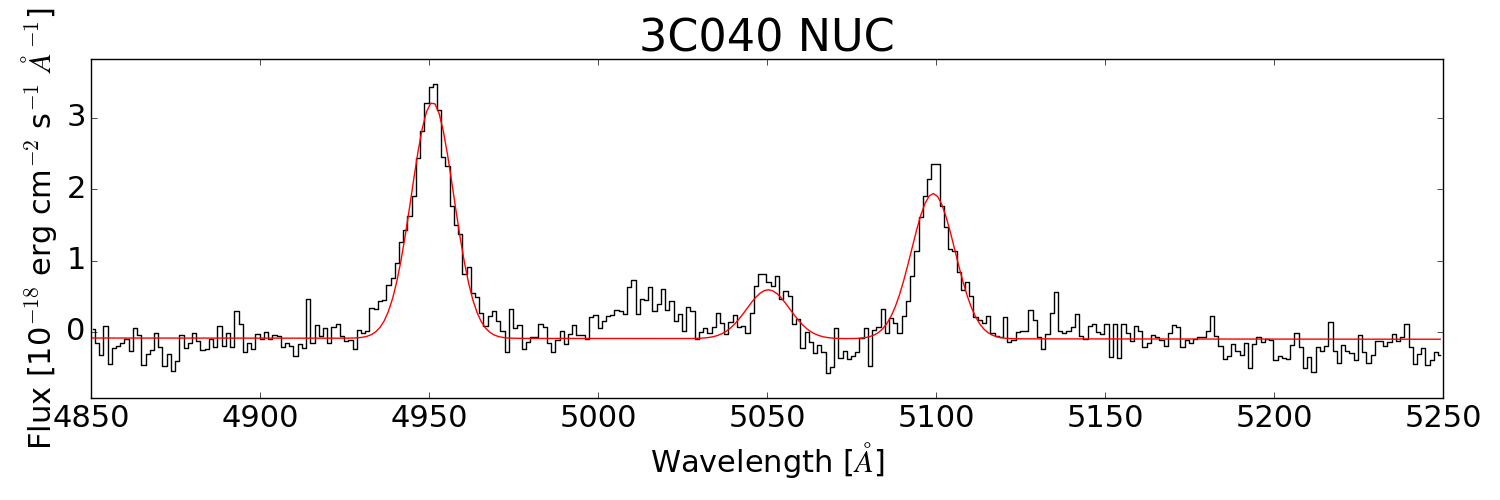}
\includegraphics[width=9cm]{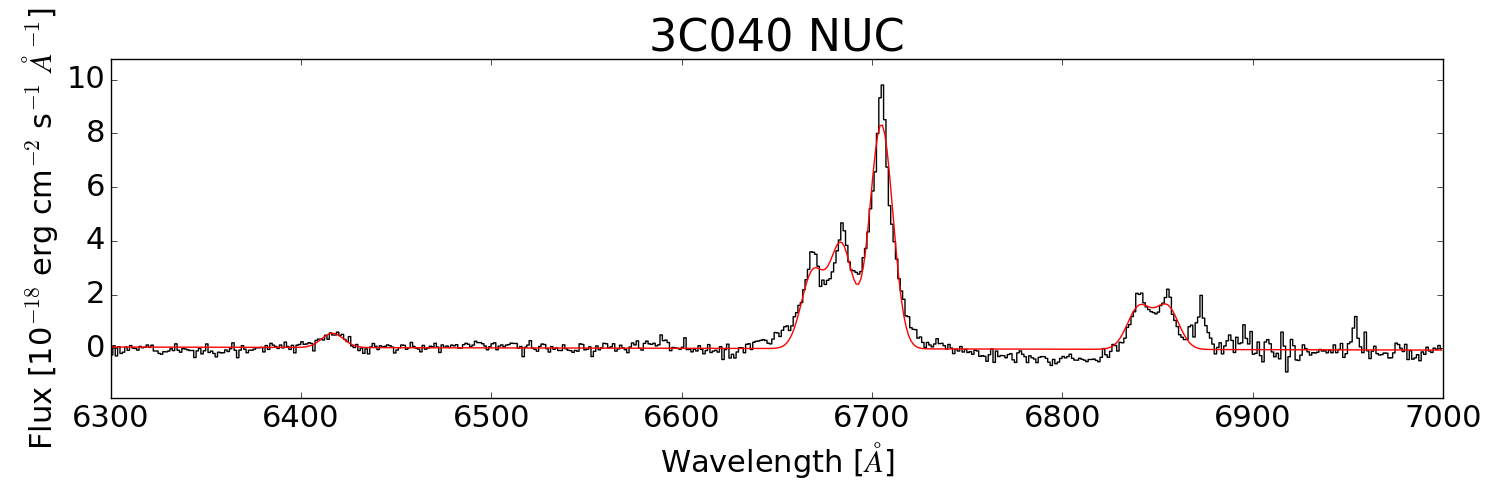}

\includegraphics[width=9cm]{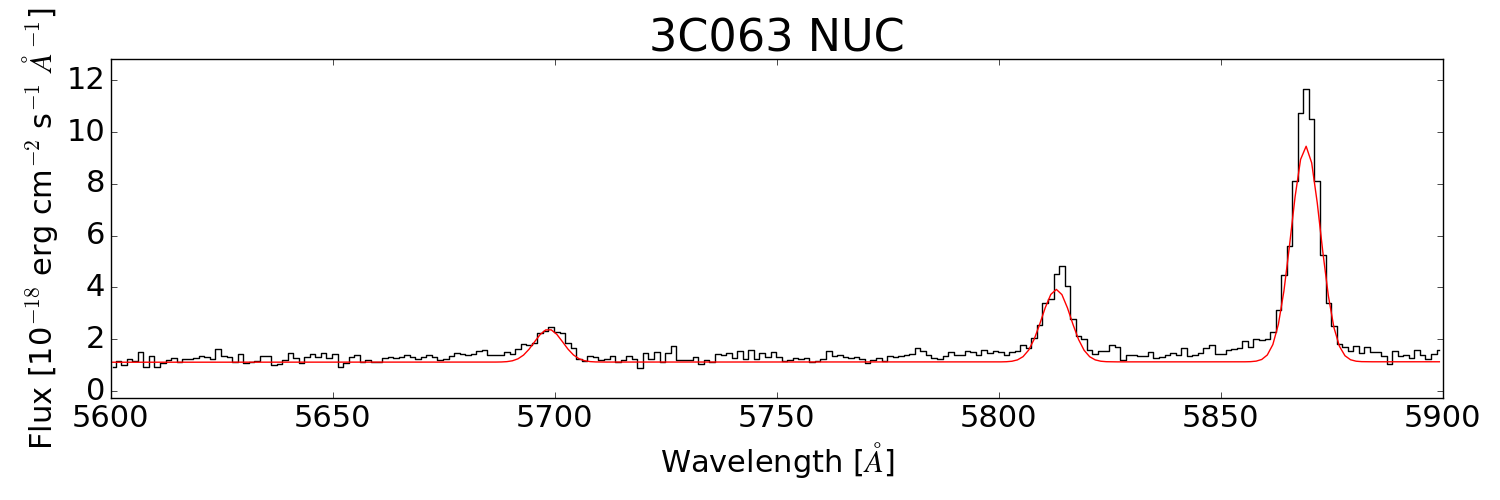}
\includegraphics[width=9cm]{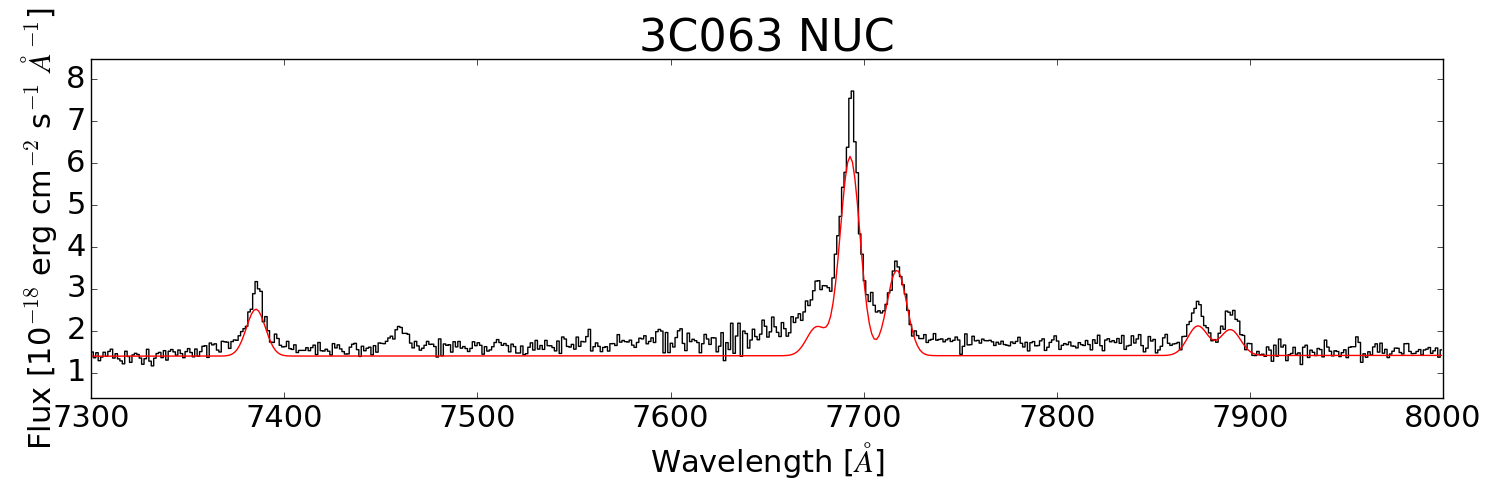}
\includegraphics[width=9cm]{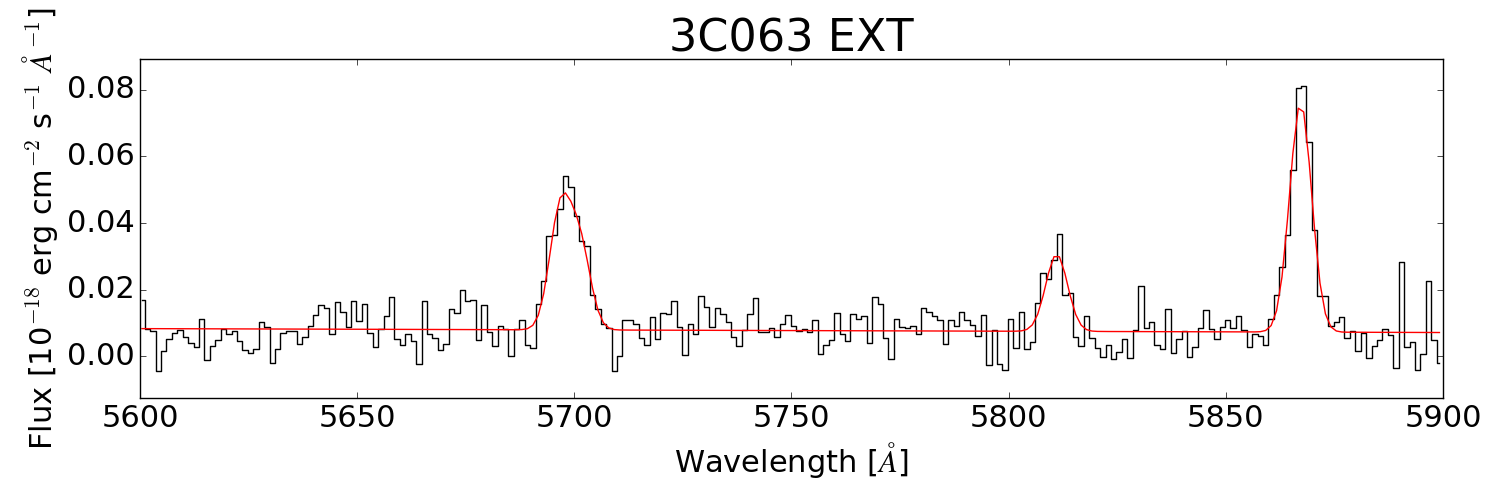}
\includegraphics[width=9cm]{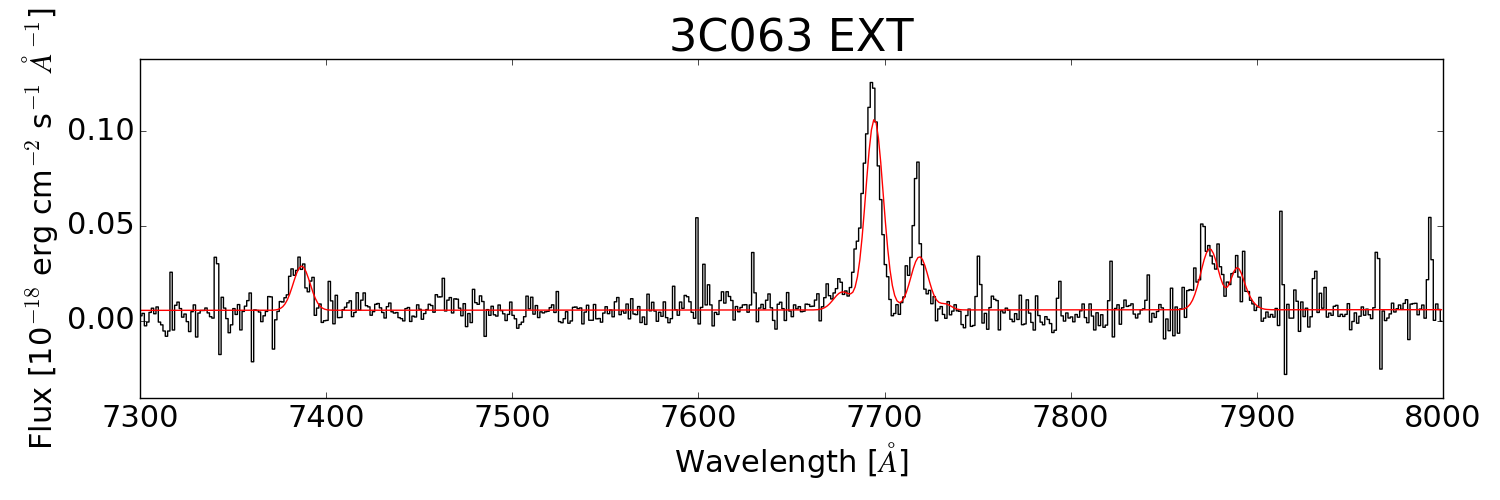}

\includegraphics[width=9cm]{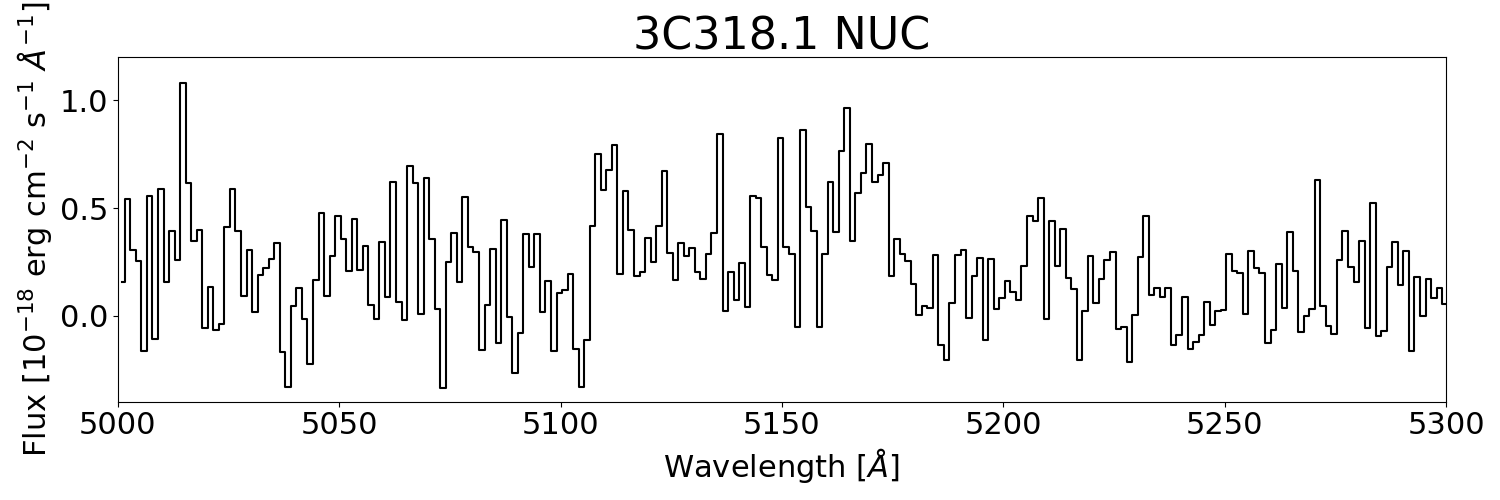}
\includegraphics[width=9cm]{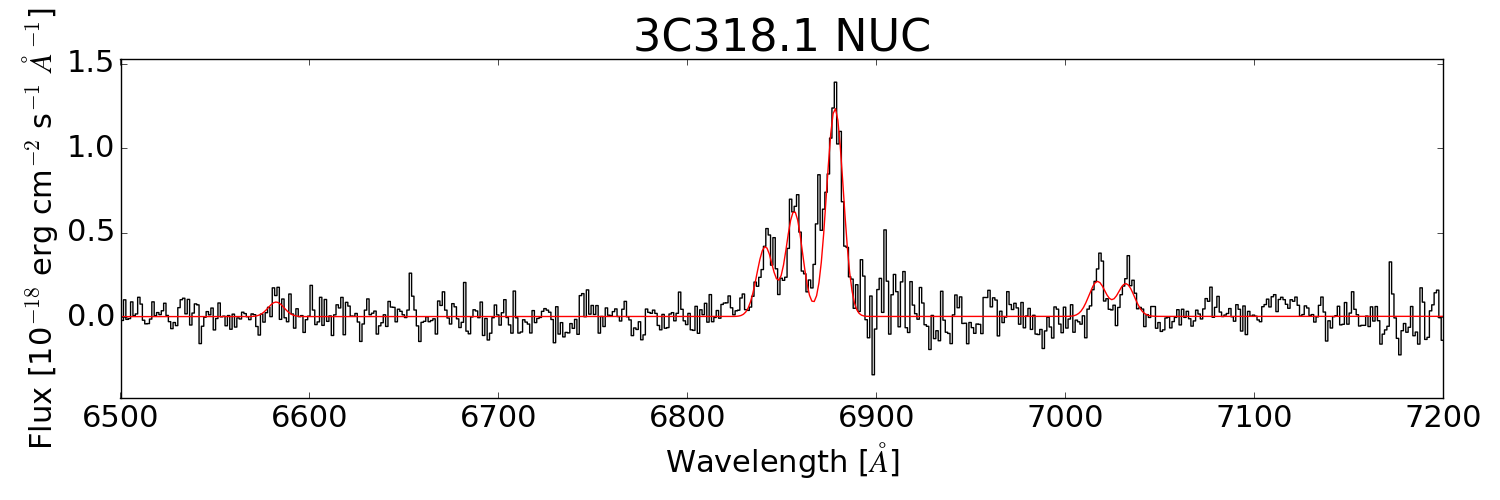}
\includegraphics[width=9cm]{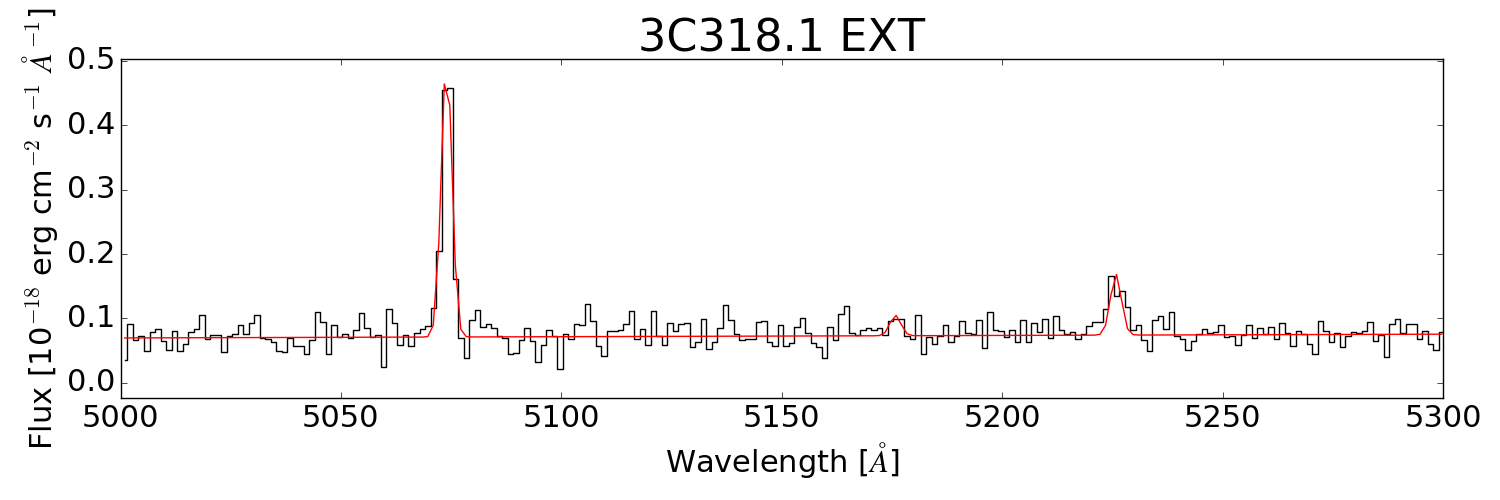}
\includegraphics[width=9cm]{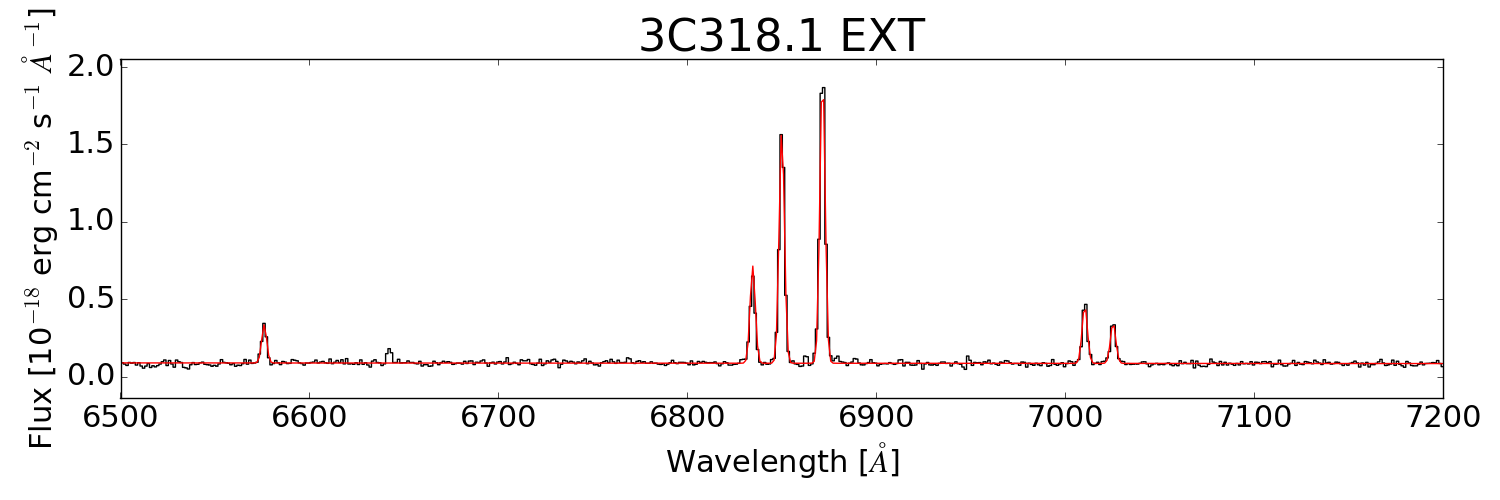}

\includegraphics[width=9cm]{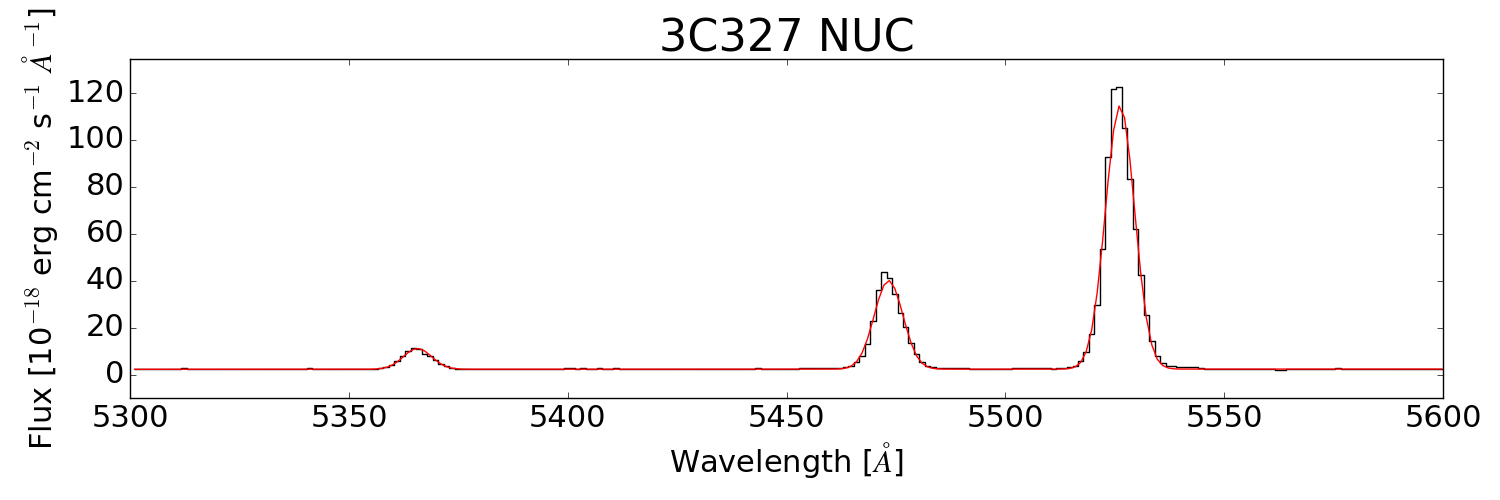}
\includegraphics[width=9cm]{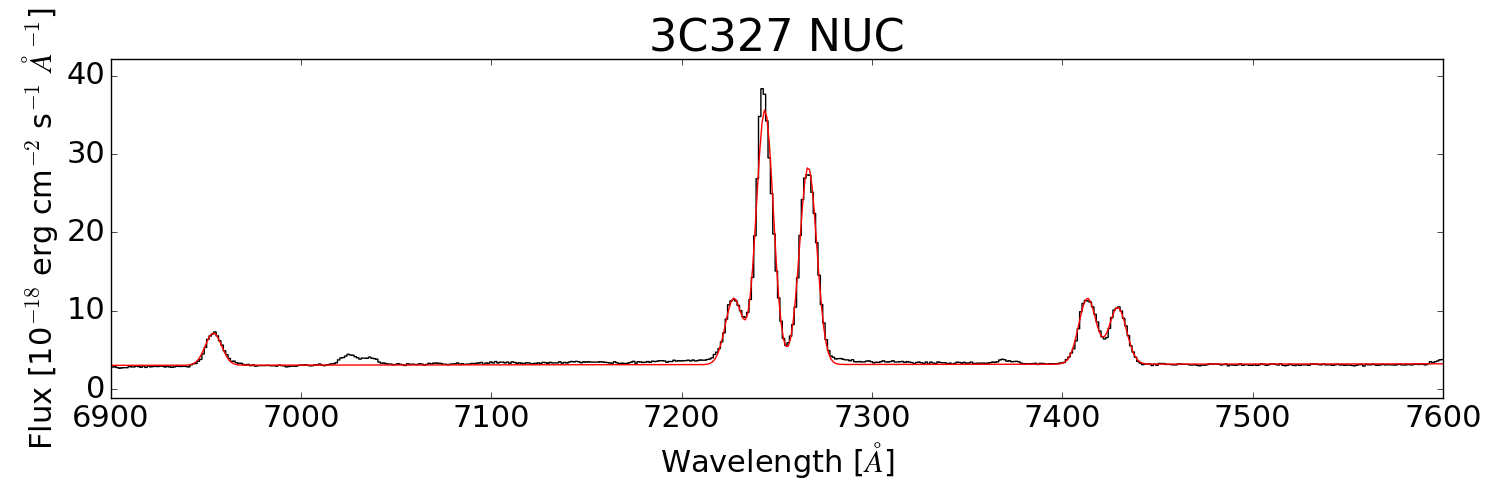}
\includegraphics[width=9cm]{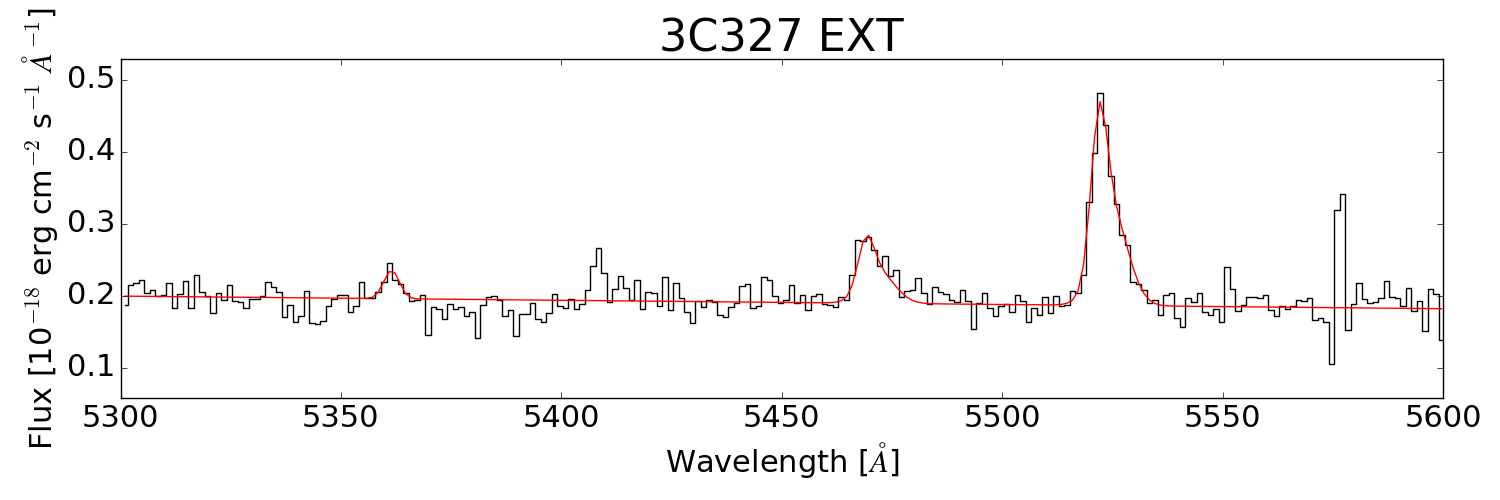}
\includegraphics[width=9cm]{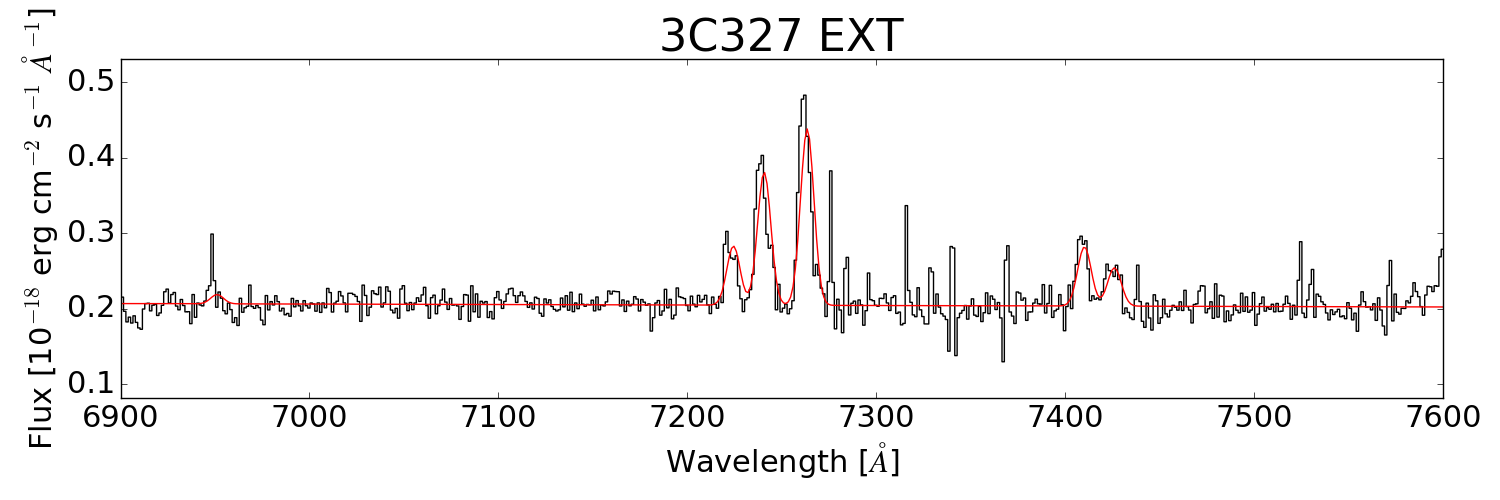}

}
\caption{continued}
\end{figure*}  
\addtocounter{figure}{-1}
\begin{figure*}  
\centering{ 
\includegraphics[width=9cm]{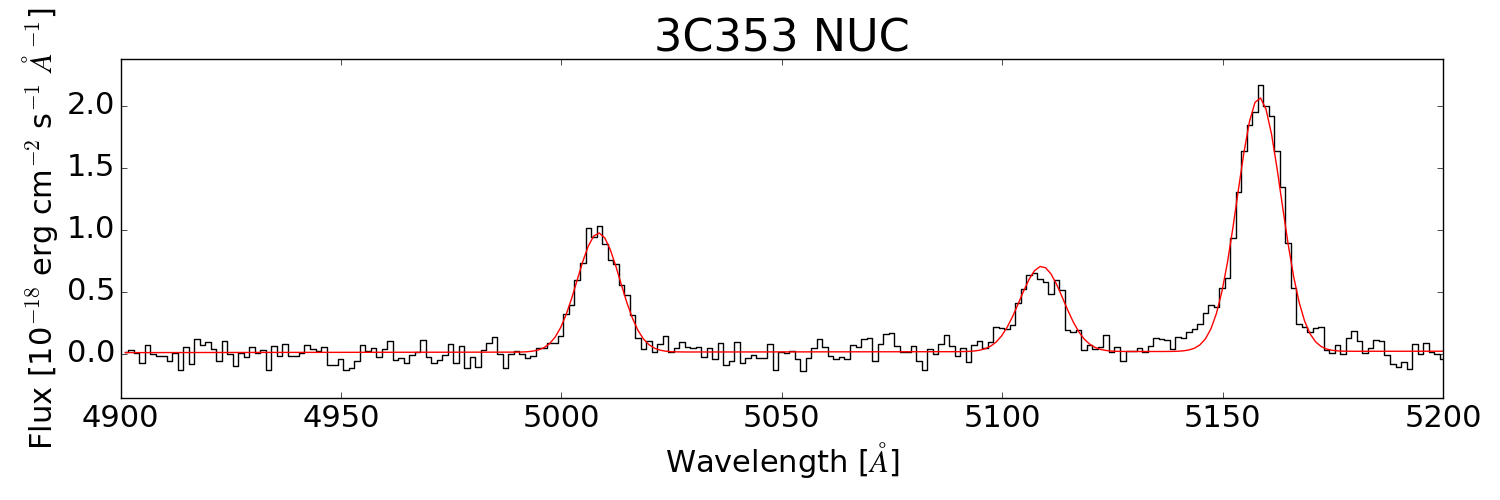}
\includegraphics[width=9cm]{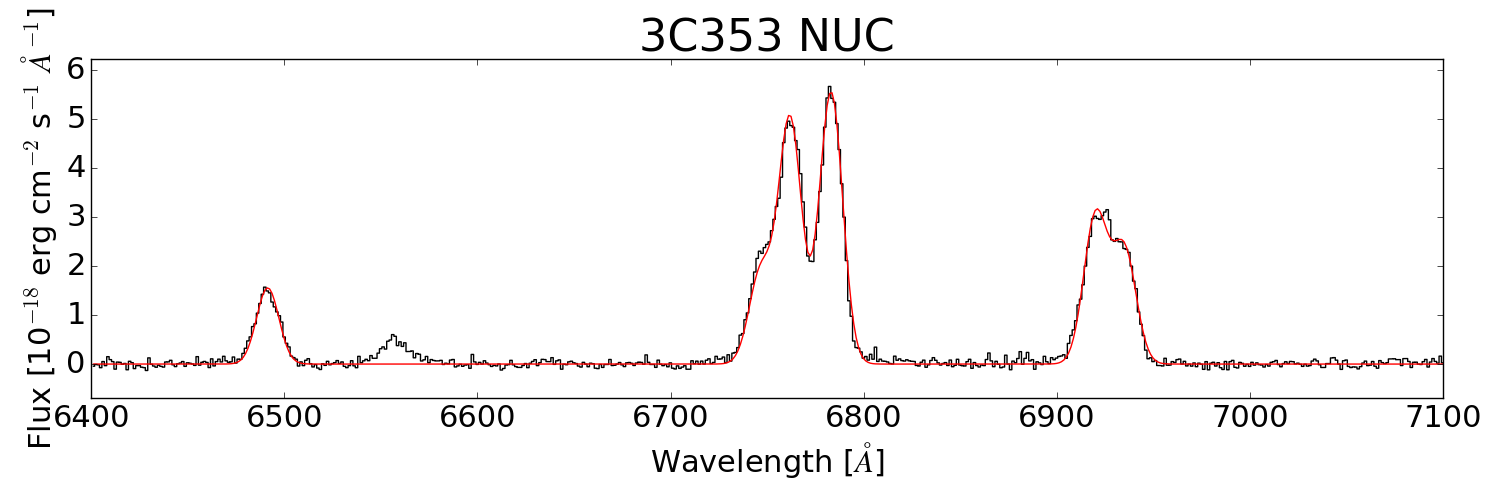}

\includegraphics[width=9cm]{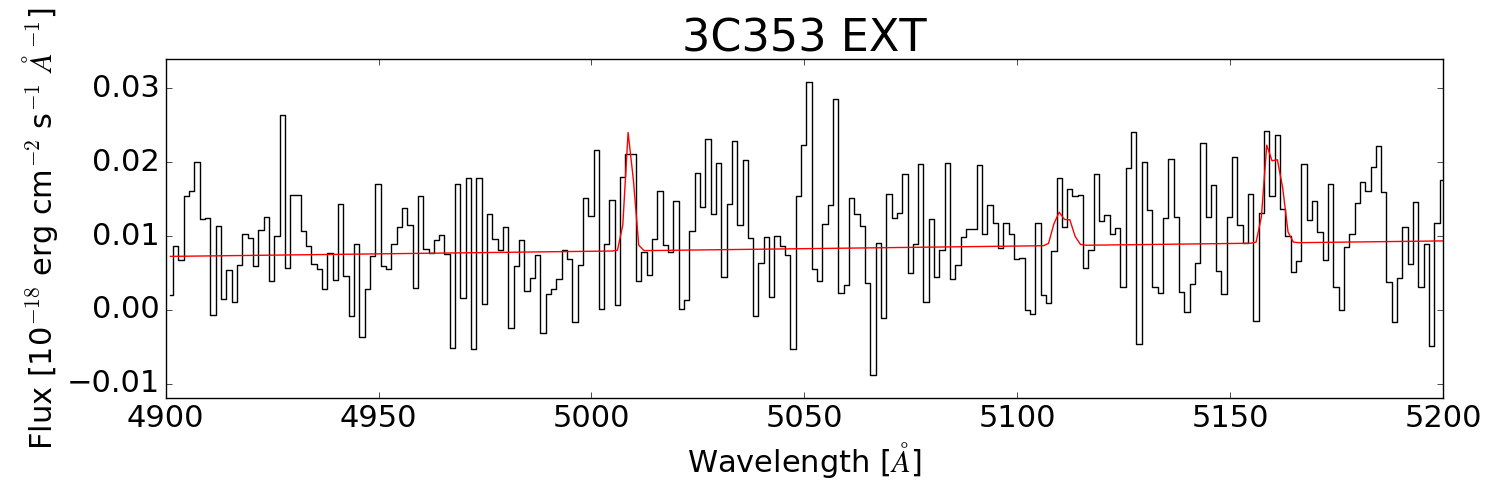}
\includegraphics[width=9cm]{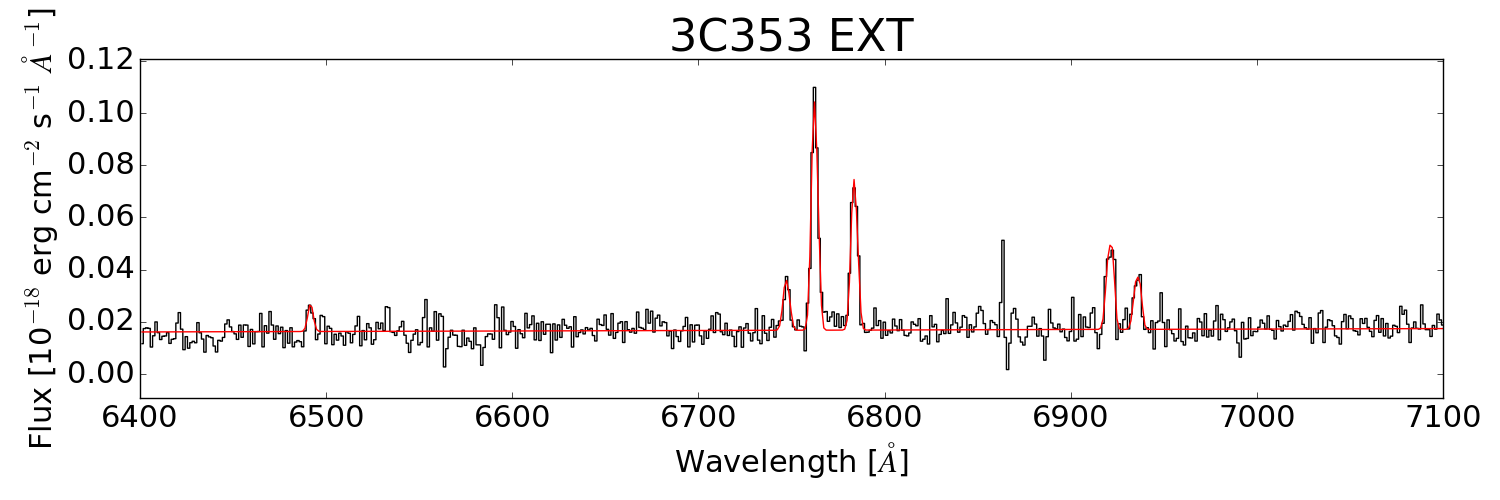}
\includegraphics[width=9cm]{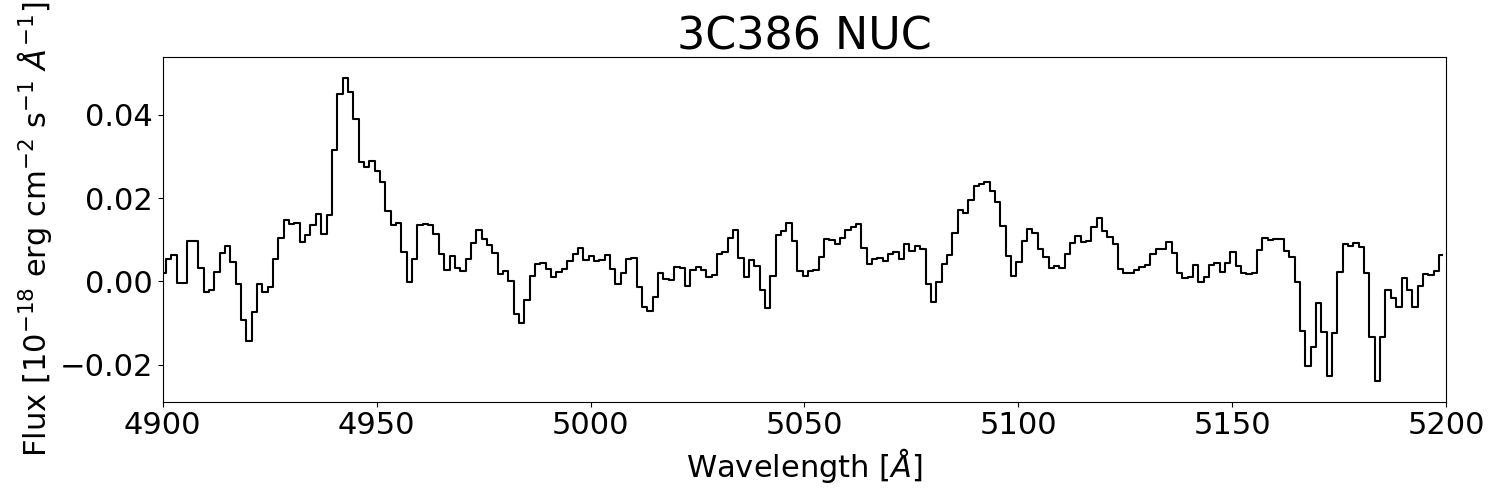}
\includegraphics[width=9cm]{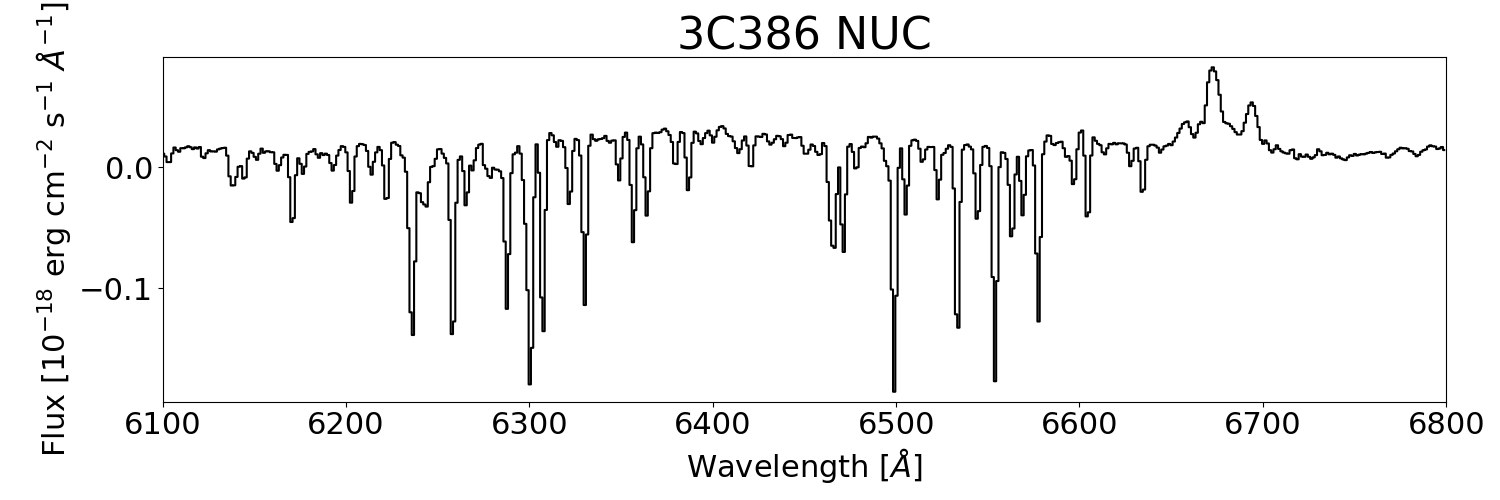}
\includegraphics[width=9cm]{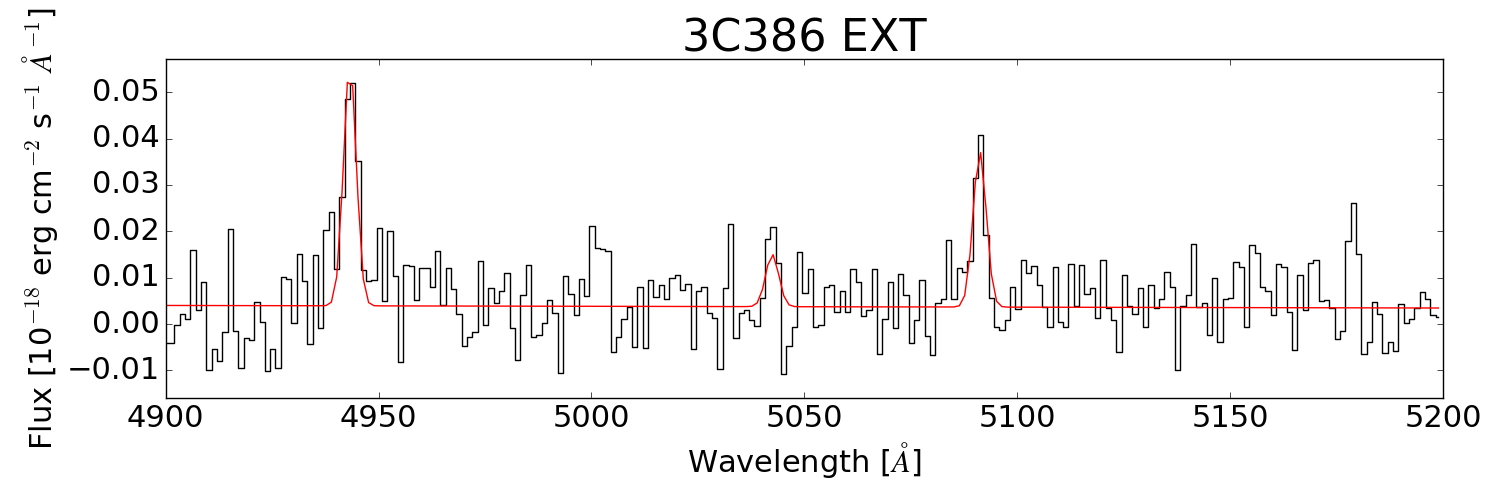}
\includegraphics[width=9cm]{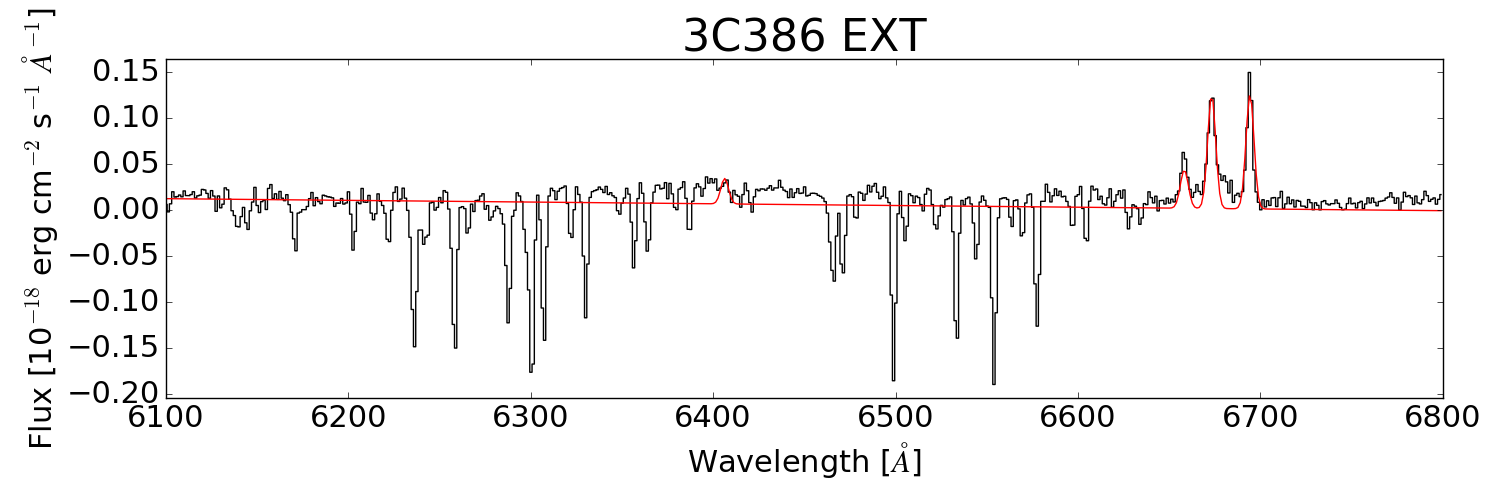}
\includegraphics[width=9cm]{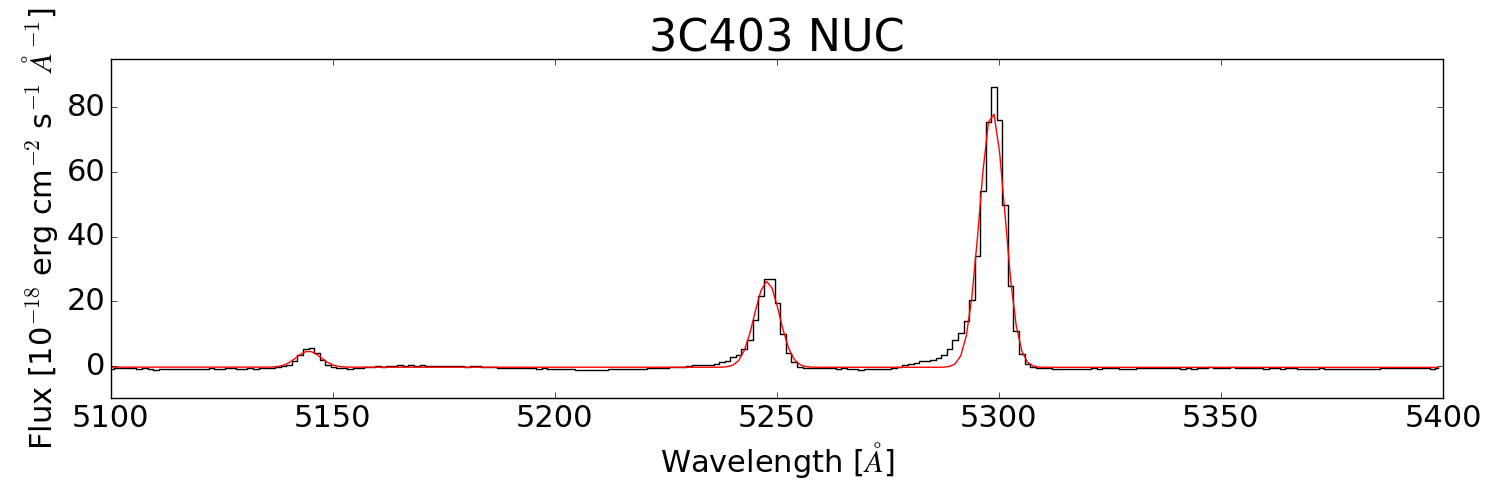}
\includegraphics[width=9cm]{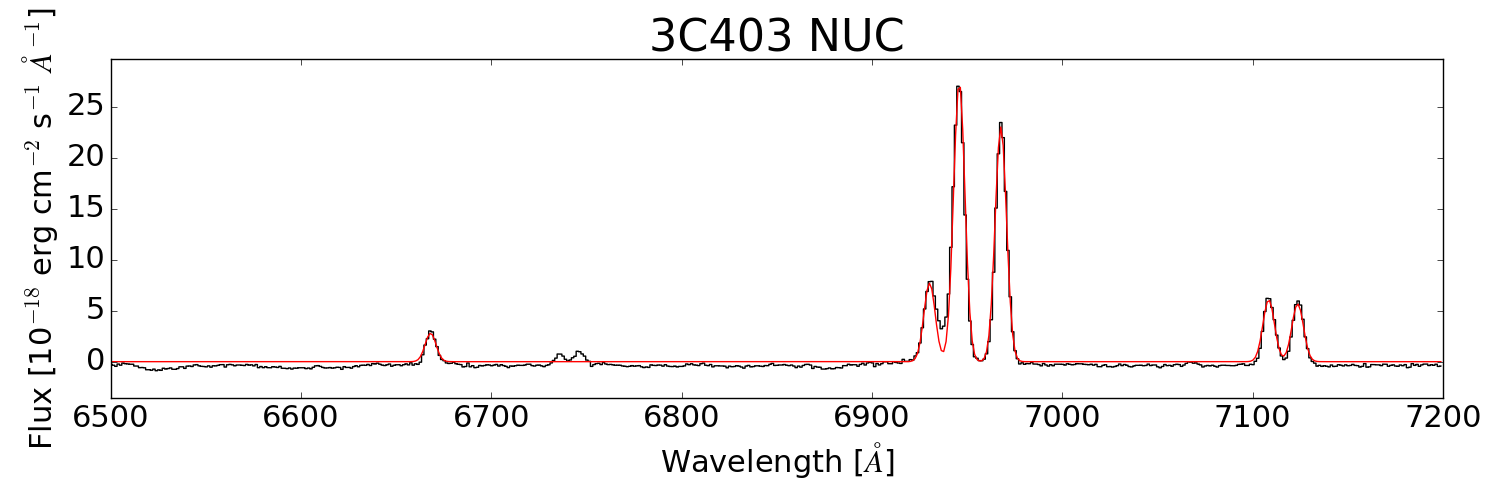}
\includegraphics[width=9cm]{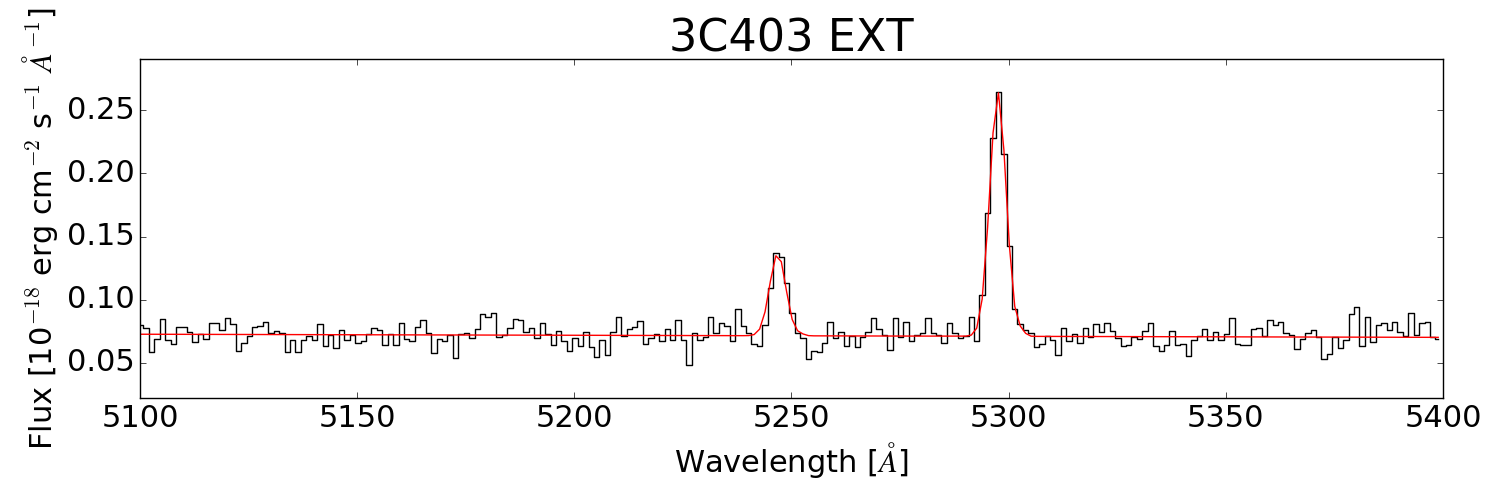}
\includegraphics[width=9cm]{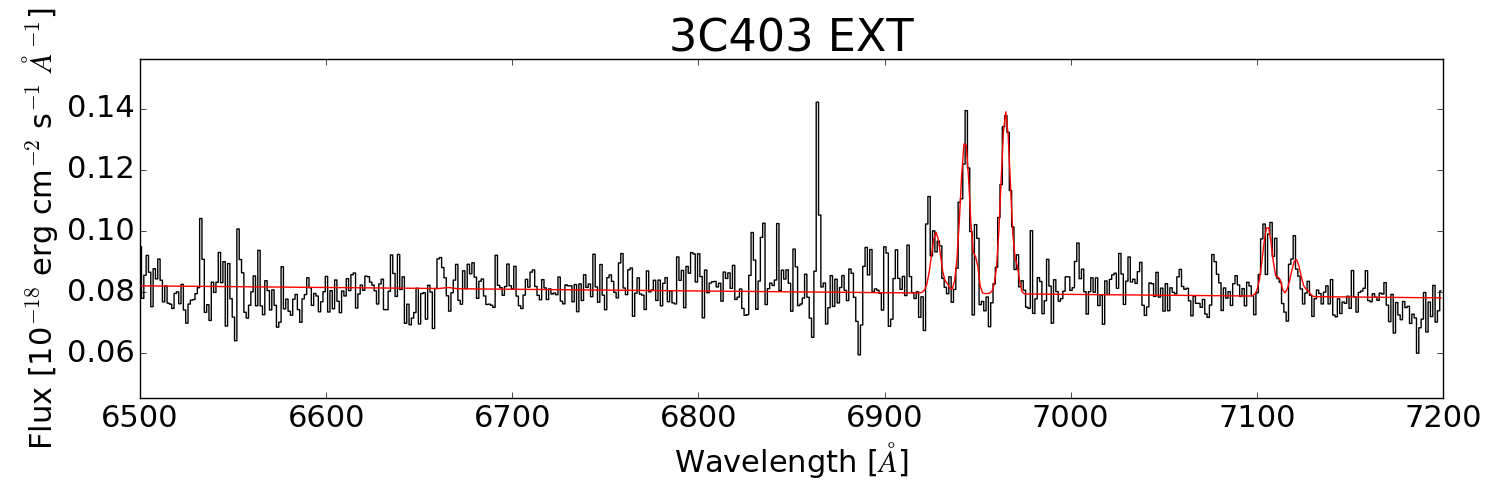}

\includegraphics[width=9cm]{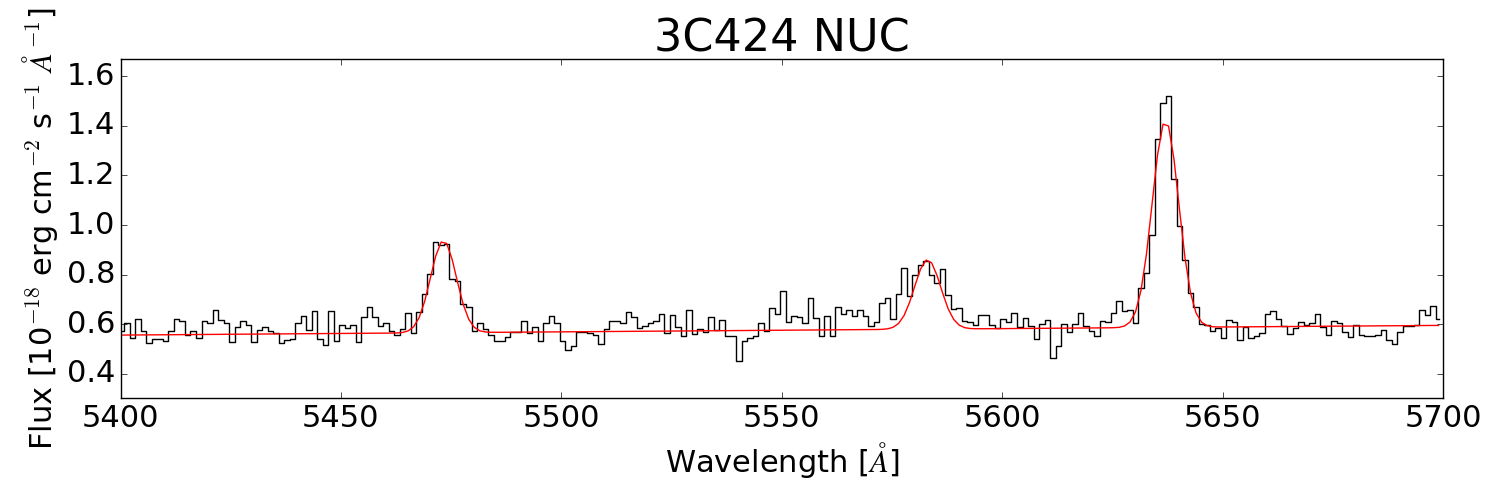}
\includegraphics[width=9cm]{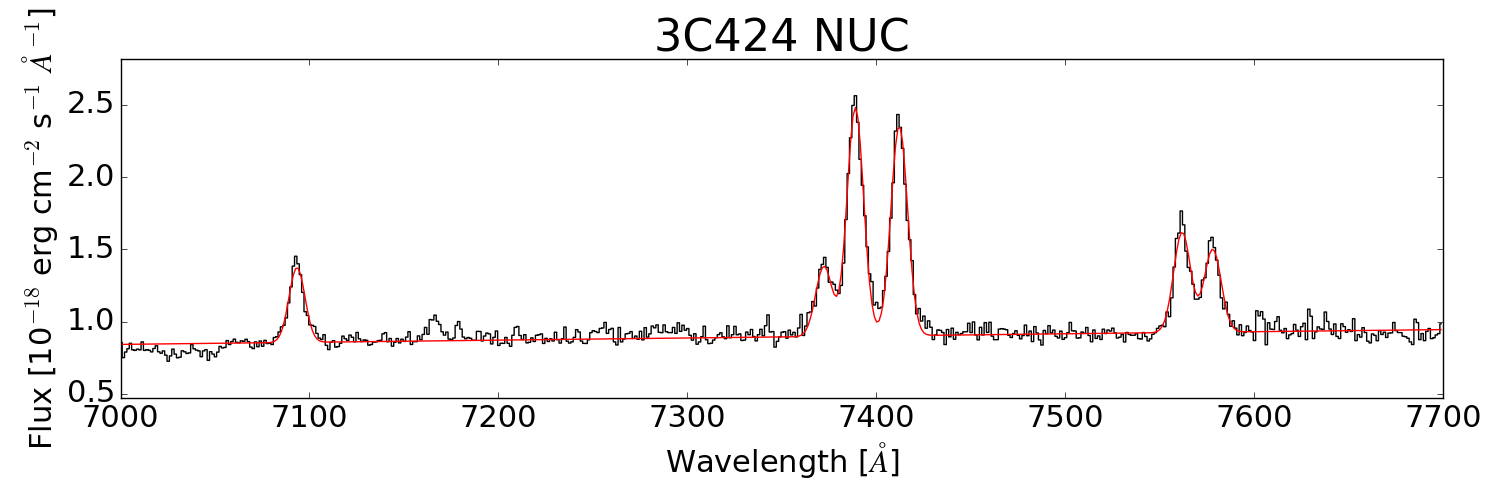}

\includegraphics[width=9cm]{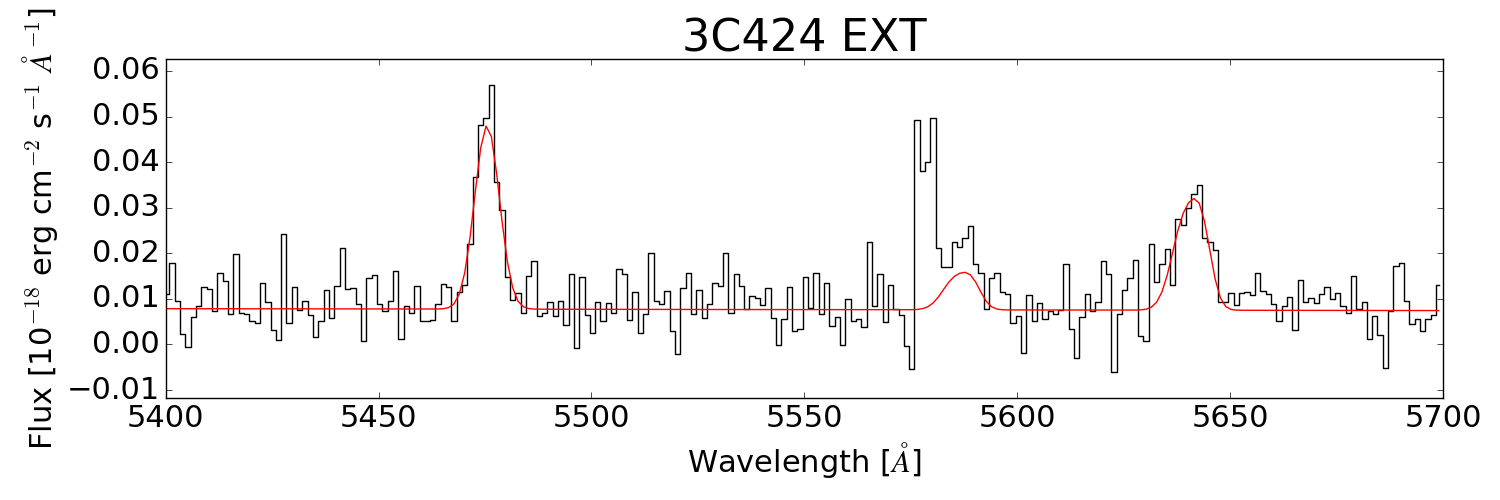}
\includegraphics[width=9cm]{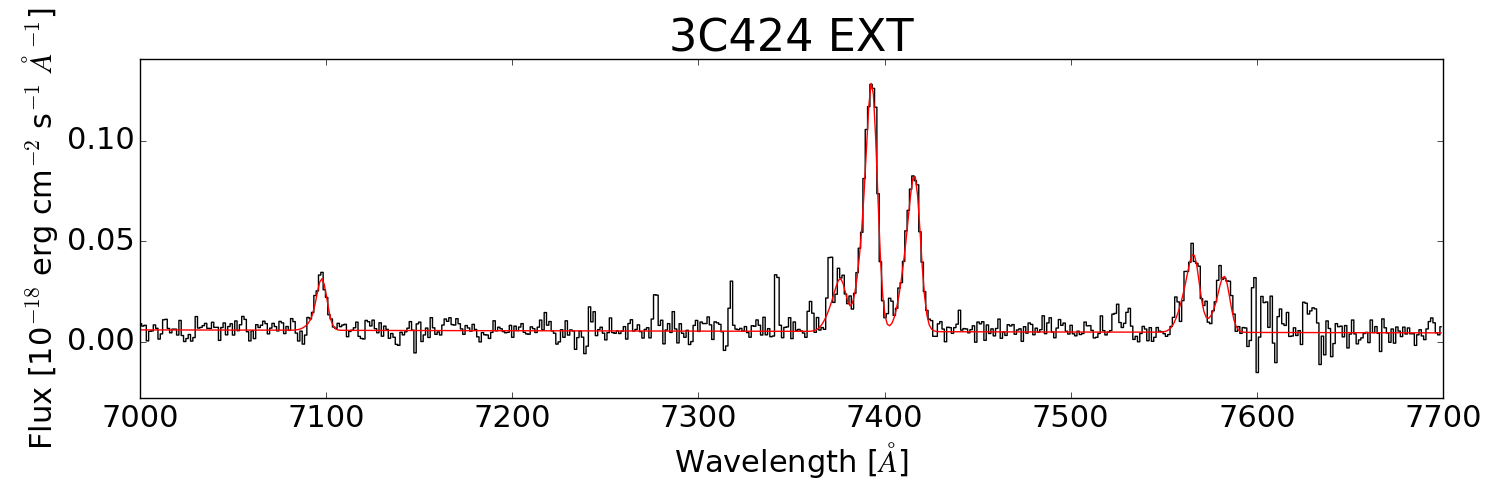}

}
\caption{continued}
\end{figure*}  

\addtocounter{figure}{-1}
\begin{figure*}  
\centering{ 
\includegraphics[width=9cm]{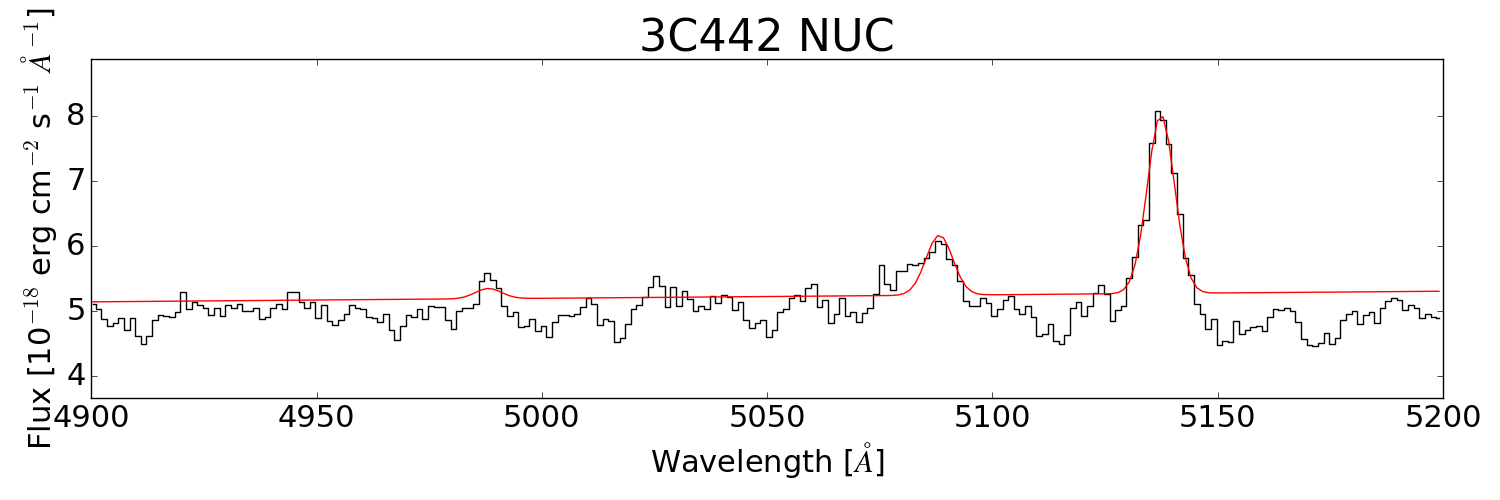}
\includegraphics[width=9cm]{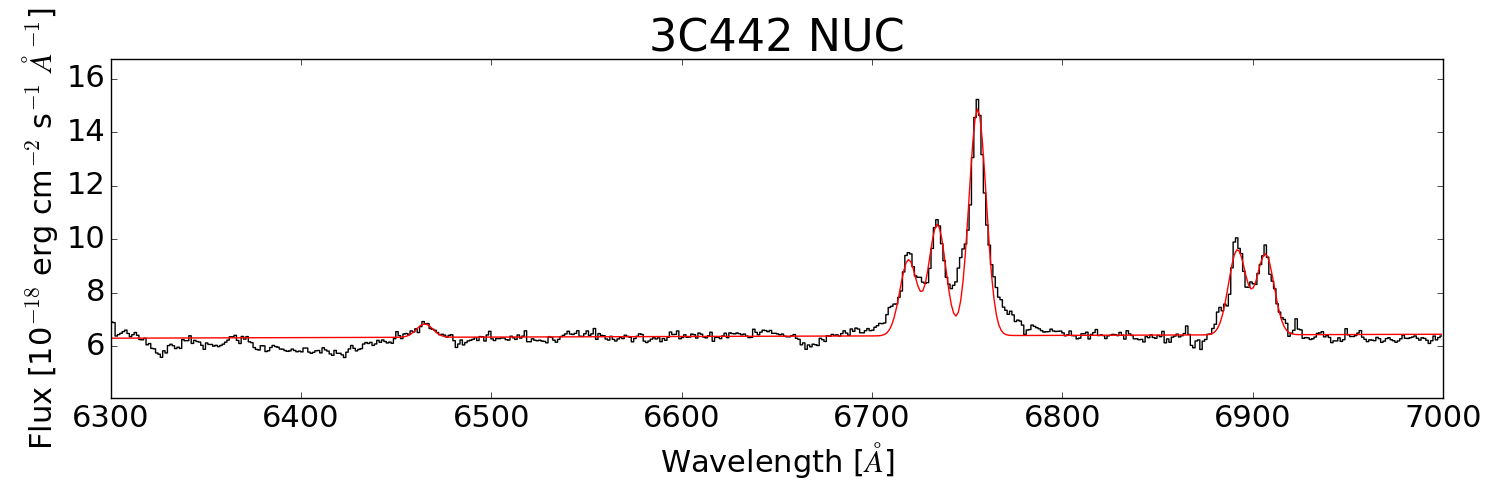}
\includegraphics[width=9cm]{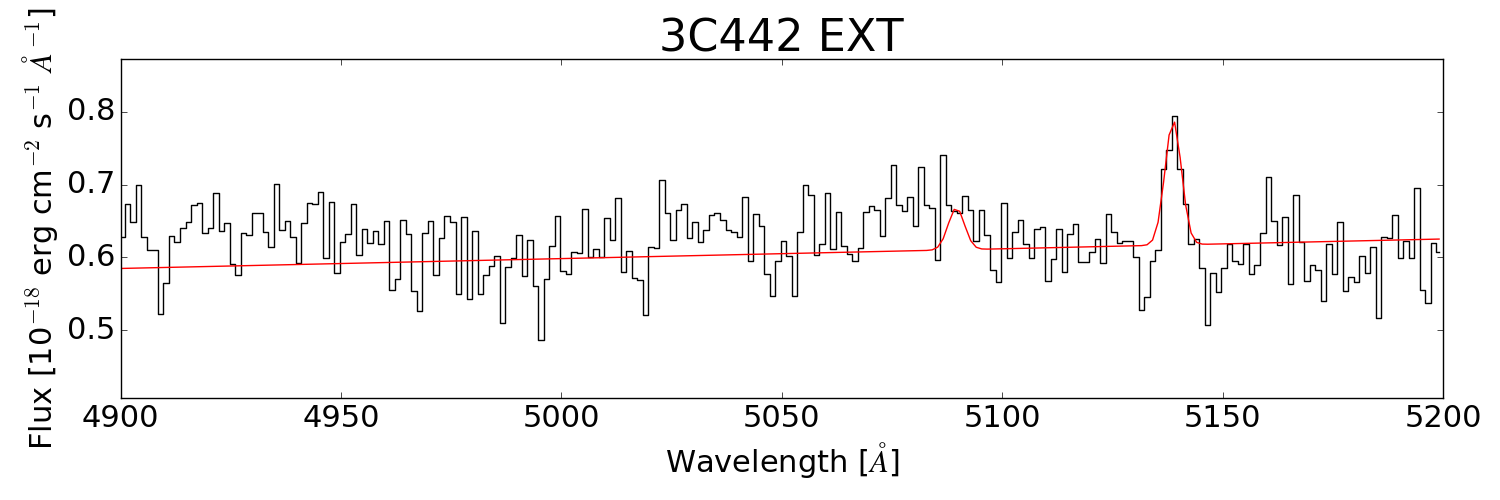}
\includegraphics[width=9cm]{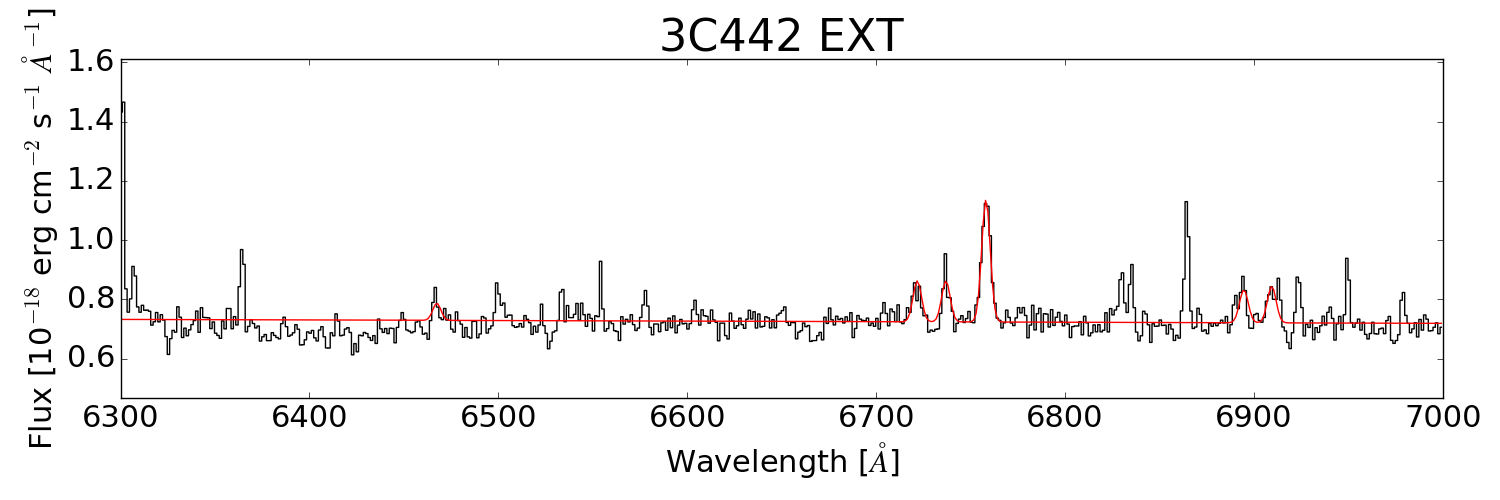}

\includegraphics[width=9cm]{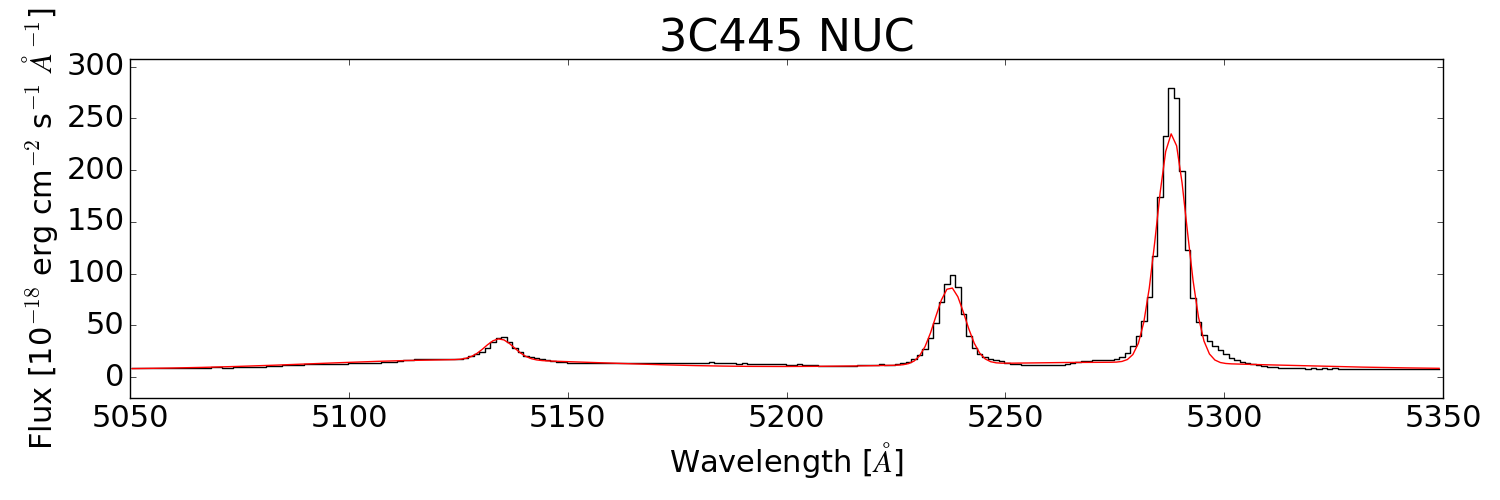}
\includegraphics[width=9cm]{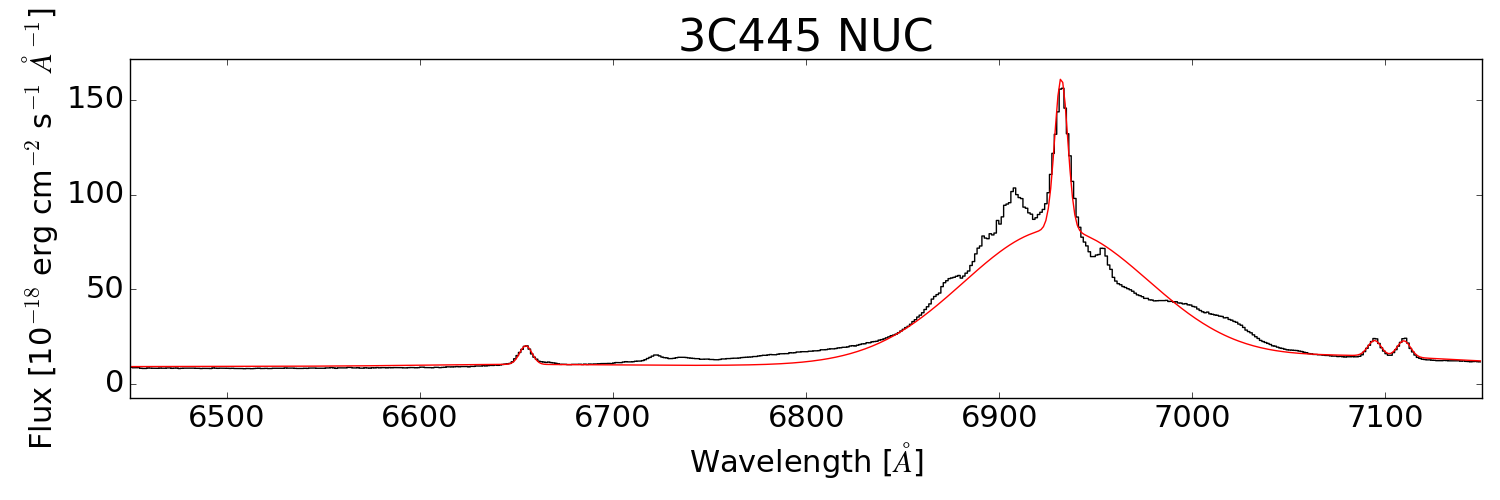}
\includegraphics[width=9cm]{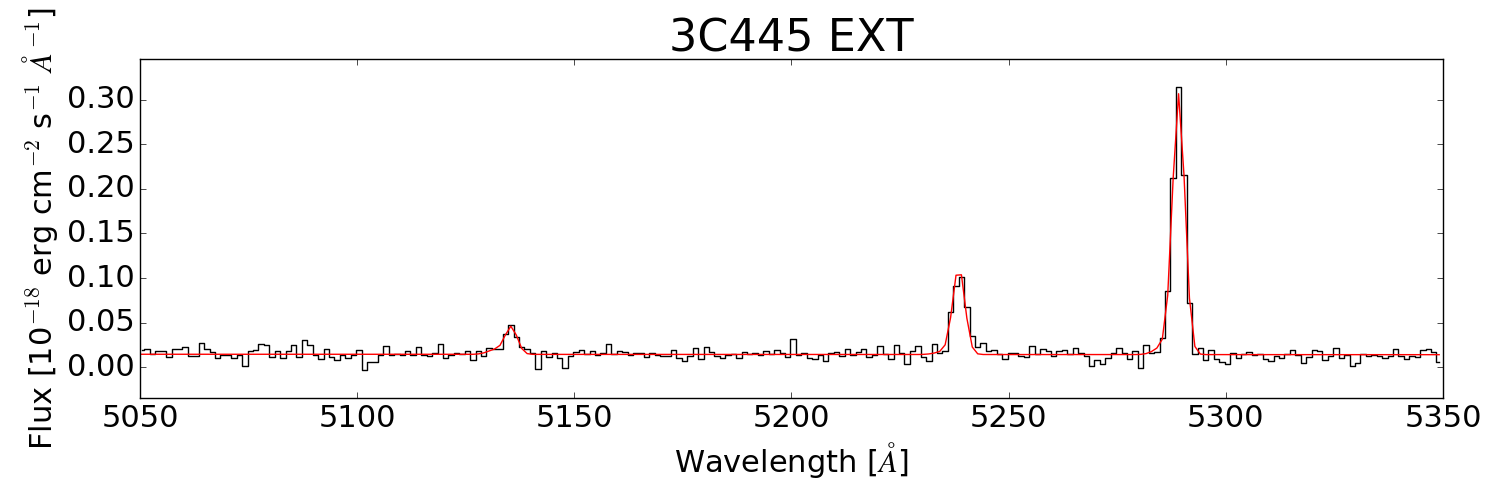}
\includegraphics[width=9cm]{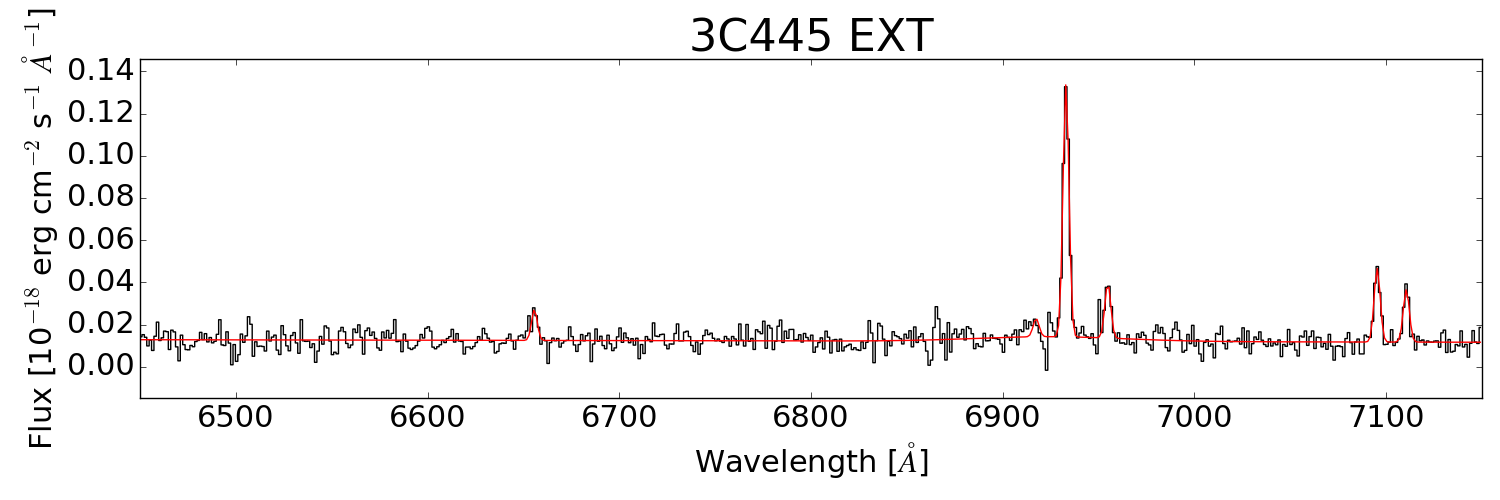}
\includegraphics[width=9cm]{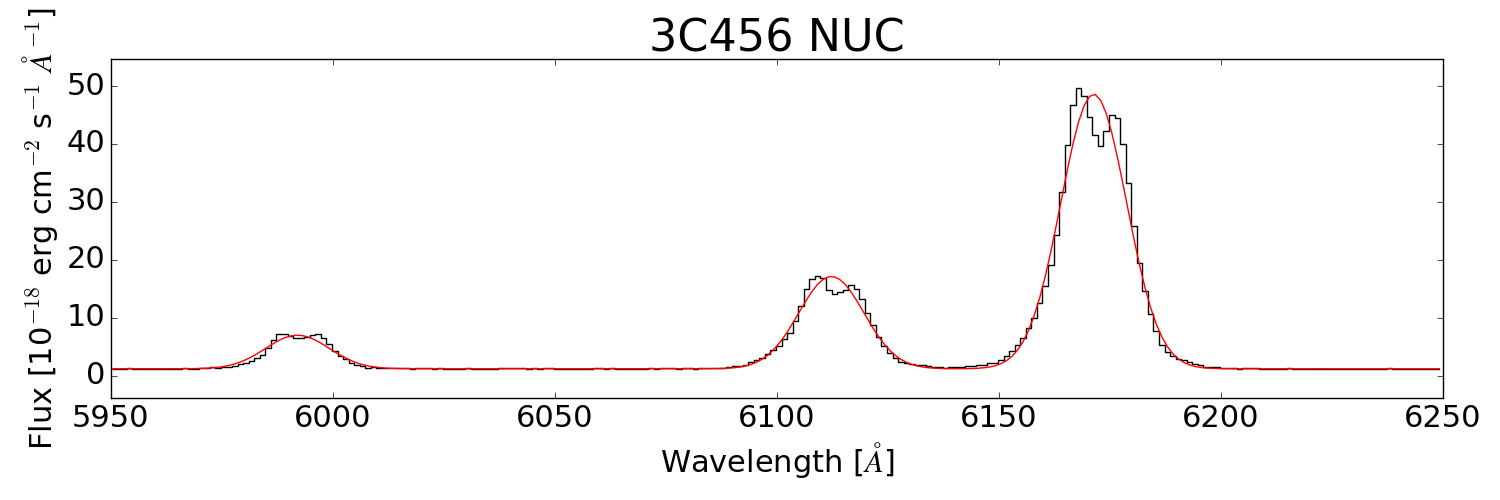}
\includegraphics[width=9cm]{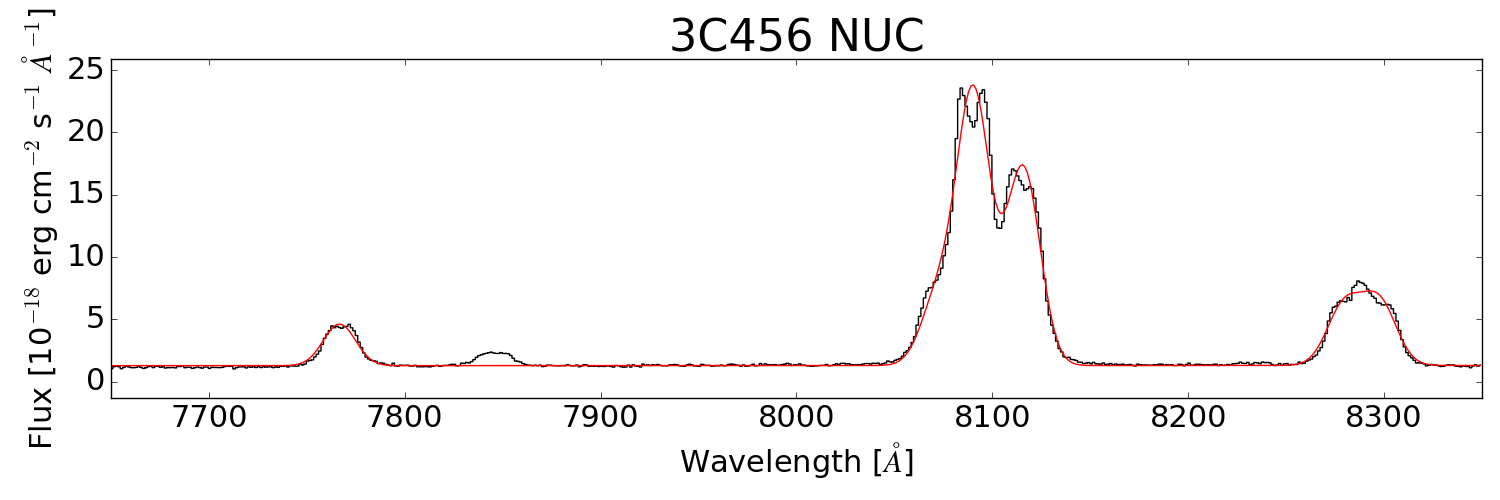}

\includegraphics[width=9cm]{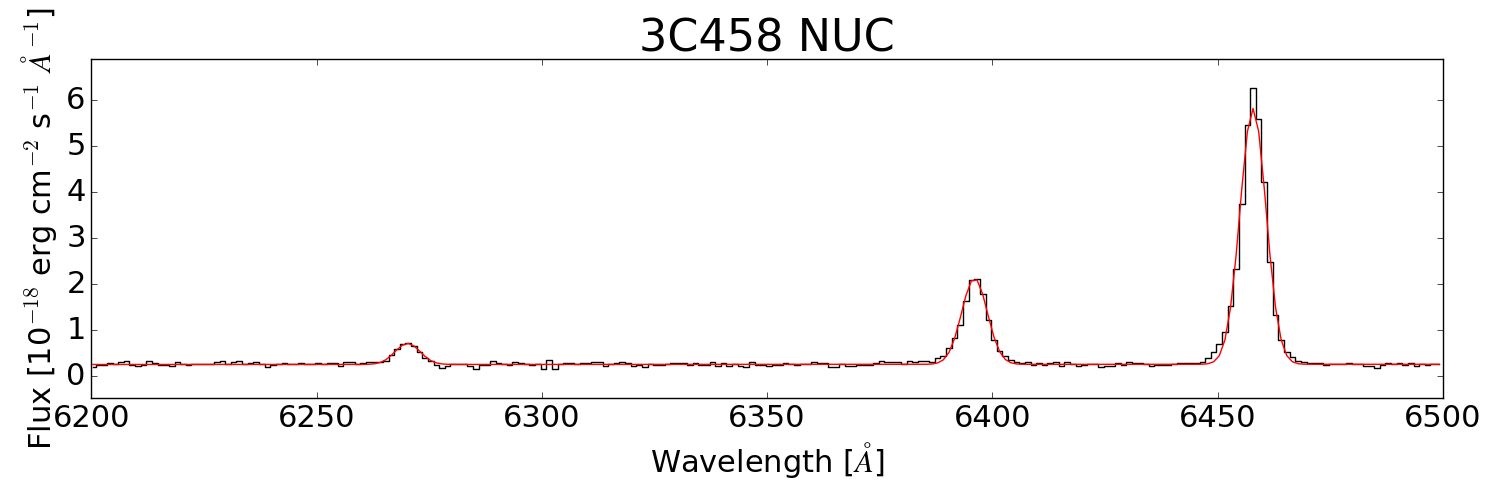}
\includegraphics[width=9cm]{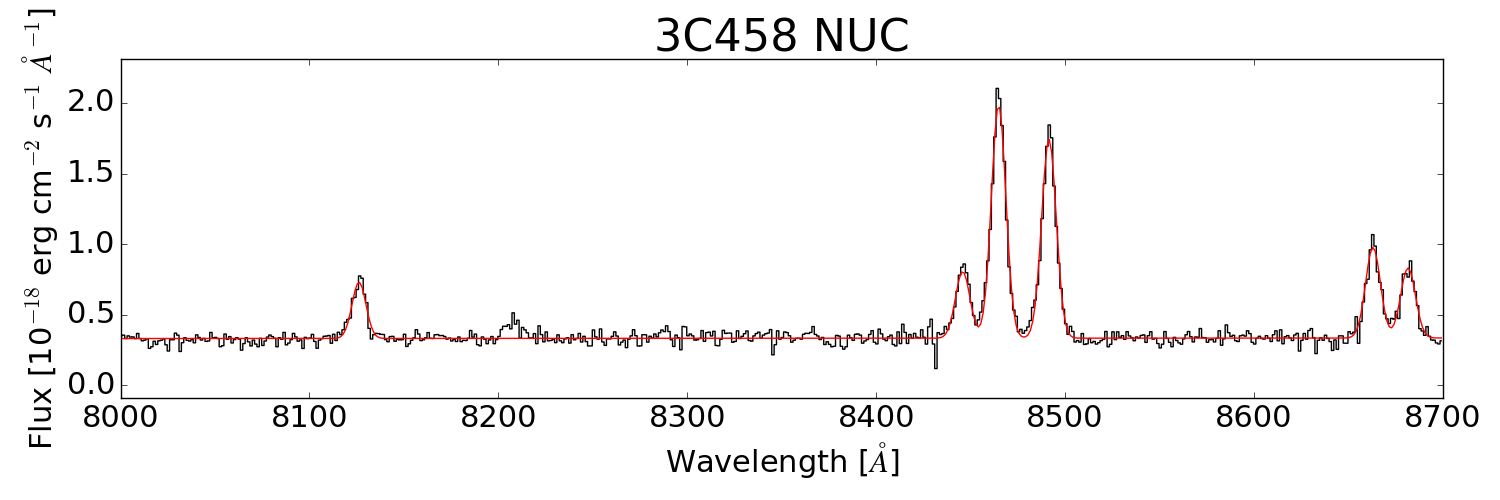}
\includegraphics[width=9cm]{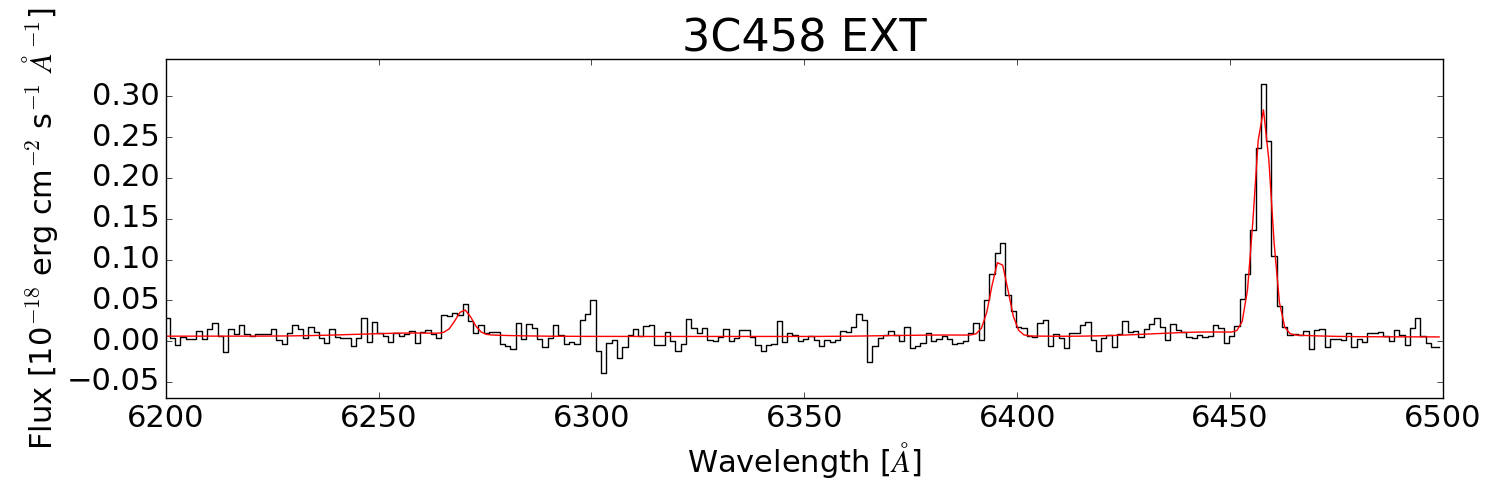}
\includegraphics[width=9cm]{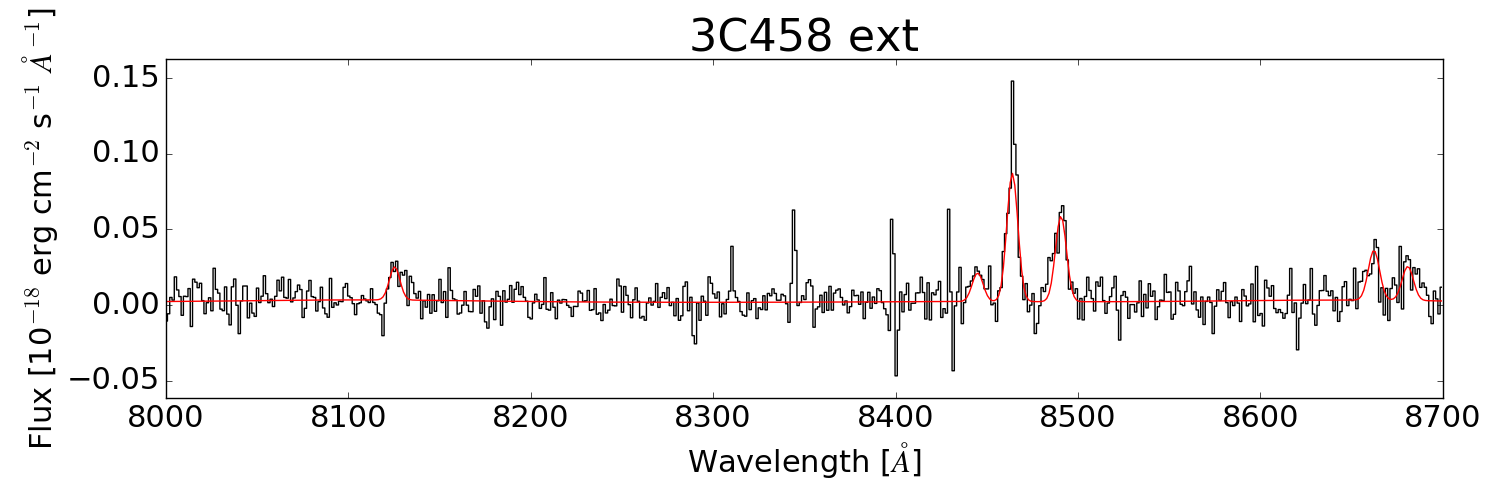}}
\caption{continued}
\end{figure*} 

\addtocounter{figure}{-1}
\begin{figure*}  
\centering{ 
\includegraphics[width=9cm]{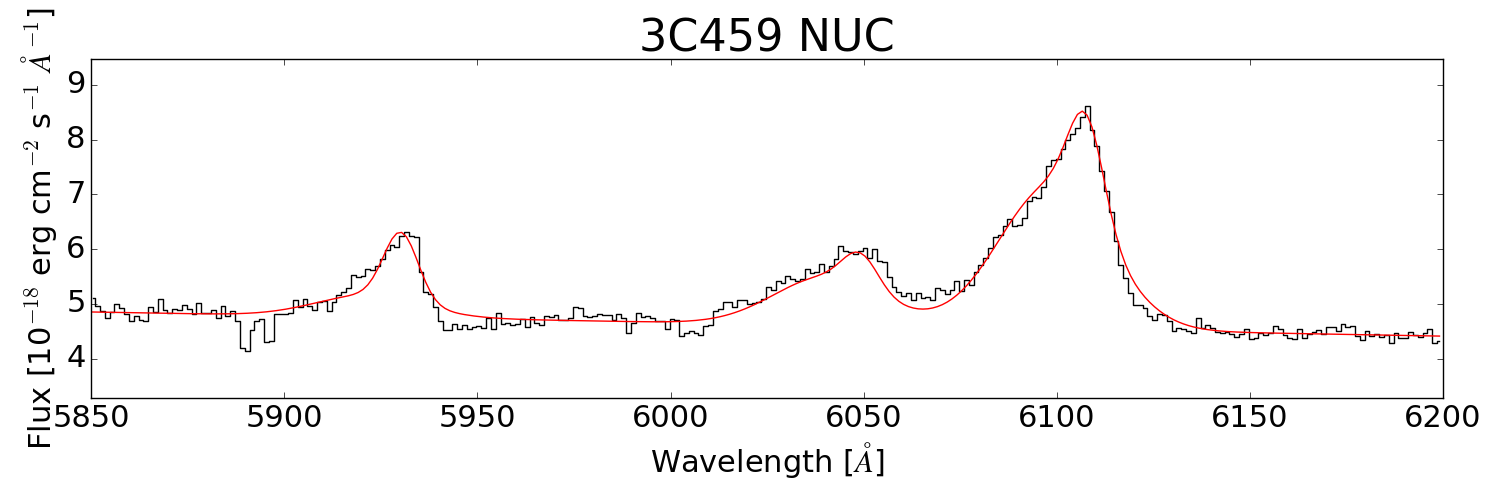}
\includegraphics[width=9cm]{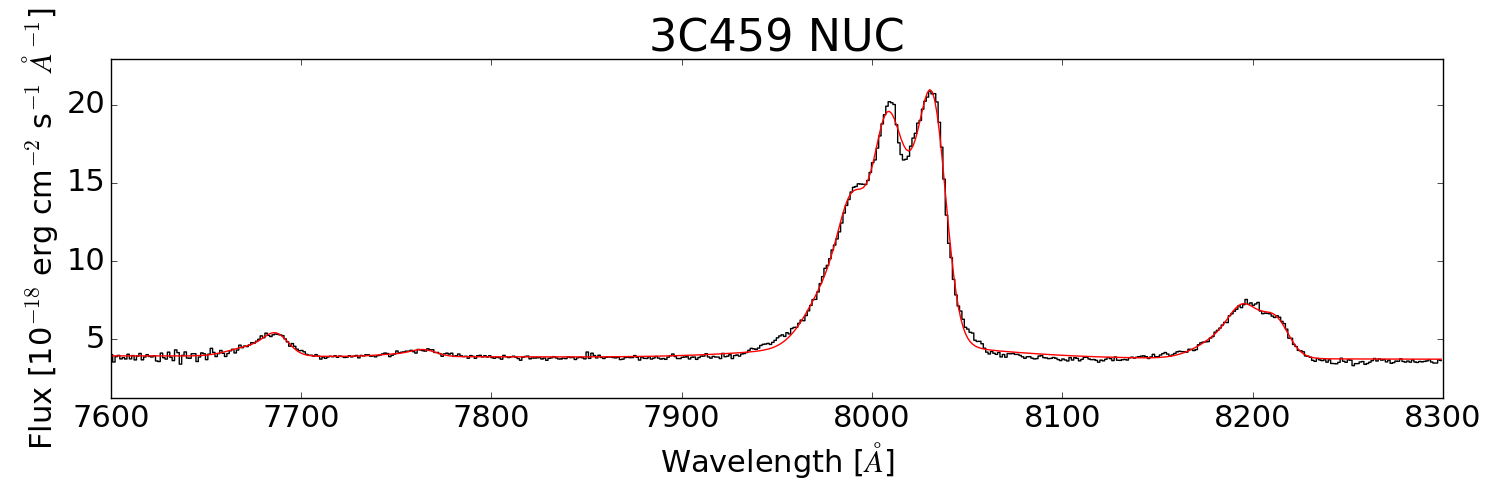}
\includegraphics[width=9cm]{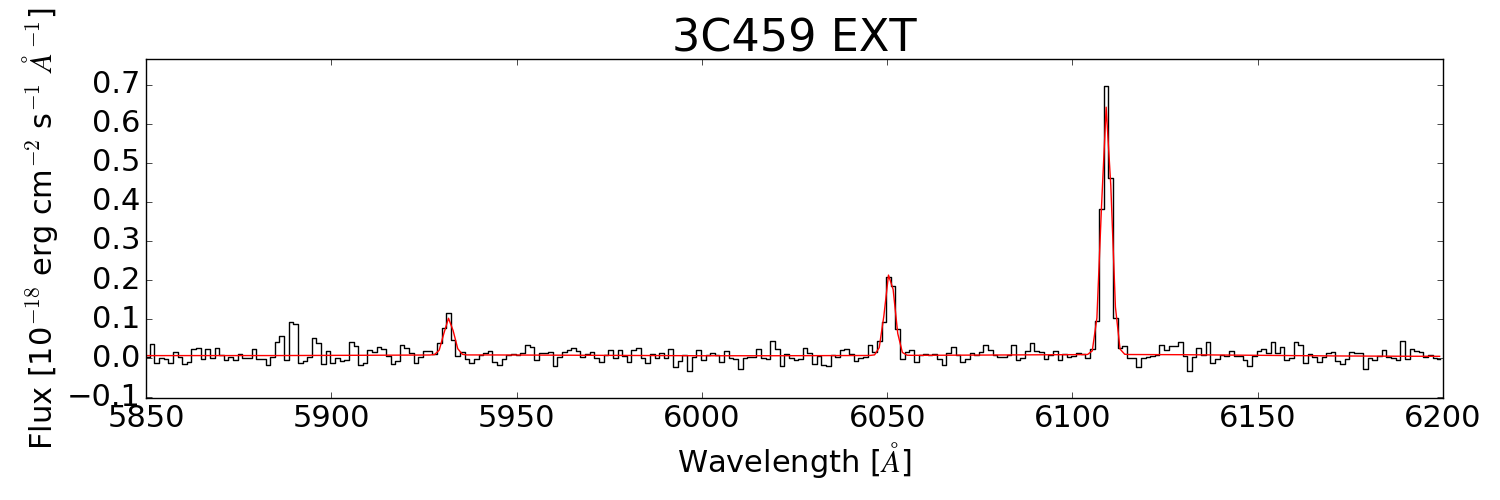}
\includegraphics[width=9cm]{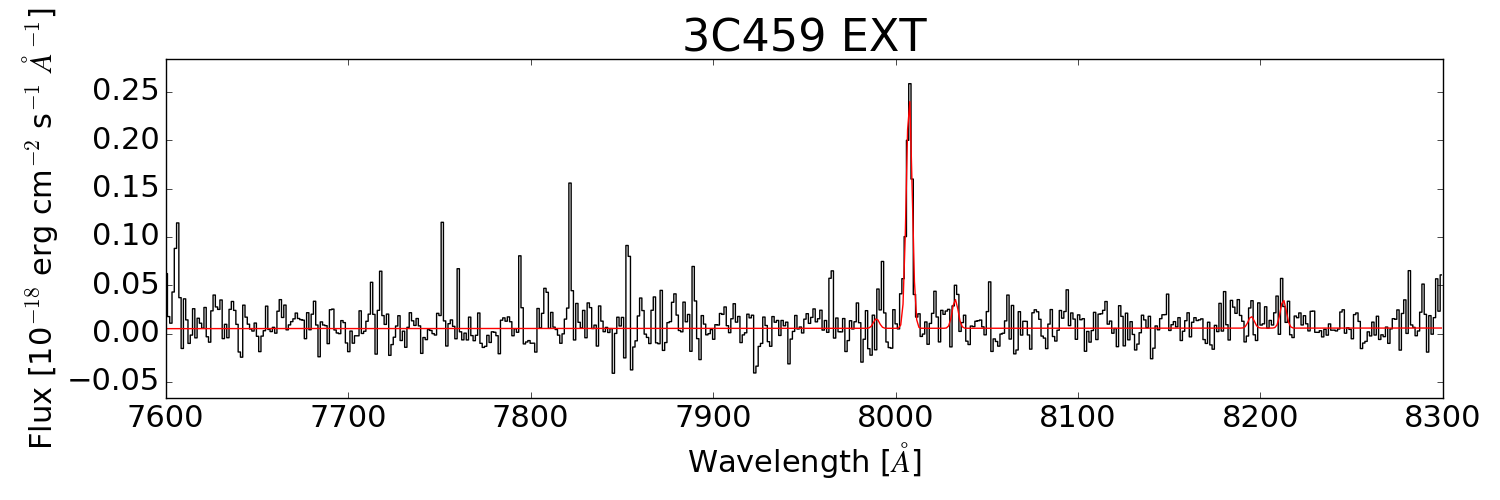}}
\caption{continued}
\end{figure*} 

\end{appendix}

\end{document}